\documentclass[11pt]{article}

\usepackage[usenames,dvipsnames,table]{xcolor}
\usepackage[utf8]{inputenc}

\pdfoutput=1 
\usepackage{jheppub} 

\usepackage{graphicx}
\usepackage{amsmath, amssymb}
\usepackage{enumitem}

\newcommand{\be}{\begin{eqnarray}}
\newcommand{\ee}{\end{eqnarray}}
\newcommand{\bea}{\begin{eqnarray}}
\newcommand{\eea}{\end{eqnarray}}

\newcommand{\nn}{\nonumber}
\newcommand{\bn}{\begin{enumerate}}
\newcommand{\en}{\end{enumerate}}

\newcommand{\PE}{\mathop{\rm PE}}

\def\Tr{\mathop{\mathrm{Tr}}\nolimits}

\def\e{\mathrm{e}}

\def\half{\frac{1}{2}}

\newcommand{\udl}[1]{\mathrm{d} #1 \,}

\newcommand{\sbfunc}[1]{s_b\left( #1\right)}

\newcommand{\Gpq}[1]{\Gamma_e\left( #1\right)}

\def\ga{\alpha}
\def\gb{\beta}

\def\Gc{\Gamma}
\def\Gd{\Delta}
\def\gd{\delta}

\def\gs{\sigma}

\def\gr{\rho}
\def\gp{\phi}

\title{4d $S$-duality wall and $SL(2,\mathbb{Z})$ relations }

\author[a,b]{Lea E.~Bottini,}
\author[c]{Chiung Hwang,}
\author[b,d]{Sara Pasquetti,}
\author[b,d]{and Matteo Sacchi}

\affiliation[a]{Mathematical Institute, University of Oxford, Woodstock Road, Oxford, OX2 6GG, United Kingdom}
\affiliation[b]{Dipartimento di Fisica, Università di Milano-Bicocca,
Piazza della Scienza 3, I-20126 Milano, Italy}
\affiliation[c]{Department of Applied Mathematics and Theoretical Physics, University of Cambridge, Cambridge CB3 0WA, United Kingdom}
\affiliation[d]{INFN, sezione di Milano-Bicocca, Piazza della Scienza 3, I-20126 Milano, Italy}

\emailAdd{lea.bottini@maths.ox.ac.uk}
\emailAdd{ch911@cam.ac.uk}
\emailAdd{sara.pasquetti@gmail.com} 
\emailAdd{m.sacchi13@campus.unimib.it}

\abstract{In this paper we present various $4d$ $\mathcal{N}=1$ dualities involving  theories obtained by gluing two $E[USp(2N)]$ blocks via the gauging of a common $USp(2N)$ symmetry with the addition of $2L$ fundamental matter chiral fields. For $L=0$ in particular the  theory has a quantum deformed moduli space with chiral symmetry breaking and  its index takes the form of a delta-function. We interpret it as the Identity wall which identifies the two surviving $USp(2N)$ of each $E[USp(2N)]$ block. All the dualities are  derived from iterative applications of the  Intriligator--Pouliot duality. This plays for us the role of the fundamental duality, from which we derive all others. We then focus on the $3d$ version of our $4d$ dualities, which now involve the $\mathcal{N}=4$ $T[SU(N)]$  quiver theory that is known to correspond to the $3d$ $S$-wall. We show how these $3d$ dualities correspond to the relations $S^2=-1$, $S^{-1}S=1$ and $T^{-1} S T=S^{-1} T S$ for the $S$ and $T$ generators of $SL(2,\mathbb{Z})$. These observations lead us to conjecture that  $E[USp(2N)]$ can also be interpreted as a $4d$ $S$-wall.
}

\begin{document} 

\maketitle
\flushbottom

\newpage
\section{Introduction}

Recently the $E_\rho^\sigma[USp(2N)]$ family of $4d$ $\mathcal{N}=1$ quiver theories labelled by $\rho$ and $\sigma$ partitions of $N$ have been introduced \cite{Hwang:2020wpd}. Upon circle compactification to $3d$ and suitable real mass deformations, these theories flow to the $3d$  $T_\rho^\sigma[SU(N)]$ family of linear $\mathcal{N}=4$ quivers introduced in \cite{Gaiotto:2008ak}.
The $E_\rho^\sigma[USp(2N)]$ theories, like their $3d$ counterparts, enjoy a mirror duality which relates pairs of theories with swapped partitions  $\rho$ and $\sigma$.

The $T_\rho^\sigma[SU(N)]$ theories can be realised on Hanany--Witten brane set-ups with D3-branes suspended between NS5 and D5-branes \cite{Hanany:1996ie}. They can also be defined as the theories  at the end of an renormalization group (RG) flow triggered by  nilpotent vacuum expectation values (VEVs), labelled by $\rho$ and $\sigma$, for the Higgs and Coulomb moment maps of the $T[SU(N)]$ theory.
The  $E_\rho^\sigma[USp(2N)]$ theories were analogously  defined by turning on  VEVs, labelled by $\rho$ and $\sigma$,   for the moment map operators of the  $E[USp(2N)]$ theory introduced in \cite{Pasquetti:2019hxf}.
The  $E[USp(2N)]$ theory has a non-abelian global symmetry group consisting of a  manifest $USp(N)_x$ and an emergent  $USp(N)_y$ factor  which are swapped by the action of mirror symmetry, and reduces to the $3d$ $T[SU(N)]$ theory, which analogously  has a manifest $SU(N)_x$ and an emergent $SU(N)_y$ acting respectively on the Higgs and Coulomb branches which are swapped by mirror symmetry \cite{Intriligator:1996ex}.

Therefore $3d$ mirror dualities belong to the very large family of $3d$ dualities which can be derived starting from $4d$  $\mathcal{N}=1$ dualities,  performing $\mathbb{S}^1$ compactifications to $3d$ and turning on various  deformations as discussed in \cite{Aharony:2013dha,Aharony:2013kma}. 

It is then natural to wonder whether other known results for  $3d$ theories have a $4d$ counterpart.
For example, $T[SU(N)]$ was identified with the $S$-duality wall \cite{Gaiotto:2008ak}, implementing the action of the $S$ element of  $SL(2,\mathbb{Z})$ and  interpolating between two copies of the $4d$ $\mathcal{N}=4$ $SU(N)$ SYM with coupling  $\tau$ and   $-\frac{1}{\tau}$.
The $T[SU(N)]$ theory is then expected to  satisfy  various relations inherited from the properties of the $SL(2,\mathbb{Z})$
 generators
$S$ and $T$ (the latter corresponding in field theory to the  insertion of a Chern-Simon (CS) coupling): $S^2=-1, S^{-1} S=1$ and $T^{-1} S T=S^{-1} T S$.
These relations have been tested  using the $\mathbb{S}^3$ partition function of the  $T[SU(N)]$ theory
 \cite{Benvenuti:2011ga,Nishioka:2011dq,Gulotta:2011si,Assel:2014awa}.
Here will  focus on whether such relations have a field theory interpretation as genuine  dualities involving the $T[SU(N)]$  theory, rather than just as matrix model identities,  and whether such dualities hold  for the $4d$ counterpart  $E[USp(2N)]$ as well.

To answer these questions we analyse the quiver theories obtained by gluing two copies of
the $E[USp(2N)]$ theory, gauging a common $USp(2N)$ symmetry group and inserting some chiral fields.

We first consider the case in which we gauge together two  $E[USp(2N)]$ theories without extra chirals and we shall denote the resulting theory by $\mathcal{T}_g$.
We interpret the resulting object as an Identity operator identifiying the two remaining $USp(2N)$ global symmetries.  Indeed we show that the index of the glued theories is proportional to a Dirac $\delta$-function
 which identifies the fugacities of the two remaining $USp(2N)$ factors. As we will see the physics behind this behaviour is a chiral symmetry breaking pattern, closely related to the familiar one of the $SU(2)$ theory with 2 flavors \cite{Seiberg:1994bz} which indeed was shown   to have an index proportional to a  $\delta$-function with support at points where chiral symmetry breaking occurs \cite{Spiridonov:2014cxa}.
Considering the $3d$ limit followed by the mass deformations, which as  we mentioned above
reduces the $E[USp(2N)]$ theory to $T[SU(N)]$, we obtain an analogous result in $3d$:
if we glue two copies of the $T[SU(N)]$ theory by gauging a common $SU(N)$ factor we get an identity operator
which identifies the two remaining $SU(N)$ symmetries. Moreover, we we will see that depending on the way the limit is taken
we can identify this relation with either the $S^2=-1$ or the $S^{-1} S=1$ properties of the  $S$-wall.

We also show how  starting from the so-called braid duality \cite{Pasquetti:2019hxf}, which involves two copies of the
$E[USp(2N)]$  theory glued with the insertion of two chirals, 
we can obtain in $3d$ a duality  related to the $T^{-1} S T=S^{-1} T S$ property. Specifically, this duality relates the gluing of two $T[SU(N)]$ tails with a CS interaction at level $-1$ to a single $T[SU(N)]$ with background CS levels $+1$ for its $SU(N)$ global symmetries and one chiral singlet in the adjoint of one of the two $SU(N)$.

It is then tempting to push further  the analogy between $3d$ and $4d$ and speculate that $E[USp(2N)]$ might play the role of an $S$-wall in $4d$. We will further investigate this possibility in an upcoming paper \cite{prl}.

In this paper we also discuss other interesting gluings of $E[USp(2N)]$ theories with some flavors in the middle.
Interestingly we are able to prove some dualities for the resulting theories, which belong to the $E_\rho^\sigma[USp(2N)]$ family. This is done by means of a procedure based on the iteration of the Intriligator--Pouliot (IP) duality \cite{Intriligator:1995ne} along the quiver, which throughout this paper  plays the role of fundamental duality. This also suggests the  possibility of deriving mirror dualities in terms of some more fundamental ones, such as IP. This will be the main result of \cite{prl}.


\section{A review of the $E[USp(2 N)]$ and $FE[USp(2N)]$ theories}
\label{sec:EUSp intro}

\begin{figure}[t]
	\centering
  	\includegraphics[scale=0.65]{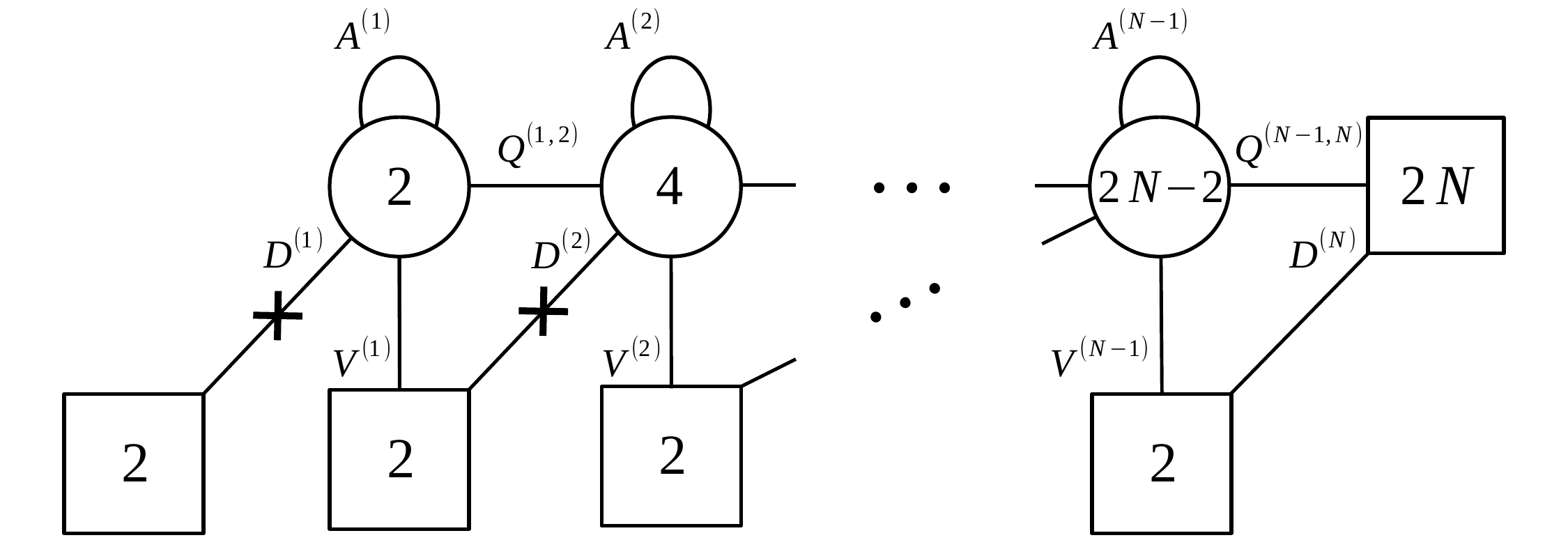} 
   \caption{The quiver diagram for $E[USp(2N)]$. Round nodes denote gauge symmetries and square nodes denote global symmetries , where the number $2 n$ insides each node represents the $USp(2n)$ group. Single lines denote chiral fields in representations of the nodes they are connecting. In particular, lines between adjacent nodes denote chiral fields in the bifundamental representation of the two nodes symmetries, while arcs denote chiral fields in the antisymmetric representation of the corresponding node symmetry. Crosses represent the singlets $\gb_n$ that flip the diagonal mesons.}
  	\label{euspfields}
\end{figure}

In this section we quickly review the $E[USp(2N)]$ theory and its dual frames. This theory was first introduced in \cite{Pasquetti:2019hxf} and later studied  in \cite{Hwang:2020wpd,Garozzo:2020pmz,Hwang:2021xyw}. We refer the reader to these references for more details, while here we will just focus on the aspects that will be relevant for us.

The $E[USp(2N)]$ theory is the $4d$ $\mathcal{N}=1$ quiver theory represented in Figure \ref{euspfields}. 
The superpotential contains a cubic coupling between the bifundamentals $Q^{(n,n+1)}$ and the antisymmetrics $A^{(n)}$, another cubic coupling between the chirals in each triangle of the quiver and finally the flip terms with the singlets $\beta_n$, denoted by cross marks in the quiver, coupled to the diagonal mesons
\begin{align}
\mathcal{W}_{E[USp(2N)]}&=\sum_{n=1}^{N-1}\Tr_{n}\left[A^{(n)}\left(\Tr_{n+1}Q^{(n,n+1)}Q^{(n,n+1)}-\Tr_{n-1}Q^{(n-1,n)}Q^{(n-1,n)}\right)\right]\nn\\
&+\sum_{n=1}^{N-1}\Tr_{y_{n+1}}\Tr_{n}\Tr_{n+1}\left(V^{(n)}Q^{(n,n+1)}D^{(n+1)}\right)+\nn\\
&+\sum_{n=1}^{N-1} \gb_n\Tr_{y_n}\Tr_{n}\left(D^{(n)}D^{(n)}\right)\, ,
\label{superpoteusp}
\end{align}
where $\Tr_n$ denotes the trace over the color indices of the $n$-th $USp(2n)$ gauge node, while $\Tr_{y_n}$ denotes the trace over the the $n$-th $SU(2)$ flavor symmetry. Notice that for $n=N$ we have the trace over the $USp(2N)_x$ flavor symmetry, which we will also denote by $\Tr_N=\Tr_x$. All the traces are defined including the $J$ antisymmetric tensor of $USp(2n)$
\be
J=\mathbb{I}_n\otimes i\,\gs_2\, .
\ee

\begin{figure}[t]
	\centering
  	\includegraphics[scale=0.65]{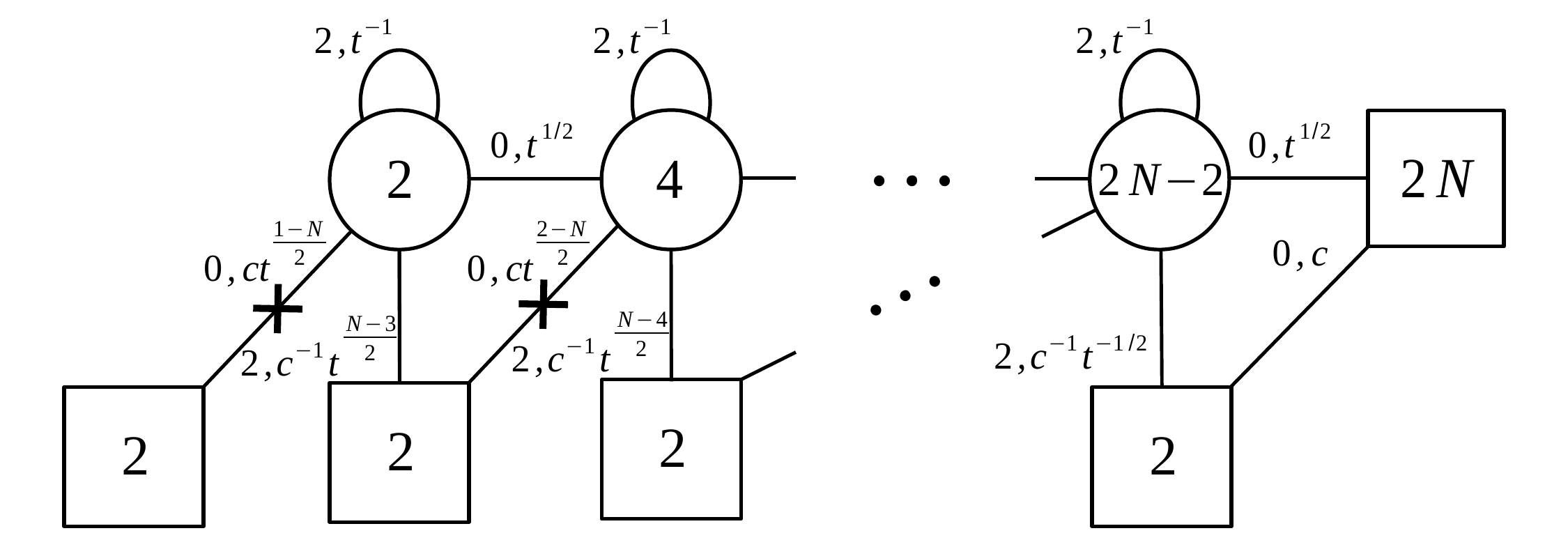} 
   \caption{Trial R-charges and charges under the abelian symmetries. The power of $c$ is the charge under $U(1)_c$, while the power of $t$ is the charge under $U(1)_t$.}
  	\label{euspfugacities}
\end{figure}

The manifest global symmetry 
\be
USp(2N)_x\times\prod_{n=1}^NSU(2)_{y_n}\times U(1)_t\times U(1)_c
\ee
is enhanced in the IR to
\be
USp(2N)_x\times USp(2N)_y\times U(1)_t\times U(1)_c\, .
\ee
This can be argued, as shown in \cite{Pasquetti:2019hxf}, in various ways, for example by checking that the gauge invariant operators form representations of the enhanced symmetry and that the expanded supersymmetric index forms characters of this symmetry. Another way to understand the enhancement is by means of a self-duality that we will review momentarily.
The charges of all the chiral fields under the two $U(1)$ symmetries as well as their trial R-charges in our conventions are summarized in Figure \ref{euspfugacities}. 

\begin{table}[t]
\centering
\scalebox{1}{
\begin{tabular}{|c|cccc|c|}\hline
{} & $USp(2N)_x$ & $USp(2N)_y$ & $U(1)_t$ & $U(1)_c$ & $U(1)_{R_0}$ \\ \hline
$\mathsf{H}$ & ${\bf N(2N-1)-1}$ & $\bf 1$ & $1$ & 0 & 0 \\
$\mathsf{C}$ & $\bf1$ & ${\bf N(2N-1)-1}$ & $-1$ & 0 & 2 \\
$\Pi$ & $\bf N$ & $\bf N$ & 0 & $+1$ & 0 \\
$B_{nm}$ & $\bf1$ & $\bf1$ & $m-n$ & $-2$ & $2n$ \\ \hline
\end{tabular}}
\caption{Trasnformation rules of the $E[USp(2N)]$ operators.}
\label{eusptable}
\end{table}

The gauge invariant operators of $E[USp(2N)]$ that will be important for us are of three main types. Here we just review their properties under the global symmetry, while we refer the reader to  \cite{Pasquetti:2019hxf} for their explicit construction:
\begin{itemize}
\item  two operators, which we denote by $\mathsf{H}$ and $\mathsf{C}$, in the traceless antisymmetric representation of $USp(2N)_x$ and $USp(2N)_y$ respectively;
\item an operator $\Pi$ in the bifundamental representation of $USp(2N)_x\times USp(2N)_y$;
\item some gauge invariant operators that are also singlets under the non-abelian global symmetries and are only charged under $U(1)_c$ and $U(1)_t$, which include the singlets $\gb_n$ and which are collectively denoted by $B_{nm}$.
\end{itemize}
%
%
%
%
%
%
%
The charges and representations of all these operators under the global symmetry are given in Table \ref{eusptable}.

In \cite{Pasquetti:2019hxf} it was shown that $E[USp(2N)]$ has a limit to the $T[SU(N)]$ theory \cite{Gaiotto:2008ak}. This limit consists of three main steps. The first one is a dimensional reduction on $\mathbb{S}^1$ so to get a $3d$ $\mathcal{N}=2$ quiver gauge theory that is identical to $E[USp(2N)]$, but with extra superpotential terms containing monopole operators \cite{Aharony:2013dha,Aharony:2013kma}. The second step is a Coulomb branch VEV that higgses the gauge groups from $USp(2n)$ to $U(n)$. One should also simultaneously give some compensating real mass deformations to keep part of the matter fields massless. The result is the $M[SU(N)]$ theory of \cite{Pasquetti:2019tix}. In this latter reference it was then shown that a further real mass deformation for the $U(1)_c$ symmetry, under which only the fields of the saw are charged, makes $M[SU(N)]$ flow to $T[SU(N)]$. In Section \ref{3dlimits} we give more details on these limits, while here we just mention that, among the operators of $E[USp(2N)]$, $\Pi$ and $B_{nm}$ become massive, while the traceless antisymmetric operators $\mathsf{H}$, $\mathsf{C}$ reduce to the moment map operators of $T[SU(N)]$. 

Our main computational tool will be the supersymmetric index \cite{Romelsberger:2005eg,Kinney:2005ej,Dolan:2008qi} (see also \cite{Rastelli:2016tbz} for a review and Appendix \ref{S3S1pf} for our conventions) of the $E[USp(2N)]$ theory. This is a function of the fugacities $x_n$, $y_n$, $t$ and $c$ in the Cartan of the global symmetry that can be expressed with the following recursive definition:
\begin{equation}
\makebox[\linewidth][c]{\scalebox{1}{$
\begin{split}
&\mathcal{I}_{E[USp(2N)]}(\vec x;\vec y;t;c)=  \\
&=\Gpq{pq\,c^{-2}t}\prod_{n=1}^N\Gpq{c\,y_N^{\pm1}x_n^{\pm1}}\oint\udl{\vec{z}_{N-1}^{(N-1)}} \Gd_{N-1}(\vec z_{N-1}^{(N-1)};pq/t) \prod_{i=1}^{N-1}\frac{\prod_{n=1}^N\Gpq{t^{1/2}z^{(N-1)}_i{}^{\pm1}x_n^{\pm1}}}{\Gpq{t^{1/2}c\,y_N^{\pm1}z^{(N-1)}_i{}^{\pm1}}} \\
&\qquad \times  \mathcal{I}_{E[USp(2(N-1))]}\left(z^{(N-1)}_1,\cdots,z^{(N-1)}_{N-1};y_1,\cdots,y_{N-1};t;t^{-1/2}c\right)\, ,
\label{indexEN}
\end{split}$}}
\end{equation}
with the base of the iteration defined as
\be
\mathcal{I}_{E[USp(2)]}(x;y;c) =\Gpq{c\,y^{\pm1}x^{\pm1}}\,.
\ee
We also defined the integration measure of the $m$-th $USp(2n)$ gauge node as
\be
\udl{\vec{z}^{(m)}_n}=\frac{1}{2^n n!}\prod_{i=1}^n\frac{\udl{z^{(m)}_i}}{2\pi i\,z^{(m)}_i}
\ee
and the contribution of the $USp(2n)$ vector and antisymmetric chiral multiplets as
\begin{align}
\label{eq:measure}
\Gd_n(\vec{z}_n^{(m)};pq/t) = \frac{\left[(p;p)_\infty (q;q)_\infty\right]^n \Gpq{pq\,t^{-1}}^n\prod_{i<j}^n\Gpq{pq t^{-1} \,z_i^{(m)}{}^{\pm1}z_j^{(m)}{}^{\pm1}}}{\prod_{i=1}^n\Gpq{z_i^{(m)}{}^{\pm2}}\prod_{i<j}^n\Gpq{z_i^{(m)}{}^{\pm1}z_j^{(m)}{}^{\pm1}}} \, .
\end{align}
As pointed out in \cite{Pasquetti:2019hxf}, the expression \eqref{indexEN} coincides up to some prefactor corresponding to singlet fields with the interpolation kernel $\mathcal{K}_c(x,y)$ studied in \cite{2014arXiv1408.0305R}, where many integral identities for this function were proven. These are naturally interpreted as dualities for $E[USp(2N)]$, which we are now going to review.

\begin{figure}[t]
	\centering
	\includegraphics[scale=1.1]{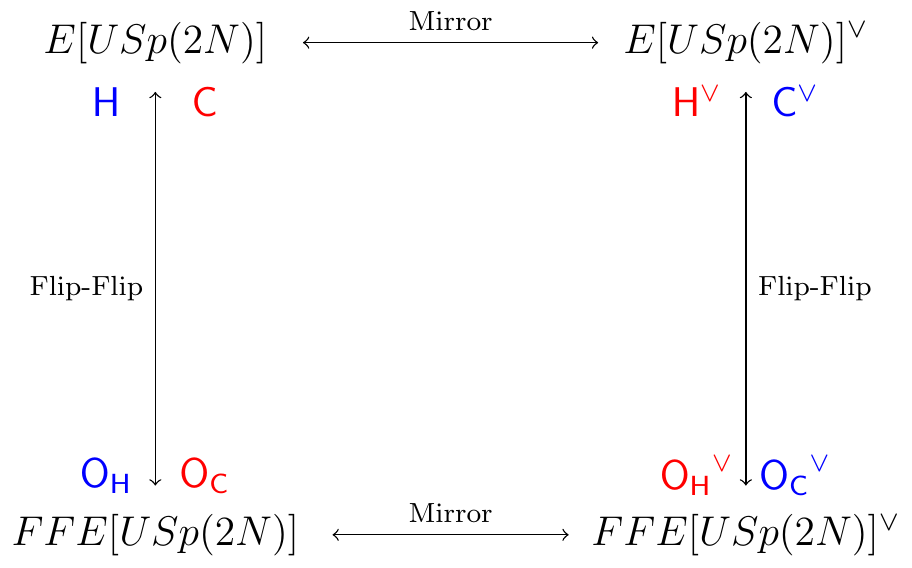}
	\caption{Duality web of the $E[USp(2N)]$ theory. On the horizontal direction we have the mirror-like duality, while on the vertical direction we have the flip-flip duality. Operators of the same color are mapped to each other across the dualities.}
	\label{euspweb}
\end{figure}

$E[USp(2N)]$ enjoys various dualities that constitute a commutative diagram schematically represented in Figure \ref{euspweb}.
First of all, we have the mirror  dual frame $E[USp(2N)]^\vee$ where the $USp(2N)_x$ and $USp(2N)_y$ symmetries are exchanged and the $U(1)_t$ fugacity is mapped to
\be
t\rightarrow\frac{pq}{t}\,.
\label{redeft}
\ee
Under this duality, $E[USp(2N)]$ is self-dual with a non-trivial map of the gauge invariant operators
\be
\mathsf{H}\quad&\leftrightarrow&\quad \mathsf{C}^\vee\nn\\
\mathsf{C}\quad&\leftrightarrow&\quad \mathsf{H}^\vee\nn\\
\Pi\quad&\leftrightarrow&\quad \Pi^\vee\nn\\
B_{nm}\quad&\leftrightarrow&\quad B_{mn}^\vee\, .
\label{opmap4dmirror}
\ee
At the level of the index we have the following identity:
\be
\mathcal{I}_{E[USp(2N)]}(\vec x;\vec y;t,c)=\mathcal{I}_{E[USp(2N)]}(\vec y;\vec x;p q/t,c)\, ,
\label{selfduality}
\ee
which has been proven in Theorem 3.1 of \cite{2014arXiv1408.0305R}. This duality reduces in the $3d$ limit to the known self-duality under mirror symmetry of $T[SU(N)]$.

The second duality that forms the diagram is called the flip-flip duality. The flip-flip dual frame $FFE[USp(2N)]$ is defined as $E[USp(2N)]$ plus two sets of singlets $\mathsf{O_H}$ and $\mathsf{O_C}$ flipping the two operators $\mathsf{H}^{FF}$ and $\mathsf{C}^{FF}$
\be
\mathcal{W}_{FFE[USp(2N)]}=\mathcal{W}_{E[USp(2N)]}+\Tr_x\left(\mathsf{O_H}\mathsf{H}^{FF}\right)+\Tr_y\left(\mathsf{O_C}\mathsf{C}^{FF}\right)\, .
\ee
In this case the $USp(2N)_x$ and $USp(2N)_y$ symmetries are left unchanged, while only the $U(1)_t$ fugacity transforms as in \eqref{redeft}. The operator map is indeed
\be
\mathsf{H}\quad&\leftrightarrow&\quad \mathsf{O}_{\mathsf{H}}\nn\\
\mathsf{C}\quad&\leftrightarrow&\quad \mathsf{O}_{\mathsf{C}}\nn\\
\Pi\quad&\leftrightarrow&\quad\Pi^{FF}\nn\\
B_{nm}\quad&\leftrightarrow&\quad B^{FF}_{mn}\, .
\ee
The  flip-flip dual frame can be reached by iteratively applying the Intriligator--Pouliot duality \cite{Intriligator:1995ne} by means of an iterative procedure as shown in \cite{Hwang:2020wpd}.
At the level of the supersymmetric index, the flip-flip duality is encoded in the following integral identity:
\begin{align}
\mathcal{I}_{E[USp(2N)]}(\vec x;\vec y;t;c)&=\prod_{n<m}^N\Gpq{t x_n^{\pm1}x_m^{\pm1}}\Gpq{p q t^{-1} y_n^{\pm1}y_m^{\pm1}}\mathcal{I}_{E[USp(2N)]}(\vec x;\vec y;p q/t;c)\,,
\label{flipflipselfduality}
\end{align}
which is proven in Proposition 3.5 of \cite{2014arXiv1408.0305R}. This duality reduces in the $3d$ limit to the flip-flip duality of $T[SU(N)]$ discussed in \cite{Aprile:2018oau}, which can also be derived by iteratively applying a more fundamental duality, in this case the Aharony duality \cite{Aharony:1997gp}, as shown in \cite{Hwang:2020wpd} (see also Appendix B of \cite{Giacomelli:2020ryy}).

For later convenience we also introduce a variant of the $E[USp(2N)]$ theory. We call this $FE[USp(2N)]$ theory since it is defined as $E[USp(2N)]$ with one extra set of singlets $\mathsf{O}_\mathsf{H}$, as well as a singlet $\beta_N$, interacting via the superpotential
\be\label{eq:FE}
\mathcal{W}_{FE[USp(2N)]}=\mathcal{W}_{E[USp(2N)]}+\Tr_x\left(\mathsf{O}_\mathsf{H}  \mathsf{H}\right) + \beta_N \Tr_x \Tr_{y_N} D^{(N)} D^{(N)} \,.
\ee
This theory has a self-dual frame which we call $FE[USp(2N)]^\vee$, that can be understood as a consequence of the combined mirror and flip-flip dualities of $E[USp(2N)]$. Across this duality the $USp(2N)_x$ and $USp(2N)_y$ symmetries are exchanged, while $U(1)_t$ and $U(1)_c$ are left unchanged. 
At the level of the supersymmetric index, this is encoded in the following integral identity:
\be
\mathcal{I}_{FE[USp(2N)]}(\vec x;\vec y;t;c)=\mathcal{I}_{FE[USp(2N)]}(\vec y;\vec x;t;c)\,,
\label{flipselfduality}
\ee
which can be easily derived from eqs.~\eqref{selfduality} and \eqref{flipflipselfduality}.
 We will refer to this as \emph{spectral duality}, since in the $3d$ limit it reduces to the spectral duality of $FT[SU(N)]$ discussed in \cite{Aprile:2018oau}.

The index of $FE[USp(2N)]$ is defined recursively as
\begin{align}
&\mathcal{I}_{FE[USp(2N)]}(\vec x;\vec y;t;c)= \nn \\
&=\Gpq{pq\,c^{-2}}\Gpq{pq\,t^{-1}}^N\prod_{n<m}^N\Gpq{pq\,t^{-1}x_n^{\pm1}x_m^{\pm1}}\prod_{n=1}^N\Gpq{c\,y_N^{\pm1}x_n^{\pm1}}\nn\\
&\times\oint\udl{\vec{z}_{N-1}^{(N-1)}} \Gd_{N-1}(\vec z_{N-1}^{(N-1)}) \prod_{i=1}^{N-1}\frac{\prod_{n=1}^N\Gpq{t^{1/2}z^{(N-1)}_i{}^{\pm1}x_n^{\pm1}}}{\Gpq{t^{1/2}c\,y_N^{\pm1}z^{(N-1)}_i{}^{\pm1}}}\nn \\
&\times  \mathcal{I}_{FE[USp(2(N-1))]}\left(z^{(N-1)}_1,\cdots,z^{(N-1)}_{N-1};y_1,\cdots,y_{N-1};t;t^{-1/2}c\right)\, ,
\label{indexFEN}
\end{align}
where now the base of the iteration is
\be
\mathcal{I}_{FE[USp(2)]}(x;y;t;c)=\Gpq{pq\,c^{-2}}\Gpq{pq\,t^{-1}}\Gpq{c\,y^{\pm1}x^{\pm1}}
\ee
and $\Gd_n(\vec{z}_n^{(m)})$ contains the contribution of the vector only and not the one of the antisymmetric chiral
\begin{align}
\label{eq:measurenoasymm}
\Gd_n(\vec{z}_n^{(m)}) = \frac{\left[(p;p)(q;q)\right]^n}{\prod_{i=1}^n\Gpq{z_i^{(m)}{}^{\pm2}}\prod_{i<j}^n\Gpq{z_i^{(m)}{}^{\pm1}z_j^{(m)}{}^{\pm1}}} \, .
\end{align}
It simply relates to the index of $E[USp(2 N)]$ as follows:
\begin{align}
\label{eq:FEvsE}
\mathcal{I}_{FE[USp(2N)]}(\vec x;\vec y;t;c) = \Gpq{pq\,c^{-2}}\Gpq{pq\,t^{-1}}^N\prod_{n<m}^N\Gpq{pq\,t^{-1}x_n^{\pm1}x_m^{\pm1}} \mathcal{I}_{E[USp(2N)]}(\vec x;\vec y;t;c) \,.
\end{align}

\section{Gluing $S$-walls without matter: the Identity wall}
\label{sec:delta}
\subsection{Gluing $E[U Sp(2N )]$ theories: the Delta-function property}
\label{sec:delta sub1}

\begin{figure}[t]
\centering
\includegraphics[width=1\textwidth]{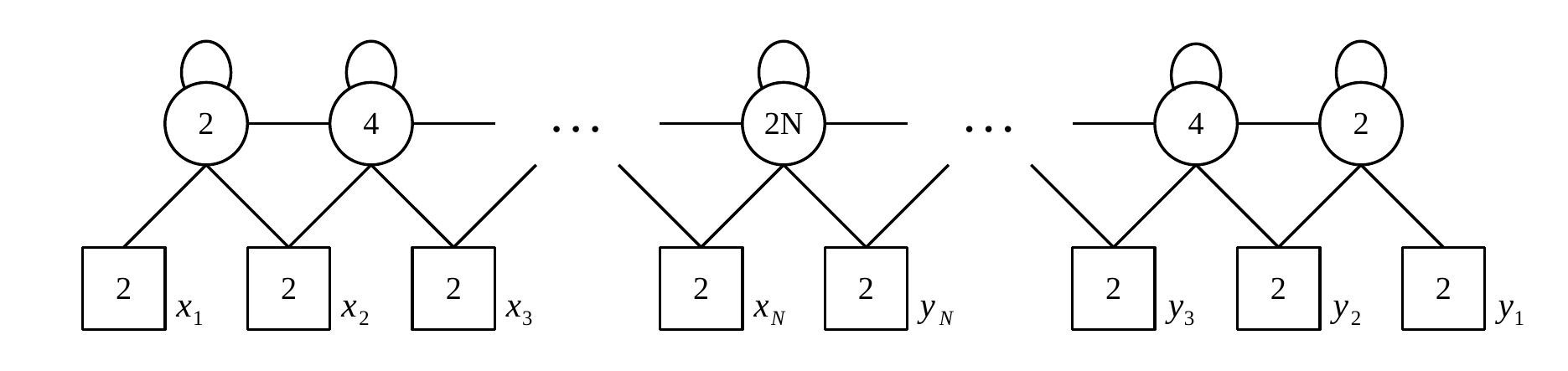}
\caption{The theory $\mathcal{T}_g$, obtained by the Lagrangian gluing of two $E[USp(2N)]$ theories. }
\label{fig:delta1}
\end{figure}

In this section we study the gluing of  two $E[USp(2N)]$ theories corresponding to commonly gauging a diagonal combination of one $USp(2N)$ symmetry. 
We begin considering the theory $\mathcal{T}_g$, obtained by
gauging a diagonal combination of the two manifest $USp(2N)$ symmetries of each $E[USp(2N)]$ block as in Figure \ref{fig:delta1}.
The gauging is done adding an antisymmetric chiral field $A$ as well as two singlets $\gb_N^{L/R}$ and a superpotential interaction of the form (the rest of the superpotential is simply the sum of those of the two $E[USp(2N)]$ theories)
\be
\label{eq:delta sup}
\gd\mathcal{W}=\Tr_z\left[ A\cdot\left(\mathsf{H}^L-\mathsf{H}^R\right)\right]+\gb_N^L\Tr_z\left[D_L^{(N)}D_L^{(L)}\right]+\gb_N^R\Tr_z\left[D_R^{(N)}D_R^{(L)}\right]\,,
\ee
where the indices $L/R$ distinguish the field and operators of the left and right $E[USp(2N)]$ blocks, for example $\mathsf{H}^L$ and $\mathsf{H}^R$ denote the $\mathsf{H}$ operators in the antisymmetric representation of their manifest $USp(2N)$ symmetries which we identify and call $USp(2N)_z$, and $\mathrm{Tr}_z$ is taken over such gauged $USp(2 N)_z$. This superpotential implies that the $U(1)_t$ symmetries of the two blocks are identified. Moreover, one can check that of the two $U(1)_c$ symmetries, one coming from each block, the diagonal combination is anomalous because of the $USp(2N)_z$ gauging, while only the anti-diagonal combination survives. We claim that the supersymmetric index of this theory satisfies the following remarkable property:
\begin{align}
\label{4ddelta}
\mathcal{I}^{N}_{g} &= \Gpq{pq c^{\pm 2}} \oint\udl{\vec{z}_N}\Gd_N(\vec{z};pq/t)\mathcal{I}_{E[USp(2N)]}(\vec{z};\vec{x};t;c)\mathcal{I}_{E[USp(2N)]}(\vec{z};\vec{y};t;c^{-1}) \nonumber \\
&= \frac{\prod_{j=1}^N 2\pi i x_j}{\Gd_N(\vec x;t)} \sum_{\sigma \in S_N} \sum_\pm \prod_{i=1}^N \delta\left(x_i-y_{\sigma(i)}^{\pm1}\right) \equiv {}_{\vec{x}}\hat{\mathbb{I}}_{\vec{y}}(t) \,.
\end{align}
Note that the summation $\sum_{\gs\in S_N} \sum_{\pm}$ spans the Weyl group of $USp(2 N)$ and  ${}_{\vec{x}}\hat{\mathbb{I}}_{\vec{y}}(t)$, the Identity operator, which identifies the Cartans of the remaining $USp(2 N)_x$ and 
$USp(2 N)_y$ symmetries,  is a normalised delta-function defined in such a way that
\begin{align}
\oint \udl{\vec z_N} \Gd_N(\vec z;t)\, {}_{\vec{z}}\hat{\mathbb{I}}_{\vec{y}}(t) = 1 \,.
\end{align}

In Subsection \ref{deltalimit} we will show how, starting from this result, we can prove that  gluing two copies of the $T[SU(N)]$ theory by gauging a diagonal $SU(N)$ symmetry we obtain the Identity operator identifying the two remaining $SU(N)$ symmetries, as expected from the identification of the $T[SU(N)]$ theory with the $S$ generator of $SL(2,\mathbb{Z})$.

%

We observe that eq.~\eqref{4ddelta} can also be understood as a limit of the identity associated with the braid duality \cite{Pasquetti:2019hxf}, which we will review in Subsection \ref{3dlimitbraid},
\be
&&\Gpq{pq c^{-2}} \Gpq{pq d^{-2}} \times\nonumber \\
&&\times \oint\udl{\vec{z}_N} \Gd_N(\vec{z};pq/t) \mathcal{I}_{E[USp(2N)]}(\vec{z};\vec{x};t;c)\mathcal{I}_{E[USp(2N)]}(\vec{z};\vec{y};t;d)\prod_{i=1}^N\Gpq{(pq)^{\frac{1}{2}}c^{-1}d^{-1}v^{\pm1}z_i^{\pm1}}=\nn\\
&&=\frac{\prod_{i=1}^N\Gpq{(pq)^{\frac{1}{2}}d^{-1}v^{\pm1}x_i^{\pm1}}\Gpq{(pq)^{\frac{1}{2}}c^{-1}v^{\pm1}y_i^{\pm1}}}{\Gpq{c^2 d^2} \Gpq{t}^N\prod_{i<j}^N\Gpq{t\,x_i^{\pm1}x_j^{\pm1}}}\mathcal{I}_{E[USp(2N)]}(\vec{x};\vec{y};t;cd)\, ,
\label{braidrelation}
\ee
which also appeared in Proposition 2.12 of \cite{2014arXiv1408.0305R}.
Setting $cd\to1$ on the l.h.s.~of \eqref{braidrelation} removes the $SU(2)_v$ doublets giving the l.h.s.~of \eqref{4ddelta}. We can then try to understand the effect of this limit on the r.h.s.~of the braid relation. We expect the factor $\mathcal{I}_{E[USp(2N)]}(\vec{x};\vec{y};t;cd)$ to become singular and give the delta appearing on the r.h.s.~of \eqref{4ddelta}. 
In fact, according to Theorem 3.7 of \cite{2014arXiv1408.0305R}, the following ratio:
\be
\frac{\mathcal{I}_{E[USp(2N)]}(\vec{x};\vec{y};t;cd)}{\prod_{i<j}^N\Gpq{pq\,t^{-1}y_i^{\pm1}y_j^{\pm1}}\prod_{i,j=1}^N\Gpq{cd\,x_i^{\pm1}y_j^{\pm1}}}
\ee
is a holomorphic function in the variables $x_i$ and $y_j$ for fixed $t$, $c$, $q$ and $p$. This means that the only poles of the index of $E[USp(2N)]$ w.r.t.~the $USp(2N)$ fugacities are the same as those of the combination of gamma-functions
\be
\prod_{i<j}^N\Gpq{pq\,t^{-1}y_i^{\pm1}y_j^{\pm1}}\prod_{i,j=1}^N\Gpq{cd\,x_i^{\pm1}y_j^{\pm1}}\,.
\ee
In particular, in the limit $cd\to 1$ the pairs of poles coming from the second product collide to the points $x_i=y_j^{\pm1}$. This implies that if we integrate the index of $E[USp(2N)]$ over the variables $x_i$ against a test function $f(\vec{x})$, the colliding poles pinch the integration contour and we should take the corresponding residues as discussed in \cite{Gaiotto:2012xa}. The total number of residues for the $N$-dimensional integral is $2^N N!$. Indeed, we can think of integrating over, say, $x_1$ first and take the $2 N$ residues at $x_1=y_j^\pm$ for $j=1,\cdots,N$. For each fixed $j$, the contribution of the vector multiplet, which should always be included in the test function $f(\vec{x})$, evaluated at the pole cancels the poles $x_i=y_j^\pm$ of the remaining $N-1$ variables $x_i$ for $i=2,\cdots,N$. Hence, for the second integration over $x_2$ we only have $2 (N-1)$ poles. Iterating this procedure, we get the $2^N N!$ poles $x_i=y_{\gs(i)}^\pm$ for $i=1,\cdots,N$ and $\gs\in S_N$. This explains the summation over the Weyl group of $USp(2N)$ on the r.h.s.~of \eqref{4ddelta}.  In the following, we show that the residue at these poles gives the delta-functions.

\subsection{Proof strategy and interpretation}\label{psi}

Our proof of the gluing property \eqref{4ddelta}  is based on two key observations.

The first one  is the  fact that  the $SU(2)$ theory  with $2$ flavors ($4$ fundamental chirals) 
has a quantum deformed moduli space with chiral symmetry breaking  
and its index is a distribution  acting on {\it test} theories:
\be
&&\oint\udl{z_1}\frac{\prod_{i=1}^4\Gpq{s_iz^{\pm1}}}{\Gpq{z^{\pm2}}}=\nn\\
&&=\frac{1}{(p;p)_\infty (q;q)_\infty}\left[\Gpq{s_1^{\pm1}s_2^{\pm1}}\left(\gd(\gp_1+\gp_3)+\gd(\gp_1+\gp_4)\right)+\Gpq{s_2^{\pm1}s_3^{\pm1}}\gd(\gp_1+\gp_2)\right]\,,\nn\\
\label{funddelta}
\ee
where we defined $s_i=\e^{2\pi i\gp_i}$, $0\le \gp_i<1$ and  $\prod_{i=1}^4s_i=1$.
This  neat result was obtained in \cite{Spiridonov:2014cxa}, starting from the $SU(2)$ theory with 6 fundamentals which is s-confining  and  Seiberg dual to a WZ model and
turning on a mass for  two chirals  to flow to the  theory with $4$ chirals.
By carefully studying the effect of the mass deformation in the integral identity  corresponding to the s-confining Seiberg duality, the authors of \cite{Spiridonov:2014cxa} argued that the index of the $SU(2)$ theory  with $4$ chirals  is a distribution with support 
at special values of the fugacities corresponding to the points in the moduli space where chiral symmetry breaking occurs.

We can also understand this result as follows.
The moduli space of the $SU(2)$ gauge theory with 4 flavors is parametrized by 
$V_{ij} = \mathrm{Tr} \, Q_{[i} Q_{j]}$ which  satisfies   
\begin{align}
\mathrm{Pf} \, V = \Lambda^4
\end{align}
with $\Lambda$ the strong coupling scale \cite{Seiberg:1994bz}. Since the moduli space doesn't include the origin $V_{ij} = 0$, the $SU(4)$ flavor symmetry is always broken.
If we try to gauge a subgroup of the $SU(4)$ flavor symmetry, say $SU(2)_x$, so to form a new quiver gauge theory, the gauged $SU(2)_x$ is Higgsed by the non-zero VEV of $V_{ij}$. This is the field theory interpretation of the result \eqref{funddelta} of \cite{Spiridonov:2014cxa}. 

The second key observation is the fact that by iterative applications of the Intriligator-Pouliot (IP)   duality \footnote{The IP duality relates $USp(2N_c)$ with $2N_f$ fundamental chirals and no superpotential  to  $USp(2N_f-2N_c-4)$ with $2N_f$ fundamental chirals and an antisymmetric matrix $X^{ab}$ of $N_f(2N_f-1)$ singlets  flipping the dual mesons $\mathcal{W}=X^{ab}q_a q_b$ \cite{Intriligator:1995ne}. See also Appendix \ref{IP}.} we can split the quiver $\mathcal{T}_g$ in Figure \ref{fig:delta1} into a product of  $SU(2)$ theories with $2$ flavors plus extra singlets.
 The steps of the derivation are then as follows:
\begin{itemize}
\item We pick one of the  $USp(2)$ gauge nodes at the two ends of the quiver. The antisymmetric chiral is just a singlet so we can apply the IP duality to this node. The dual gauge node is still a $USp(2)$ node and  the antisymmetric of the adjacent $USp(4)$ node becomes massive. Since now also the $USp(4)$ node has no antisymmetric we can apply the IP duality to it and we can iterate the procedure. Indeed at each application of the duality the antisymmetric chiral field of the adjacent node becomes massive.

\item The ranks of the gauge groups are left invariant by the action of the dualisations until we reach the middle $USp(2N)$ gauge node. This node sees $2\times 2(N-1)+4=4N$ fundamentals, so its rank is decreased to $USp(2N-4)$. The subsequent $USp(2N-2)$ node will now see $2\times 2(N-2)+4=4N-4$ fundamentals, so its rank is also decreased to $USp(2N-6)$ after the dualization.

\item We keep going towards the other end of the tail with the dualisations, each time decreasing the ranks of the groups of the second half of the quiver by two units. When we reach the $USp(4)$ node this confines since it only sees 8 fundamentals. The quiver then splits into two parts: one is the gluing of two $E[USp(2(N-1))]$ blocks 
and the other is an  $USp(2)$  theory with $4$ fundamentals and some singletes yielding a one-dimensional delta-function according to \eqref{funddelta}.

\item We then iterate this procedure. At the $n$-th iteration we will have the gluing of two $E[USp(2(N-n))]$ blocks plus $n$ copies of $SU(2)$ with $4$ fundamentals, which will give the product of $n$ one-dimensional delta functions. After the $N$-th iteration we will be left only with $N$ copies of $SU(2)$ with $4$ chirals, which will give us the $N$-dimensional delta-function \eqref{4ddelta} that identifies the two $USp(2N)$ global symmetries, as expected.
\end{itemize}

In the following we will explicitly apply this procedure at the level of the supersymmetric index so to prove \eqref{4ddelta} for the simplest cases of $N=1,2$. One caveat is that, as we will see, this procedure only captures the residues corresponding to one particular representative of $S_N \in W_{USp(2N)}$, which for definiteness we take to be $x_i=y_{\sigma(i)}^{\pm1} = y_i^{\pm1}$. Nevertheless, this procedure indeed gives us the correct value of the residue at this particular set of poles. Furthermore, the full set of poles can be correctly obtained by a slightly refined argument using a regularized version of the integral on the l.h.s.~of \eqref{4ddelta}, as we show in detail in Appendix \ref{sec:induction} for arbitrary $N$.

Our identity \eqref{4ddelta}  for the quiver theory $\mathcal{T}_g$  with  global symmetry $USp(2 N)_x \times USp(2 N)_y \times U(1)_c \times U(1)_t$
can be considered as the generalization of \eqref{funddelta} for the $SU(2)$ theory with $4$ chirals, to which indeed it reduces for $N=1$ (modulo some singlets).
The moduli space of $\mathcal{T}_g$ is parameterized by various gauge invariant chiral operators. Among these, we have a gauge invariant operator $\Tr_z\,\Pi_L\Pi_R$ in the bifundamental representation of $USp(2 N)_x \times USp(2 N)_y$,
where $\Pi_L$ and $\Pi_R$ are bifundamental operators between $USp(2 N)_z \times USp(2 N)_x$ and between $USp(2 N)_z \times USp(2 N)_y$ respectively (see Table \ref{eusptable}). As in the $N = 1$ case, we expect that the moduli spaced is deformed at the quantum level by
\begin{align}\label{vevpipi}
\left<\mathrm{Tr}_z \, \Pi_L \Pi_R\right> \neq 0\,,
\end{align}
which leads to the breaking of $USp(2 N)_x \times USp(2 N)_y$ to its diagonal subgroup. Therefore, if either $USp(2 N)_x$ or $USp(2 N)_y$ is gauged, it should be Higgsed by this non-zero VEV, which exactly corresponds to our delta function identity \eqref{4ddelta}.

\subsubsection{Explicit computation for $N=1$}

\label{sec:N=1}

For $N=1$ the $E[USp(2)]$ theory is just a WZ model of an $SU(2)\times SU(2)$. Hence, the $\mathcal{T}_g$ theory for $N=1$ is just an $SU(2)$ gauge theory with $4$ chirals and some flipping fields. Explicitly, its supersymmetric index is
\be
\mathcal{I}_{g}^{N=1}=\Gpq{pq\,c^{\pm2}}\Gpq{pq\,t^{-1}}\oint\udl{z_1} \Delta_1 (z)\Gpq{c\,x^{\pm1}z^{\pm1}}\Gpq{c^{-1}y^{\pm1}z^{\pm1}}\,.
\ee
The identity \eqref{4ddelta} is then just a direct application of \eqref{funddelta} in this case. Specifically, we apply it identifying
\be
\vec{s}=(cx,cx^{-1},c^{-1}y,c^{-1}y^{-1})\,.
\ee
This gives
\be
\mathcal{I}_{g}^{N=1}&=&\frac{\Gpq{pq\,c^{\pm2}}\Gpq{pq\,t^{-1}}}{(p;p)_\infty (q;q)_\infty}\left[\Gpq{c^{\pm2}}\Gpq{x^{\pm2}}\left(\gd(X+Y)+\gd(X-Y)\right)+\right.\nn\\
&+&\left.\Gpq{xy^{-1}}\Gpq{x^{-1}y}\Gpq{c^2x^{-1}y^{-1}}\Gpq{c^2xy}\gd(2C)\right]\,,
\ee
where we defined
\be
x=\mathrm{e}^{2\pi iX},\qquad y=\mathrm{e}^{2\pi iY},\qquad c=\mathrm{e}^{2\pi iC},\qquad 0\le X,Y,C<1\,.
\ee
The last term containing $\gd(2C)$ vanishes, because this delta implies that the contribution of one of the singlets in the prefactor becomes zero
\be
\Gpq{pq\,c^{\pm2}}=\Gpq{pq}=0\,.
\ee
Using that
\be\label{deltachange}
\gd(X\pm Y)=\gd\left(\frac{1}{2\pi i}\log(xy)\right)=2\pi ix\gd(x-y^{\pm1})\,,
\ee
we find
\be
\label{eq:deltaN=1}
\mathcal{I}_{g}^{N=1}=\frac{2\pi ix\Gpq{x^{\pm2}}}{(p;p)_\infty (q;q)_\infty\Gpq{t}}\left[\gd(x-y)+\gd(x-y^{-1})\right]\,,
\ee
which is exactly \eqref{4ddelta} for $N=1$.

\subsubsection{Explicit computation for $N=2$}

\label{sec:N=2}

For $N=2$ the supersymmetric index of the theory is
\be
\mathcal{I}_{g}^{N=2}=\Gpq{pq c^{\pm 2}}\Gpq{pq\,t^{-1}}^2\oint\udl{\vec{z}_2} \Gd_2(\vec z_2;pq/t) \mathcal{I}_{E[USp(4)]}(\vec{z};\vec{x};t;c)\mathcal{I}_{E[USp(4)]}(\vec{z};\vec{y};t;c^{-1})\,,\nn\\
\ee
where the supersymmetric index of $E[USp(4)]$ is defined by
\be
\mathcal{I}_{E[USp(4)]}(\vec{z};\vec{x};t;c)&=&\Gpq{pq\,c^{-2}t}\Gpq{pq\,t^{-1}}\prod_{i=1}^2\Gpq{c\,x_2^{\pm1}z_i^{\pm1}}\times\nn\\
&\times&\oint\udl{u_1} \Delta_1 (u)\Gpq{c\,t^{-\frac{1}{2}}x_1^{\pm1}u^{\pm1}}\Gpq{pq\,c^{-1}t^{-\frac{1}{2}}x_2^{\pm1}u^{\pm1}}\prod_{i=1}^2\Gpq{t^{\frac{1}{2}}z_i^{\pm1}u^{\pm1}}\,.\nn\\
\ee

The first step of the derivation consist of applying the IP duality to the $SU(2)$ node of one of the two $E[USp(4)]$ blocks, which is equivalent to the flip-flip duality for $E[USp(4)]$
\be
\mathcal{I}_{E[USp(4)]}(\vec{z};\vec{x};t;c)=\Gpq{t\,z_1^{\pm1}z_2^{\pm1}}\Gpq{pq\,t^{-1}x_1^{\pm1}x_2^{\pm1}}\mathcal{I}_{E[USp(4)]}(\vec{z};\vec{x};pq\,t^{-1};c)\,.
\ee
After this, the full index of the theory explicitly reads
\be
\mathcal{I}_{g}^{N=2}&=&\Gpq{pq\,t^{-1}}^2\Gpq{pq\,t^{-1}x_1^{\pm1}x_2^{\pm1}}\Gpq{pq\,c^{\pm2}}\Gpq{pq\,c^2t}\Gpq{(pq)^2c^{-2}t^{-1}}\times\nn\\
&\times&\oint\udl{\vec{z}_2}\udl{u_1}\udl{w_1} \Delta_1 (u) \Delta_1 (w)\Gd_2 (\vec z_2)\prod_{i=1}^2\Gpq{c\,x_1^{\pm1}z_i^{\pm1}}\Gpq{c^{-1}y_2^{\pm1}z_i^{\pm1}}\times\nn\\
&\times&\Gpq{(pq)^{\frac{1}{2}}t^{-\frac{1}{2}}u^{\pm1}z_i^{\pm1}} \Gpq{t^{\frac{1}{2}}w^{\pm1}z_i^{\pm1}}\Gpq{(pq)^{\frac{1}{2}}t^{\frac{1}{2}}c^{-1}x_1^{\pm1}u^{\pm1}}\times\nn\\
&\times&\Gpq{(pq)^{-\frac{1}{2}}t^{\frac{1}{2}}c\,x_2^{\pm1}u^{\pm1}}\Gpq{c^{-1}t^{-\frac{1}{2}}y_1^{\pm1}w^{\pm1}}\Gpq{pq\,c\,t^{-\frac{1}{2}}y_2^{\pm1}w^{\pm1}}\,,
\ee
where $\Gd_n(\vec z_n)$ was defined in \eqref{eq:measurenoasymm}.
Notice that the $USp(4)$ gauge node has no antisymmetric anymore, so we can proceed applying the IP duality to it. Since this node sees only 8 fundamentals, it confines. At the level of the supersymmetric index, this amounts to applying the following evaluation formula:
\begin{align}
&\oint\udl{\vec{z}_2}\Gd_2 (\vec z_2)\prod_{i=1}^2\Gpq{c\,x_1^{\pm1}z_i^{\pm1}}\Gpq{c^{-1}y_2^{\pm1}z_i^{\pm1}}\Gpq{(pq)^{\frac{1}{2}}t^{-\frac{1}{2}}u^{\pm1}z_i^{\pm1}}\Gpq{t^{\frac{1}{2}}w^{\pm1}z_i^{\pm1}}=\nn\\
&\qquad\qquad\qquad=\Gpq{c^{\pm2}}\Gpq{x_1^{\pm1}y_2^{\pm1}}\Gpq{(pq)^{\frac{1}{2}}t^{-\frac{1}{2}}c\,x_1^{\pm1}u^{\pm1}}\Gpq{t^{\frac{1}{2}}x_1^{\pm1}w^{\pm1}}\times\nn\\
&\qquad\qquad\qquad\times\Gpq{(pq)^{\frac{1}{2}}t^{-\frac{1}{2}}c^{-1}y_2^{\pm1}u^{\pm1}}\Gpq{t^{\frac{1}{2}}c^{-1}y_2^{\pm1}w^{\pm1}}\Gpq{(pq)^{\frac{1}{2}}u^{\pm1}w^{\pm1}}\,.
\end{align}
Plugging this back into $\mathcal{I}_{g}^{N=2}$ and simplifying the contribution of the massive fields we get
\be
\mathcal{I}_{g}^{N=2}&=&\Gpq{pq\,t^{-1}}^2\Gpq{pq\,t^{-1}x_1^{\pm1}x_2^{\pm1}}\Gpq{x_1^{\pm1}y_2^{\pm1}}\Gpq{pq\,c^2t}\Gpq{(pq)^2c^{-2}t^{-1}}\times\nn\\
&\times&\oint\udl{u_1} \Delta_1 (u)\Gpq{(pq)^{\frac{1}{2}}t^{-\frac{1}{2}}c^{-1}y_2^{\pm1}u^{\pm1}}\Gpq{(pq)^{-\frac{1}{2}}t^{\frac{1}{2}}c\,x_2^{\pm1}u^{\pm1}}\times\nn\\
&\times&\oint\udl{w_1} \Delta_1 (w)\Gpq{t^{\frac{1}{2}}c\,x_1^{\pm1}w^{\pm1}}\Gpq{c^{-1}t^{-\frac{1}{2}}y_1^{\pm1}w^{\pm1}}\,.
\ee

The expression that we found takes the factorized form of two independent $SU(2)$ gauge theories with $4$ chirals plus some singlets. Hence, we can evaluate each of them separately using \eqref{funddelta}. This is done in the same way as we did for the $N=1$ case and it gives the results
\be
&&\Gpq{(pq)^2t^{-1}c^{-2}}\oint\udl{u_1} \Delta_1 (u)\Gpq{(pq)^{\frac{1}{2}}t^{-\frac{1}{2}}c^{-1}y_2^{\pm1}u^{\pm1}}\Gpq{(pq)^{-\frac{1}{2}}t^{\frac{1}{2}}c\,x_2^{\pm1}u^{\pm1}}=\nn\\
&&\qquad\qquad\qquad=\frac{2\pi iy_2\Gpq{y_2^{\pm2}}\Gpq{pq\,c^{-2}t^{-1}}}{(p;p)_\infty (q;q)_\infty}\left[\gd(x_2-y_2)+\gd(x_2-y_2^{-1})\right]
\ee
and
\be
&&\Gpq{pq\,t\,c^2}\oint\udl{w_1} \Delta_1 (w)\Gpq{t^{\frac{1}{2}}c\,x_1^{\pm1}w^{\pm1}}\Gpq{c^{-1}t^{-\frac{1}{2}}y_1^{\pm1}w^{\pm1}}=\nn\\
&&\qquad\qquad\qquad=\frac{2\pi ix_1\Gpq{x_1^{\pm2}}\Gpq{c^2t}}{(p;p)_\infty (q;q)_\infty}\left[\gd(x_1-y_1)+\gd(x_1-y_1^{-1})\right]\,,
\ee
where as in the $N=1$ case the term with the third delta on the r.h.s.~of \eqref{funddelta} disappears because it is multiplied by a singlet that gives vanishing contribution after imposing the constraint of the delta. Plugging these two expressions into $\mathcal{I}_{g}^{N=2}$ and simplifying the contribution of massive fields we get
\begin{align}
\mathcal{I}_{g}^{N=2}&=\frac{\prod_{i=1}^22\pi ix_i}{\Gd_2(\vec{x};t)}\left[\gd(x_1-y_1)+\gd(x_1-y_1^{-1})\right]\left[\gd(x_2-y_2)+\gd(x_2-y_2^{-1})\right]\,,
\end{align}
which is exactly the term $x_i=y_i$ of \eqref{4ddelta} for $N=2$. 

As we mentioned, this computation only captures the contribution at $x_i = y_i^{\pm1}$ for $i = 1,2$ because we lose the information of the other poles $x_1 = y_2^{\pm1}, \, x_2 = y_1^{\pm1}$ when the $USp(4)$ gauge node is confined such that we obtain two factorized $SU(2)$ theories with 4 fundamentals. 
In the Appendix \ref{sec:induction}, we will show that if we slightly deform the quiver theory by introducing a fictitious $U(1)$ whose holonomy plays the role of a regulator, the theory is not factorized anymore because there is a bifundamental field between two $SU(2)$ gauge nodes, which becomes massive if the fictitious $U(1)$ is killed. Nevertheless, this bifundamental field plays an essential role in capturing the contribution of the other pole we neglected in the derivation we just did so that the l.h.s of \eqref{4ddelta} correctly produces all the contributions on the r.h.s..

\subsection{Non-Lagrangian gluings of $FE[USp(2N)]$ blocks}
\label{sec:non-Lagrangian}

So far we have been considering  Lagrangian gluings, obtained by gauging  manifest $USp(2 N)$ symmetries. However, thanks to the mirror property of the $E[USp(2N)]$ theory we can consider the gauging of emergent symmetries as well.

\begin{figure}[t]
\centering
\includegraphics[width=\textwidth]{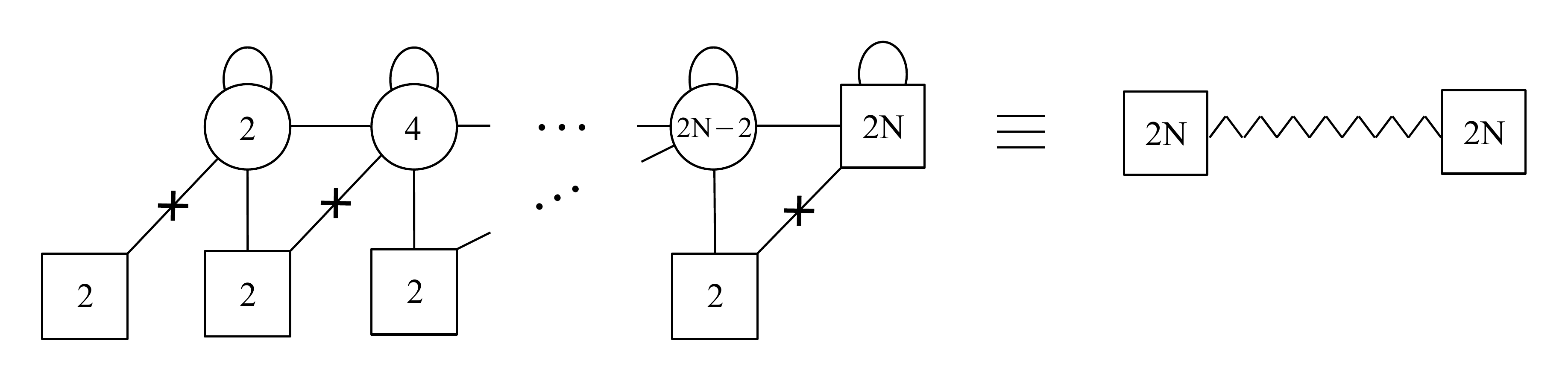}
\caption{Quiver diagram of $FE[USp(2N)]$ on the left and its shorthand representation on the right. }
\label{fig:FEUSp quiver}
\end{figure}

In fact in the following, rather than working with the  $E[USp(2N)]$ theory, we will use the $FE[USp(2 N)]$ 
theory introduced in Section \ref{sec:EUSp intro} to simplify the notation. As we will explain better momentarily, the Lagrangian gluing of two $E[USp(2N)]$ theories and the gluing of two $FE[USp(2 N)]$ lead precisely to the same theory $\mathcal{T}_g$ up to massive fields that are integrated out at low energies, provided that the gauging prescription is suitably modified between the two cases. 
The corresponding quiver diagram is shown in Figure \ref{fig:FEUSp quiver}, where we also introduce a shorthand depiction that explicitly displays both of the $USp(2N)$ symmetries that the theory possesses in the IR.

\begin{figure}[t]
\centering
\includegraphics[width=\textwidth]{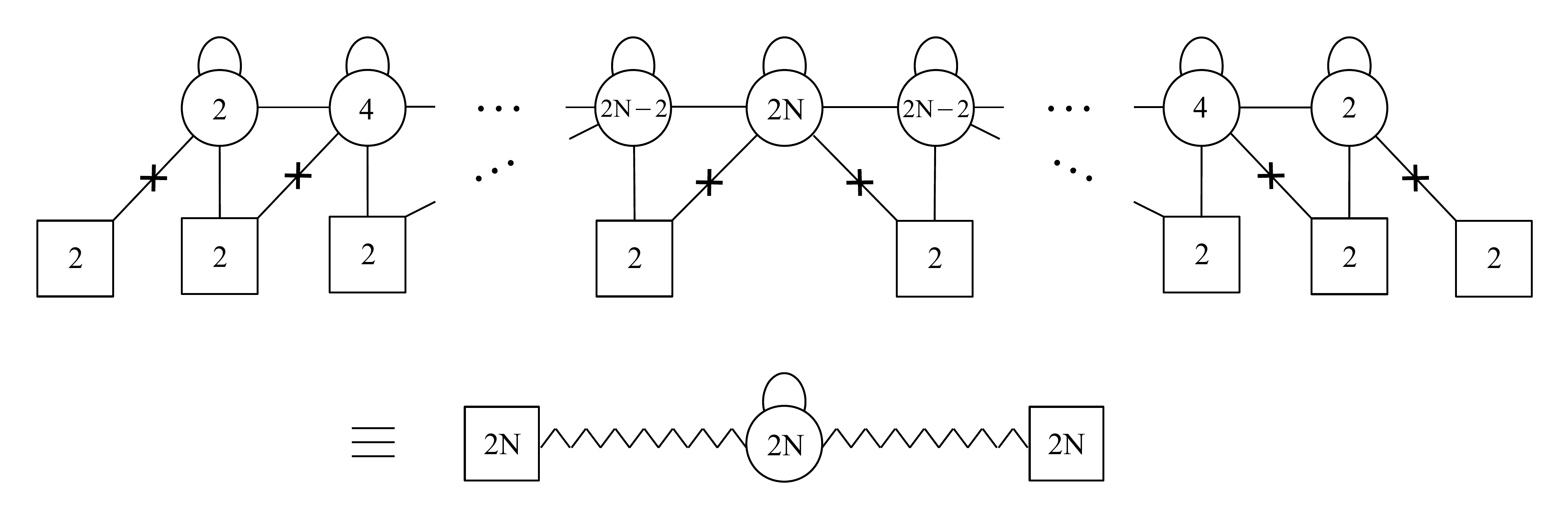}
\caption{The Lagrangian gluing of two $FE[USp(2N)]$ theories that gives the delta and the corresponding shorthand representation. The latter is independent from whether we are doing a Lagrangian or a non-Lagrangian gluing thanks to the spectral duality of $FE[USp(2N)]$.}
\label{fig:delta quiver}
\end{figure}

The gluing is now defined as the gauging of the diagonal combination $USp(2N)_z$ of the manifest $USp(2 N)$ symmetries of the two $FE[USp(2 N)]$ blocks, with an extra antisymmetric chiral field $\Phi$  which couples to $\mathsf{O_H}^L$ and $\mathsf{O_H}^R$ via the superpotential
\begin{align}\label{eq:fdelta sup}
\mathrm{Tr}_z \left[\Phi \cdot \left(\mathsf{O_H}^L-\mathsf{O_H}^R\right)\right]
\end{align}
where $\mathrm{Tr}_z$ is taken over the gauged $USp(2 N)_z$. Since this extra superpotential makes both $\Phi$ and $\mathsf{O_H}^L-\mathsf{O_H}^R$ massive, we can integrate them out, and the remaining massless field $\mathsf{O_H}^L+\mathsf{O_H}^R$ can be identified with $A$ in equation \eqref{eq:delta sup}. Moreover, the singlets $\gb_N^{(L/R)}$ are already included in the definition of $FE[USp(2N)]$. In other words, the gluing of two $FE[USp(2N)]$ theories with the superpotential \eqref{eq:fdelta sup} is equivalent to the gluing of two $E[USp(2N)]$ theories with the superpotential \eqref{eq:delta sup}, which is the theory that we dubbed $\mathcal{T}_g$. This theory is represented by the quiver diagram in Figure \ref{fig:delta quiver}.
The identity \eqref{4ddelta} can be then rewritten as follows:
\begin{align}
\label{dfhh}
\mathcal{I}_g^{N} &=\oint\udl{\vec{z}_N}\Gd_N(\vec{z};t)\mathcal{I}_{FE[USp(2N)]}(\vec{z};\vec{x};t;c)\mathcal{I}_{FE[USp(2N)]}(\vec{z};\vec{y};t;c^{-1})= {}_{\vec{x}}\hat{\mathbb{I}}_{\vec{y}}(t)\,,
\end{align}
where we are still calling the integral on the l.h.s.~$\mathcal{I}_g^{N}$ since it coincides exactly with the index of the theory $\mathcal{T}_g$ that we defined in \eqref{4ddelta}. Notice also that on the r.h.s.~we have precisely the Identity operator as defined in \eqref{4ddelta}.


So far we have just presented theory $\mathcal{T}_g$ in another equivalent way using $FE[USp(2 N)]$ blocks, while we still gauge the manifest symmetries of the blocks, which we call the Lagrangian gluing.
Using the spectral duality of $FE[USp(2 N)]$, on the other hand, one can extend the result \eqref{dfhh} for $\mathcal{T}_g$ to several non-Lagrangian dual frames where an emergent $USp(2 N)$ symmetry is gauged. For example, we can use the property \eqref{flipselfduality} of the left block to rewrite the identity \eqref{dfhh} as follows:
\begin{align}
\label{dfch}
\mathcal{I}^N_{g}  &=\oint\udl{\vec{z}_N}\Gd_N(\vec{z};t)\mathcal{I}_{FE[USp(2N)]}(\vec{x};\vec{z};t;c)\mathcal{I}_{FE[USp(2N)]}(\vec{z};\vec{y};t;c^{-1}) = {}_{\vec{x}}\hat{\mathbb{I}}_{\vec{y}}(t)\, .
\end{align}
Notice that the integral is not exactly the one defining the index $\mathcal{I}^N_{g}$ of the theory $\mathcal{T}_g$ defined in \eqref{4ddelta}, but they coincide up to spectral duality \eqref{flipselfduality}.
It corresponds to a dual frame where the diagonal combination of the emergent $USp(2 N)$ of the left block and the manifest $USp(2 N)$ of the right block is gauged, with an extra antisymmetric field $\Phi$ coupled to $\mathsf C^L$ and $\mathsf{O_H}^R$ as follows:
\begin{align}
\mathrm{Tr}_z \left[\Phi \cdot \left(\mathsf C^L-\mathsf{O_H}^R\right)\right] .
\end{align}
Similarly, we can use the spectral duality of the right block rather than the left block, which leads to
\begin{align}
\label{dfhc}
\mathcal{I}^N_{g} &=\oint\udl{\vec{z}_N}\Gd_N(\vec{z};t)\mathcal{I}_{FE[USp(2N)]}(\vec{z};\vec{x};t;c)\mathcal{I}_{FE[USp(2N)]}(\vec{y};\vec{z};t;c^{-1})  = {}_{\vec{x}}\hat{\mathbb{I}}_{\vec{y}}(t)\,.
\end{align}
This corresponds to another dual frame where the diagonal combination of the manifest $USp(2 N)$ of the left block and the emergent $USp(2 N)$ of the right block is gauged, with an extra antisymmetric field $\Phi$ coupled to $\mathsf{O_H}^L$ and $\mathsf C^R$ as follows:
\begin{align}
\mathrm{Tr}_z \left[\Phi \cdot \left(\mathsf{O_H}^L-\mathsf C^R\right)\right] .
\end{align}
Lastly, we can use the spectral duality of both blocks simultaneously, which results in the identity
\begin{align}
\label{dfcc}
\mathcal{I}^N_{g}  &=\oint\udl{\vec{z}_N}\Gd_N(\vec{z};t)\mathcal{I}_{FE[USp(2N)]}(\vec{x};\vec{z};t;c)\mathcal{I}_{FE[USp(2N)]}(\vec{y};\vec{z};t;c^{-1})  = {}_{\vec{x}}\hat{\mathbb{I}}_{\vec{y}}(t)\, .
\end{align}
This corresponds to a dual frame where the diagonal combination of the emergent $USp(2 N)$ symmetries of both blocks is gauged, with an extra antisymmetric field $\Phi$ coupled to $\mathsf C^L$ and $\mathsf C^R$ as follows:
\begin{align}
\mathrm{Tr}_z \left[\Phi \cdot \left(\mathsf C^L-\mathsf C^R\right)\right] \,.
\end{align}

\subsection{The $4d$ $S$-wall and Identity wall}

As we mentioned the theory $\mathcal{T}_g$ obtained from gluing two  $FE[USp(2 N)]$ (or  $E[USp(2 N)]$) blocks in any of the ways we just described has a quantum deformed  vacuum moduli space and displays chiral symmetry breaking. The result that its index is a delta distribution acting on test functions corresponds to this fact: if we glue $\mathcal{T}_g$ to some theory $\mathcal{T}_{test}$ exhibiting a $USp(2 N)_a$ symmetry by gauging the diagonal combination of $USp(2 N)_a$  and one of its $USp(2 N)$ symmetries, say $USp(2N)_x$, (adding an antisymmetric and the appropriate superpotential) the VEV of the bifundamental operator $\mathrm{Tr}_z \, \Pi_L \Pi_R$ will Higgs the new gauged node. In other words, the theory $\mathcal{T}_{test}$ glued with the theory $\mathcal{T}_g$ is dual to the original theory $\mathcal{T}_{test}$. Hence, we can interpret the theory $\mathcal{T}_g$ as the \emph{Identity wall}, which also explains our notation ${}_{\vec{x}}\hat{\mathbb{I}}_{\vec{y}}$ for its index.

\begin{figure}[t]
\centering
\includegraphics[width=\textwidth]{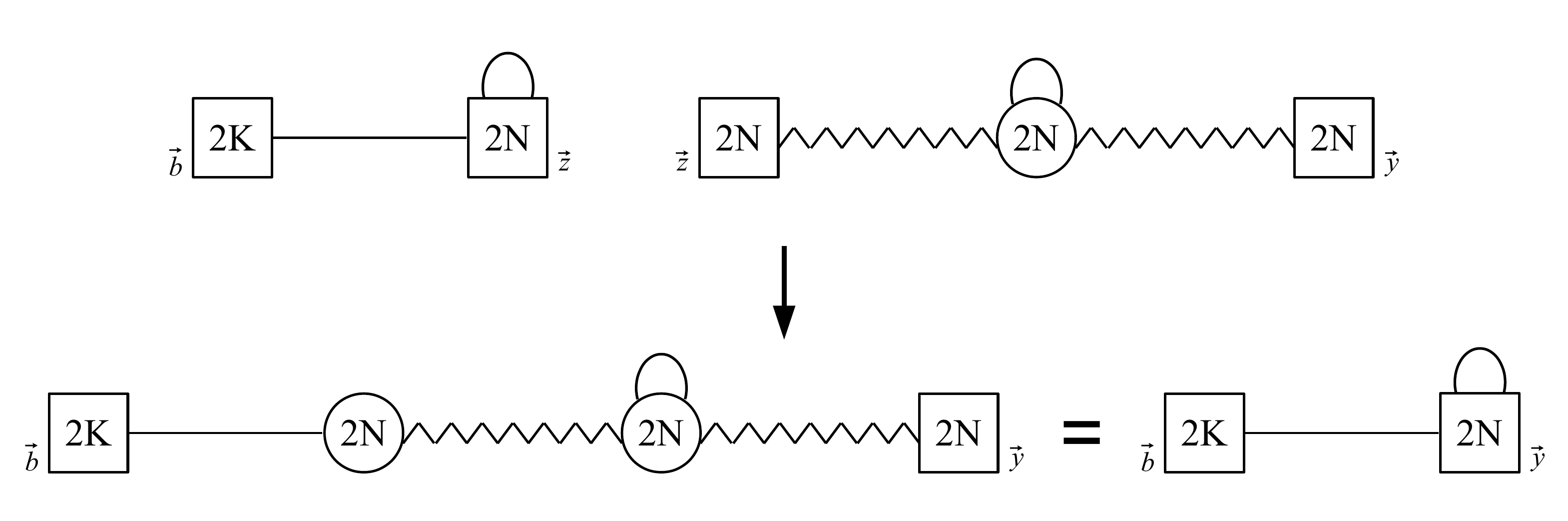}
\caption{Gluing theory $\mathcal{T}_g$ to a test theory $\mathcal{T}_{test}$. The resulting theory is dual to the original test theory $\mathcal{T}_{test}$. We thus identify $\mathcal{T}_g$ with the Identity wall.}
\label{fig:delta+fund}
\end{figure}

In order to exemplify these statements, let us consider the simple case where our test theory $\mathcal{T}_{test}$  is just a Wess-Zumino model exhibiting the $USp(2 N)_a$ symmetry, e.g., a bifundamental chiral $P$ of $USp(2 N)_a \times USp(2 K)_b$ coupled to an $USp(2 N)_a$ antisymmetric chiral $A$ as follows:
\begin{align}
\label{eq:delta ex}
\mathrm{Tr}_a \left[\mathrm{Tr}_b \left(P^2\right) \cdot A\right]\,,
\end{align}
where $\mathrm{Tr}_a$ is taken over $USp(2 N)_a$, while $\mathrm{Tr}_b$ is taken over $USp(2 K)_b$.
As we discussed, there are many equivalent dual frames of theory $\mathcal{T}_g$. For example, let us choose the frame corresponding to the identity \eqref{dfch}, where the diagonal combination of the emergent $USp(2 N)$ of the left block and the manifest $USp(2 N)$ of the right block is gauged. Thus, the non-abelian global symmetry of  $\mathcal{T}_g$ is $USp(2 N)_x \times USp(2 N)_y$, the manifest $USp(2 N)_x$ of the left block and the emergent $USp(2 N)_y$ of the right block, respectively. We can gauge the diagonal combination  of $USp(2 N)_a$ and $USp(2 N)_x$, which we call $USp(2 N)_z$, in the presence of an extra antisymmetric chiral $\Phi$ as before. $\Phi$ couples to $A$ and $\mathsf{O_H}^L$, the antisymmetric operators of $USp(2 N)_a$ and $USp(2 N)_x$ respectively, as follows:
\begin{align}
\mathrm{Tr}_z \left[\Phi \cdot \left(A-\mathsf{O_H}^L\right)\right] .
\end{align}
This quadratic superpotential makes both $\Phi$ and $A-\mathsf{O_H}^L$ massive. Once we integrate them out, we are left with the theory shown on the l.h.s of the second line in Figure \ref{fig:delta+fund}, with the superpotential between $P$ and $\mathsf{O_H}^L$
\begin{align}
\mathrm{Tr}_z \left[\mathrm{Tr}_b \left(P^2\right) \cdot \mathsf{O_H}^L\right]
\end{align}
since $A$ in \eqref{eq:delta ex} is now identified with $\mathsf{O_H}^L$.
From the discussion we made around eq.~\eqref{vevpipi} we know that the $USp(2N)_z$ gauge node is fully Higgsed and identified with $USp(2N)_y$. What we obtain is the theory on the bottom right of Figure \ref{fig:delta+fund}, which is precisely the original theory $\mathcal{T}_{test}$: a WZ model of $P$ and $A$ with the superpotential
\begin{align}
\mathrm{Tr}_y \left[\mathrm{Tr}_b \left(P^2\right) \cdot A\right]\,,
\end{align}
where $P$ is now in the bifundamental representation of $USp(2 K)_b \times USp(2 N)_y$ and $A$ is in the antisymmetric representation of $USp(2 N)_y$.

As we will discuss in  Subsection \ref{deltalimit} we have a completely analogue  result in $3d$, where by
taking two copies of the $T[SU(N)]$ theory and by  gauging a diagonal combination of the $SU(N)$ symmetries 
we obtain an Identity wall identifying the two remaining $SU(N)$ symmetries.
More precisely by taking two slightly  different $3d$ limits of the $4d$ delta property \eqref{4ddelta} one can recover both the relations $S^2=-1$ and $S^{-1}S=1$ enjoyed by $T[SU(N)]$. 
In Subsection \ref{3dlimitbraid} we will also  show that the braid duality of $E[USp(2N)]$ that we already mentioned, reduces in $3d$  to a duality involving the  gluing of $T[SU(N)]$ with CS interactions which can be interpreted  as the relation $T^{-1} S T=S^{-1} T S$ satisfied by the $SL(2,\mathbb{Z})$ generators $S$ and $T$.
Because of these observations one could expect that  the $FE[USp(2N)]$ block could play the role of an $S$-duality wall in $4d$. We will further investigate this interpretation in an upcoming paper \cite{prl}. 
We stress again the fact that we can equivalently identify $S$ with $E[USp(2N)]$ or $FE[USp(2N)]$ by just modifying the gluing/gauging prescription as we explained in Subsection \ref{sec:non-Lagrangian}.
We can then rephrase the delta-function property in this language saying that concatenating two $S$-walls these annihilate each other giving a trivial Identity wall.
Since $S^2=-1$ and $S^{-1}S=1$ are actually degenerate in $4d$ the duality group would be $PSL(2,\mathbb{Z})=SL(2,\mathbb{Z})/\mathbb{Z}_2$ rather than the full $SL(2,\mathbb{Z})$ as in $3d$.

\subsection{The asymmetric $S$-wall}

\begin{figure}[t]
\centering
\includegraphics[width=\textwidth]{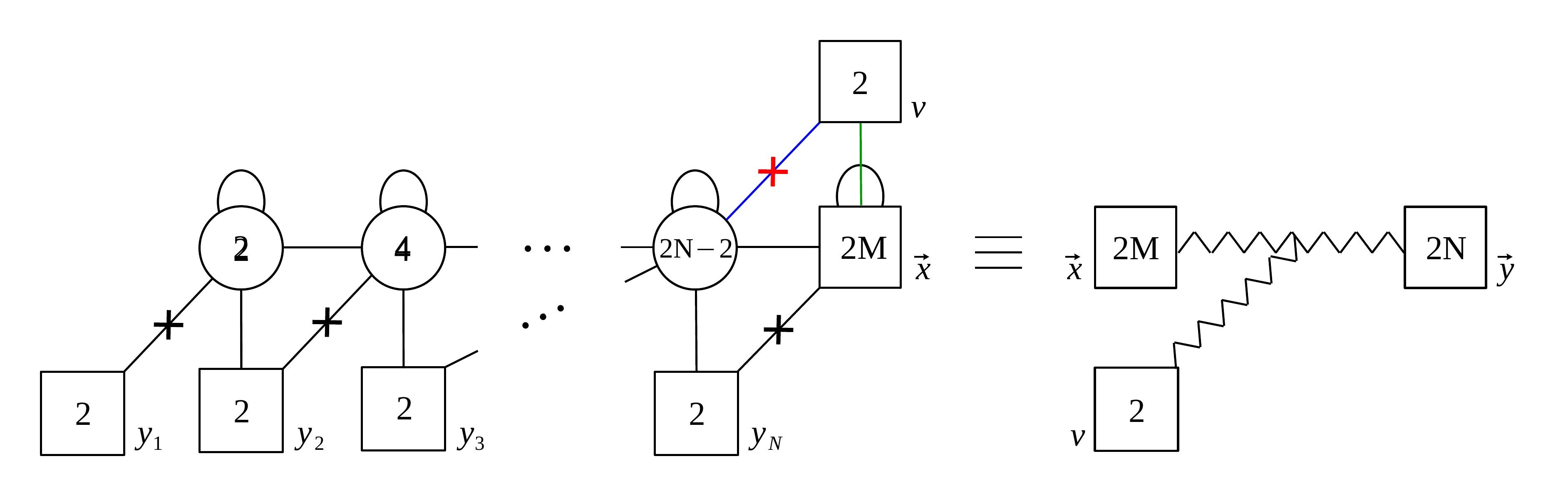}
\caption{ The quiver diagram  for the deformed $FE[USp(2 N)]$ theory and the compact form of the asymmetric $S$-wall. Each colored line denotes a set of chiral multiplets in the same representation of the non-abelian global symmetry, but with different charges under the abelian symmetries. The blue line denotes $\tilde Q^{(N-1,2)}_i$, the green line denotes $R_i$ and the red cross denotes $(\mathsf{O_H})_{M+i,M+j}^{\pm,\pm}$ in \eqref{eq:def fields}.}
\label{fig:linear def}
\end{figure}

In this subsection, we introduce an asymmetric  $S$-wall, by turning on a deformation which partially breaks the global symmetry of the  $FE[USp(2 N)]$ theory. We introduce an extra superpotential
\begin{align}
\delta \mathcal W_\text{def} = \mathrm{Tr}_y \left[\mathsf J \cdot \mathsf{O_H}\right]\,,
\end{align}
where the antisymmetric matrix $\mathsf J$ is given by
\begin{align}
\mathsf J = \frac12 \left[\mathbb J_2 \otimes \left(\mathbb O_M \oplus \mathbb J_{N-M}\right)-\mathbb J_2^T \otimes \left(\mathbb O_M \oplus \mathbb J_{N-M}^T\right)\right]\,, \qquad M < N
\end{align}
with the $K$-dimensional empty matrix $\mathbb O_K$ and the $K$-dimensional Jordan matrix $\mathbb J_K$. This superpotential can be rewritten in terms of the components of $\mathsf{O_H}$ as follows:
\begin{align}
\label{eq:linear def}
\delta \mathcal W_\text{def} = \sum_{j = 1}^{N-M-1} (\mathsf{O_H})_{M+j,M+j+1}^{+,-}
\end{align}
where an $USp(2 N)_x$ flavor index is labeled by $(i,\alpha)$ with $i = 1,\dots,N$ and $\alpha = \pm$. For example, the antisymmetric invariant symbol $J = i\sigma_2 \otimes \mathbb I_N$ of $USp(2 N)$ has the indices in the following form:
\begin{align}
(J)_{i,j}^{\alpha,\beta} = (i \sigma_2)^{\alpha,\beta} \otimes (\mathbb I_N)_{i,j} \,.
\end{align}
This extra superpotential breaks the $USp(2 N)_x$ global symmetry down to its subgroup $USp(2 M)_x \times SU(2)_v$. Thus, the fields of $FE[USp(2 N)]$ charged under the $USp(2 N)_x$ symmetry are reorganized into the representations of the unbroken symmetry $USp(2 M)_x \times SU(2)_v$ as follows:
\begin{align}
\begin{aligned}
\label{eq:def fields}
Q^{(N-1,N)} \quad &\longrightarrow \quad \underbrace{\mathbf{(2 M,1)}}_{\tilde Q^{(N-1,M)}} \oplus \underbrace{\mathbf{(1,2)}^{\oplus N-M}}_{\tilde Q_{i = 1,\dots,N-M}^{(N-1,2)}} \,, \\
\mathsf{O_H} \quad &\longrightarrow \quad \underbrace{\mathbf{(M(2M-1),1)}}_{\mathsf{\tilde O_H}} \oplus \underbrace{\mathbf{(2 M,2)}^{\oplus N-M}}_{R_{i = 1,\dots,N-M}} \oplus \underbrace{\mathbf{(1,1)}^{\oplus (N-M) (2 N-2 M-1)}}_{(\mathsf{O_H})_{M+i,M+j}^{\pm,\pm}} \,.
\end{aligned}
\end{align}
For instance, field $R_i$ in the bifundamental of $USp(2 M)_x \times SU(2)_v$ is defined by
\begin{align}
\left(R_i^+,R_i^-\right)_j^{\beta} = \left((\mathsf{O_H})_{j,M+i}^{\beta,+},(\mathsf{O_H})_{j,N+1-i}^{\beta,-}\right) \,, \qquad i = 1,\dots,N-M
\end{align}
with the $SU(2)_v$ index $\pm$ and the $USp(2 M)_x$ index $(j,\beta)$. The corresponding quiver diagram 
and the compact form are given in  Figure \ref{fig:linear def}.

Note that this deformed $FE[USp(2 N)]$ theory is closely related to the theory denoted by $E^{[N-M,1^{M}]}[USp(2N)]$ in \cite{Hwang:2020wpd}. Since $\mathsf O_{\mathsf{H}}$ couples to the gauge invariant operator $\mathsf H$, see \eqref{eq:FE}, the extra superpotential \eqref{eq:linear def} gives rise to a non-vanishing VEV of $\mathsf H$ triggering an RG flow to $E^{[N-M,1^{M}]}[USp(2N)]$ with some additional gauge singlets.

Clearly we can consider also an $S$-wall with a deformation of the emergent $USp(2N)_y$ global symmetry. This is actually provided by the spectral duality. Indeed the $FE[USp(2 N)]$ theory deformed by the superpotential \eqref{eq:linear def} is dual to the $FE[USp(2 N)]$ theory deformed by the superpotential
\begin{align}
\label{eq:mass def}
\delta \mathcal W_\text{def} = \sum_{j = 1}^{N-M-1} (\mathsf{C})_{M+j,M+j+1}^{+,-} \,,
\end{align}
where $\mathsf C$ is the gauge invariant operator in the antisymmetric representation of the emergent $USp(2 N)_y$ symmetry of the $FE[USp(2 N)]$ theory. This deformation breaks $USp(2 N)_y$ down to its $USp(2 M)_y \times SU(2)_v$. Indeed, it corresponds to a massive deformation for some of the chirals in the saw, see Subsection 3.2 of \cite{Hwang:2020wpd} for more details.

\subsection{Gluing asymmetric $S$-walls}

We  now consider gluing an $S$-wall to an asymmetric $S$-wall. For convenience, here we choose a particular frame where we glue the emergent $USp(2 N)$ of the $S$-wall  with the manifest $USp(2 N)$ of the asymmetric $S$-wall
which has  the emergent $USp(2 N)$ symmetry  broken to $USp(2 M) \times SU(2)$ by the superpotential \eqref{eq:mass def}. We will refer to the resulting theory as theory $\tilde{\mathcal{T}}_{g}$.

At the level of the index, such deformation corresponds to the specialization of part of the variables $\vec y$ in the form of a geometric progression: $y_{M+i} \rightarrow t^{\frac{N-M+1}{2}-i} v$ for $i = 1, \dots, N-M$ with $M \leq N$. Then the identity \eqref{dfch} reduces to:
\begin{align}
\label{spec4ddelta}
\mathcal{I}_{g}^{N,M} &=\oint\udl{\vec{z}_N}\Gd_N(\vec{z};t)\mathcal{I}_{FE[USp(2N)]}(\vec{x};\vec{z};t;c) \nonumber \\
&\qquad \mathcal{I}_{FE[USp(2N)]}(\vec{z};\vec{y},t^{\frac{N-M-1}{2}}v,t^{\frac{N-M-3}{2}}v,\cdots,t^{-\frac{N-M-1}{2}}v;t;c^{-1})=\qquad\qquad\nn\\
&= \frac{\prod_{j=1}^N 2\pi i x_j}{\Gd_N(\vec x;t)}\sum_{\gs\in S_N} \sum_{\pm} \prod_{i=1}^M\gd\left(x_{\sigma(i)}^{\pm1}-y_i\right) \prod_{j=1}^{N-M}\gd\left(x_{\sigma(M+j)}^{\pm1}-t^{\frac{N-M-1}{2}-j} v\right)\equiv {}_{\vec{x}}\hat{\mathbb{I}}_{\vec{y},v}(t)\,.
\end{align}
Notice that we can recover the standard delta-property \eqref{4ddelta} for $M=N$.

\begin{figure}[t]
\centering
\includegraphics[width=\textwidth]{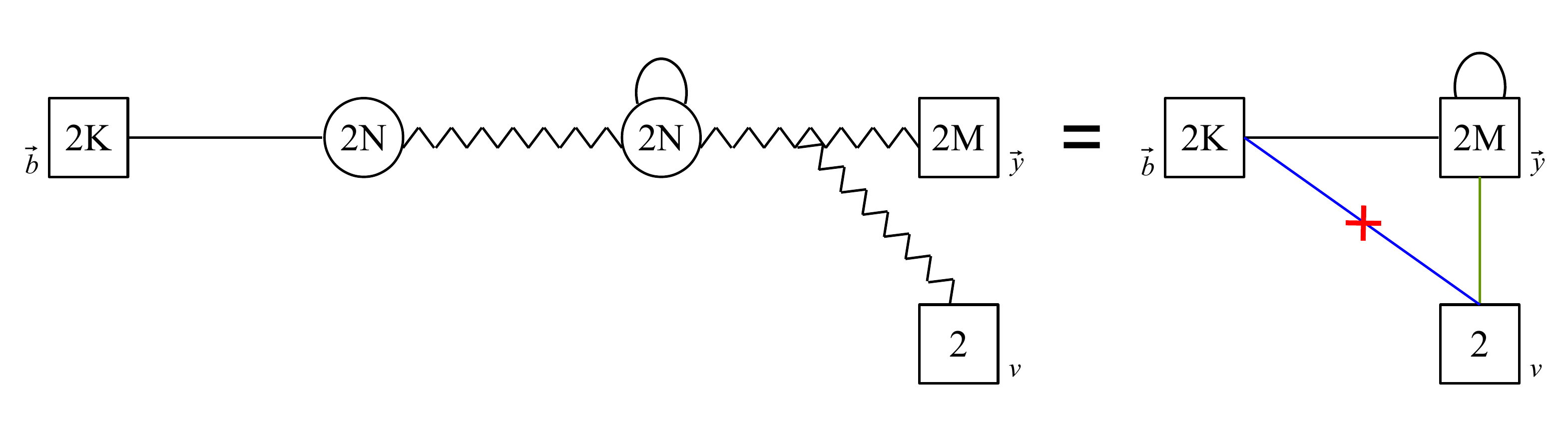}
\caption{\label{fig:spec_delta+fund} On the left, the diagram representing the gluing of theory $\tilde{\mathcal{T}}_{g}$, which we identify with the deformed Identity wall, and a test theory $\mathcal{T}_{test}$. On the right, the dual theory $\tilde{\mathcal{T}}_{test}$. Each colored line denotes a set of chiral multiplets in the same representation of the non-abelian global symmetry, but with different charges under the abelian symmetries. Especially, the blue line denotes $Q_i$, the green line denotes $R_i$ and the red cross denotes $(A)_{M+j,M+k}^{\alpha,\beta}$ in \eqref{eq:spec_def fields}.}
\end{figure}

We can now think of gluing  theory $\tilde{\mathcal{T}}_{g}$ to a test theory $\mathcal{T}_{test}$ exhibiting the $USp(2 N)_a$ global symmetry. The delta-function property of $\tilde{\mathcal{T}}_{g}$, which we call the deformed Identity wall, then leads to a duality between theory $\mathcal{T}_{test}$ glued with theory $\tilde{\mathcal{T}}_{g}$ and the deformed $\tilde{\mathcal{T}}_{test}$ whose $USp(2 N)_a$ symmetry is broken to $USp(2 M)_y \times SU(2)_v$. Again let us consider the simplest case where the test theory $\mathcal{T}_{test}$ is given by a WZ model with the superpotential
\begin{align}
\label{eq:spec_delta ex}
\mathrm{Tr}_a \left[\mathrm{Tr}_b \left(P^2\right) \cdot A\right]\,,
\end{align}
where $P$ is in the bifundamental representation of $USp(2 N)_a \times USp(2 K)_b$ and $A$ is in the antisymmetric representation of $USp(2 N)_a$. Now we gauge the diagonal combination $USp(2 N)_z \subset USp(2 N)_a \times USp(2 N)_x$, where $USp(2N)_x$ comes from $\tilde{\mathcal{T}}_{g}$, with an extra antisymmetric chiral $\Phi$ coupled to $A$ and $\mathsf{O_H}^L$ as follows:
\begin{align}
\mathrm{Tr}_z \left[\Phi \cdot \left(A-\mathsf{O_H}^L\right)\right] .
\end{align}
Since $\Phi$ and $A-\mathsf{O_H}^L$ are massive, if we integrate them out we are left with the theory corresponding to the left diagram in Figure \ref{fig:spec_delta+fund} with the superpotential between $P$ and $\mathsf{O_H}^L$
\begin{align}
\label{eq:bifund sup}
\mathrm{Tr}_z \left[\mathrm{Tr}_b \left(P^2\right) \cdot \mathsf{O_H}^L\right]
\end{align}
as $A$ in \eqref{eq:spec_delta ex} is identified with $\mathsf{O_H}^L$.
As before, the $USp(2N)_z$ is fully Higgsed, but this time it is identified with $USp(2M)_y\times SU(2)_v$. We accordingly decompose the fields as
\begin{align}
\begin{aligned}
\label{eq:spec_def fields}
P \qquad &\longrightarrow \qquad \underbrace{\mathbf{(2 K,2 M,1)}}_{\tilde P} \oplus \underbrace{\mathbf{(2 K,1,2)}^{\oplus N-M}}_{Q_{i = 1,\dots, N-M}} \,, \\
A \qquad &\longrightarrow \qquad \underbrace{\mathbf{(1,M(2M-1),1)}}_{\tilde A} \oplus \underbrace{\mathbf{(1,2 M,2)}^{\oplus N-M}}_{R_{i = 1,\dots, N-M}} \oplus \underbrace{\mathbf{(1,1,1)}^{\oplus (N-M) (2 N-2 M-1)}}_{(A)^{\alpha,\beta}_{M+j,M+k}} \,,
\end{aligned}
\end{align}
where $(\alpha,\beta) = (\pm,\pm)$ if $1 \leq j < k \leq N-M$ and $(\alpha,\beta) = (+,-)$ if $1 \leq j = k \leq N-M$. The result of the Higgsing is the WZ model $\tilde{T}_{test}$ on the right of Figure \ref{fig:spec_delta+fund} with the superpotential
\begin{align}
&\mathrm{Tr}_y \left[\mathrm{Tr}_b \left(\tilde P^2\right) \cdot \tilde A\right]+\sum_{i = 1}^{N-M} \mathrm{Tr}_y \mathrm{Tr}_b \left[\tilde P Q^\pm_i R^\mp_i\right]+\sum_{i = 1}^{N-M} \mathrm{Tr}_b \left[Q^-_i Q^+_i\right] (A)^{+,-}_{M+i,M+i} \nonumber \\
&+\sum_{j < k} \epsilon_{\alpha \beta} \epsilon_{\gamma \delta} \, \mathrm{Tr}_b \left[Q^\alpha_j Q^\gamma_k\right] (A)^{\beta,\delta}_{M+j,M+k}+\sum_{i = 1}^{N-M-1} (A)^{+,-}_{M+i,M+i+1} \,.
\end{align}
where the last term comes from the deformation \eqref{eq:mass def}
and is the responsible for the breaking of $USp(2N)$ to $USp(2M)_y\times SU(2)_v$.

\section{Gluing $S$-walls with matter}
\subsection{Gluing with two fundamental chirals}
\label{g2chir}

We will now consider the gluing of two $S$-walls with additional matter fields.
 We start  considering the case where we gauge the diagonal combination $USp(2 N)_z$ of two manifest $USp(2 N)$ symmetries of the $S$-walls with an antisymmetric field $\Phi$ and two extra fundamental fields $(P^+,P^-)$ interacting via the superpotential
\begin{align}
\label{eq:braid sup 1}
\mathrm{Tr}_z \left[\Phi \cdot \left(\mathsf{O_H}^L-\mathsf{O_H}^R\right)\right]+ \mathrm{Tr}_z \left[P^+ P^- \cdot \left(\mathsf{O_H}^L+\mathsf{O_H}^R\right)\right]
\end{align}
where $\mathrm{Tr}_z$ is taken over the gauged $USp(2 N)_z$ and $\mathsf{O_H}^L, \, \mathsf{O_H}^R$ are the gauge singlet fields of the two $FE[USp(2 N)]$ blocks in the antisymmetric representation of $USp(2 N)_z$. 

\begin{figure}[t]
\centering
\includegraphics[width=1\textwidth]{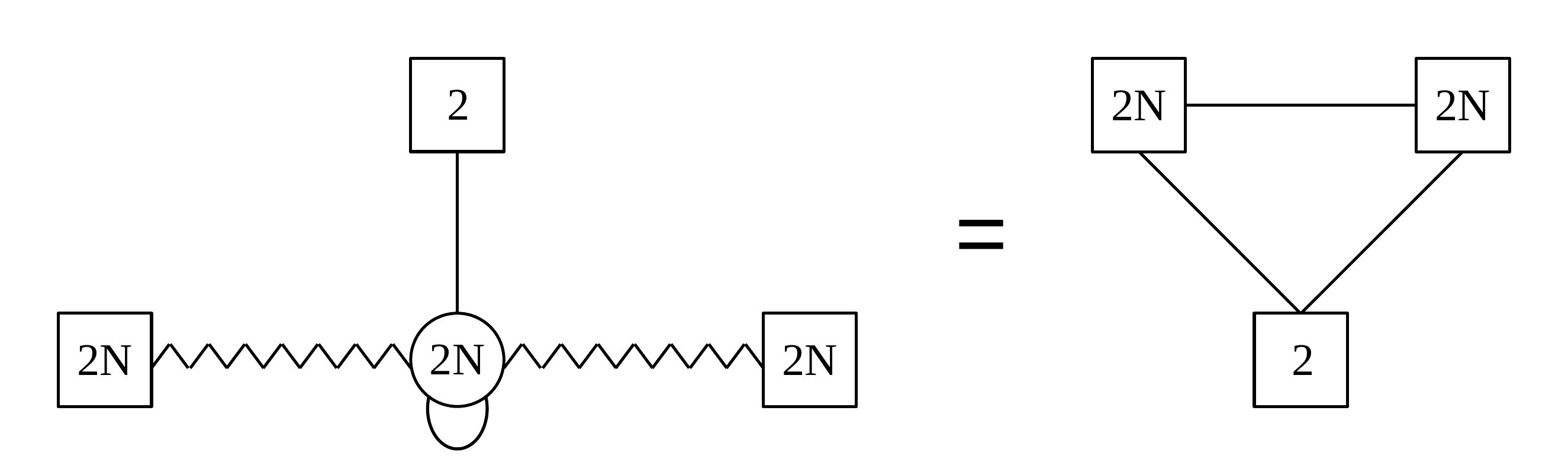}
\caption{The theory obtained by gluing two $S$-walls with  2 chirals is dual to a WZ model. }
\label{fig:triangle}
\end{figure}

Also with this gluing the ``$U(1)_t$" symmetries of the two  $FE[USp(2 N)]$  blocks are identified because of the first term in the superpotential.
Of the  two ``$U(1)_c$'' symmetries for the $FE[USp(2 N)]$ blocks, which we call $U(1)_c$ and $U(1)_d$, one combination is broken and correspondingly the fugacities of $U(1)_c$ and $U(1)_d$  satisfy
\begin{align}
c d = (p q/t)^\frac12\,.
\end{align}
The total global symmetry is
\begin{align}
USp(2 N)_x \times USp(2 N)_y \times SU(2)_v \times U(1)_c \times U(1)_t
\end{align}
where $USp(2 N)_x$ and $USp(2 N)_y$ are the emergent $USp(2 N)$ symmetries of each $FE[USp(2 N)]$ block
and $(P^+,P^-)$ is a doublet  of $SU(2)_v$.

Notice that the first term in \eqref{eq:braid sup 1} gives mass to both $\Phi$ and $\mathsf{O_H}^L-\mathsf{O_H}^R$, so they can be integrated out, and we are left with
 \begin{align}
\label{eq:braid sup 2}
\delta \mathcal W_\text{fund} = \mathrm{Tr}_z \left[\epsilon_{\alpha\beta} P^\alpha P^\beta \cdot A\right] \,,\quad
{\rm where }\quad  A = \mathsf{O_H}^L + \mathsf{O_H}^R\,.
\end{align}

As  depicted in Figure \ref{fig:triangle}, we will show that the dual is a simple WZ theory.
In terms of the supersymmetric index, this duality corresponds to the following identity:
\begin{align}
\label{eq:equal ranks}
& \oint \udl{\vec z_N} \Gd_N(\vec{z};t) \prod_{i = 1}^N \Gpq{t^\frac12 v^{\pm1} z_i^{\pm1}} \mathcal I_{FE[USp(2 N)]}(\vec z;\vec x;t;c) \mathcal I_{FE[USp(2 N)]}(\vec z;\vec y;t;(pq/t)^\frac12 c^{-1}) \nonumber \\
&= \prod_{i,j = 1}^N \Gpq{(pq/t)^\frac12 x_i^{\pm1} y_j^{\pm1}} \prod_{i = 1}^N \Gpq{t^\frac12 c v^{\pm1} x_i^{\pm1}} \prod_{j = 1}^N \Gpq{(pq)^\frac12 c^{-1} v^{\pm1} y_j^{\pm1}}\,,
\end{align}
where $v$ is the fugacity of $SU(2)_v$ acting on $(P^+,P^-)$.

\begin{figure}[t]
\centering
\includegraphics[width=1\textwidth]{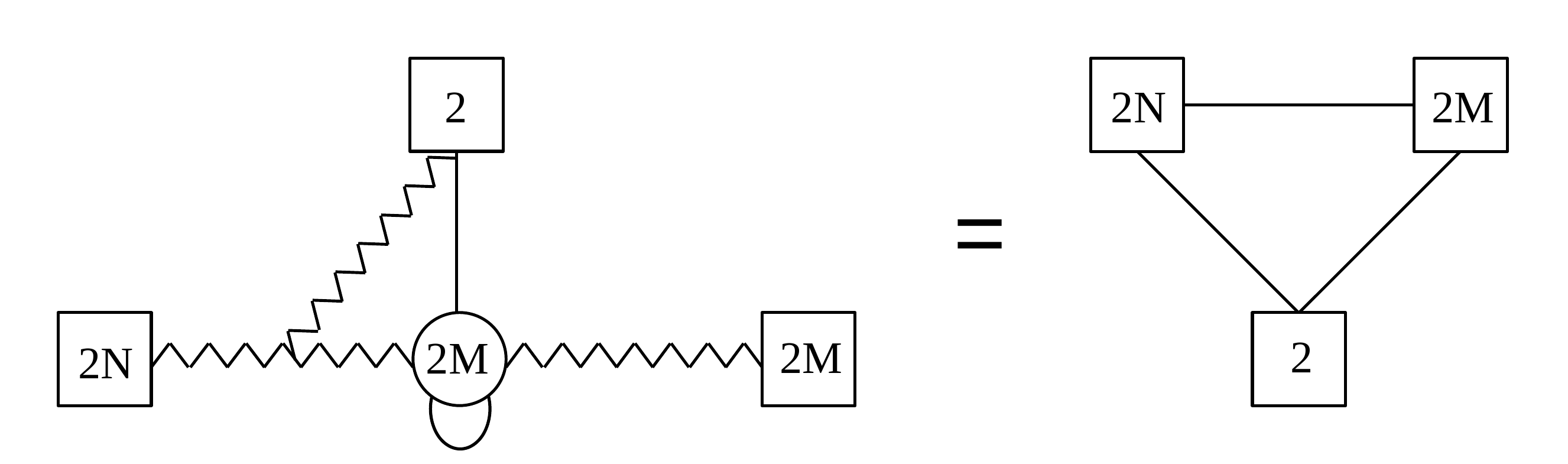}
\caption{The theory obtained by gluing two $S$-walls of different length  with 2 chirals is dual to a WZ model.}
\label{fig:specialized triangle}
\end{figure}
\begin{figure}[t]
\centering
\includegraphics[width=0.88\textwidth]{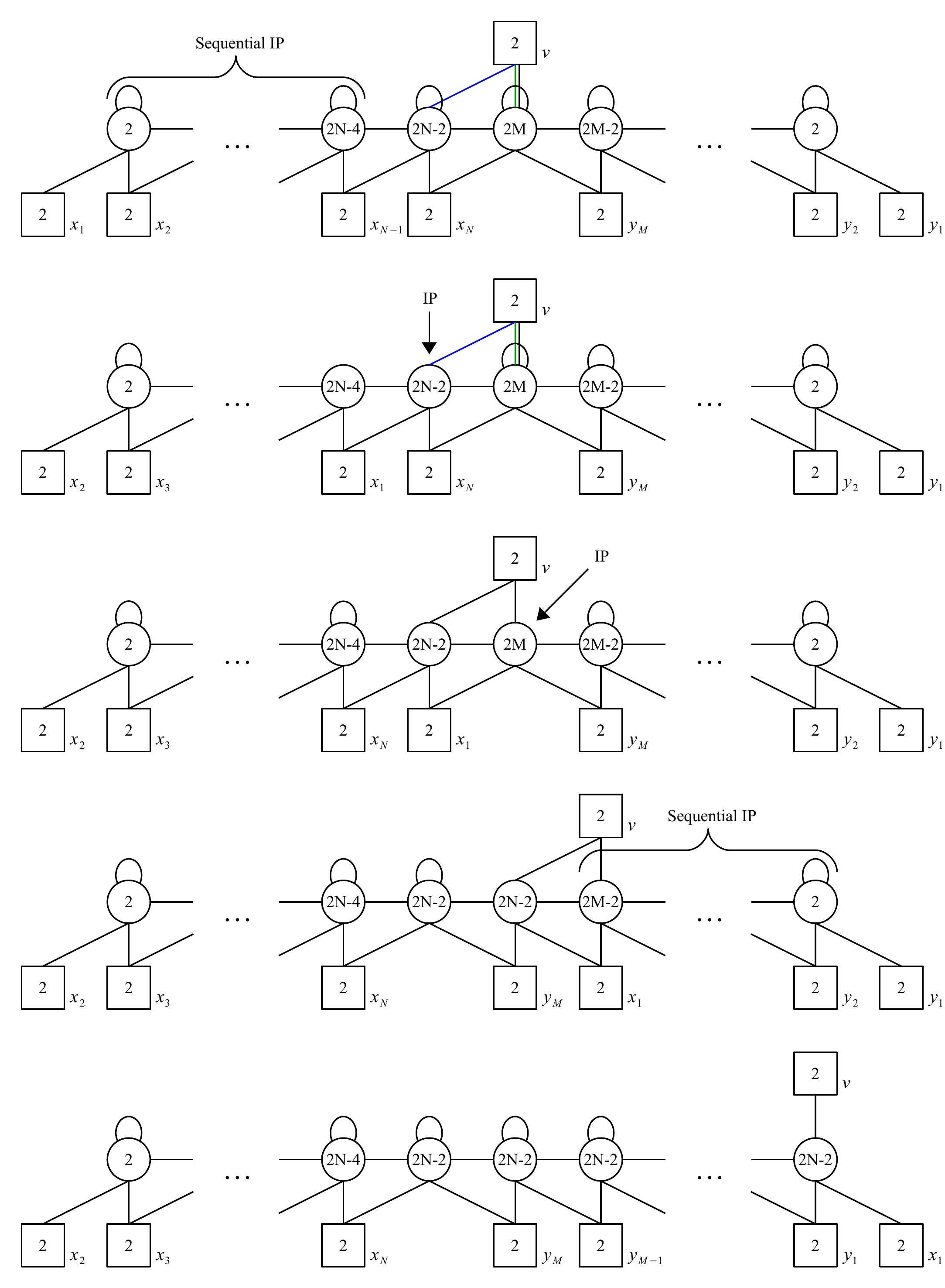}
\caption{The first steps for the derivation of the duality associated to the gluing of two $S$-walls of different lengths with the insertion of 2 chirals. We avoid drawing gauge singlets not to clutter the picture.}
\label{fig:NS5 dual 1}
\end{figure}

The duality in  Figure \ref{fig:triangle} can be  generalized to the case where we glue $S$-walls of different  lengths $N > M$
as shown in compact form in Figure \ref{fig:specialized triangle}.
More precisely as shown in the first line in  Figure \ref{fig:NS5 dual 1} we take one asymmetric $S$-wall, the  $FE[USp(2 N)]$ theory deformed by the superpotential \eqref{eq:linear def},
which breaks $USp(2 N)$ to $USp(2 M) \times SU(2)_v$, and the $S$-wall $FE[USp(2 M)]$ exhibiting two $USp(2 M)$ symmetries and gauge the diagonal combination, which we call $USp(2 M)_z$, of $USp(2 M)$ of the deformed $FE[USp(2 N)]$ and the manifest $USp(2 M)$ symmetry of $FE[USp(2 M)]$. We then introduce the doublet of fundamental chirals $(P^+,P^-)$ with the superpotential
\begin{align}
\label{wf}
\delta \mathcal W_\text{fund} = \epsilon_{\alpha \beta} \mathrm{Tr}_z R_{N-M}^\alpha P^\beta
\end{align}
where we recall that the fields $R_i$  (depicted in  green in Figure  \ref{fig:NS5 dual 1}) were defined  in \eqref{eq:def fields}. 
Notice that the superpotential \eqref{wf} has both the effects of fixing the abelian charges of $P$ and of identifying the $SU(2)$  symmetry rotating   $P$ with $SU(2)_v$ acting on $R_{N-M}$ (and on the other $R_i$).
From Figure  \ref{fig:NS5 dual 1} we also see that the fields $\tilde Q^{(N-1,2)}_i$ and $R_i$ respectively represented by the blue and green line form cubic superpotential with the horizontal bifundamental $\tilde Q^{(N-1,M)}$
 inherited from the cubic superpotential of $FE[USp(2 N)]$ involving $\mathsf{O_H}^L$.

At the level of the index, the  deformation \eqref{eq:linear def} is translated into the specialization of the variables $\vec z_N$ of $\mathcal I_{E[USp(2 N)]}(\vec x;\vec z;t;c)$ in \eqref{eq:equal ranks} to $\vec z_N = (z_1,\dots,z_M,t^\frac{N-M-1}{2} v,\dots,t^{-\frac{N-M-1}{2}} v)$. Then we can integrate over the remaining $\vec z_M = (z_1,\dots,z_M)$ and obtain the following generalized identity:\footnote{
Notice that the index of $FE[USp(2 N)]$ includes the contribution of $\mathsf{O_H}^L$ which, 
due to the specialization $z_{M+i} \rightarrow t^{\frac{N-M+1}{2}-i} v$ for $i = 1, \dots, N-M$,
is decomposed as:
\begin{align}
\label{boh}
\prod_{i < j}^N \Gpq{pq t^{-1} z_i^{\pm1} z_j^{\pm1}} \quad \longrightarrow \quad \left\{\begin{array}{l}
\prod_{i < j}^M \Gpq{pq t^{-1} z_i^{\pm1} z_j^{\pm1}} , \\
\prod_{i = 1}^M \prod_{j = 1}^{N-M} \Gpq{pq t^{-1} z_i^{\pm1} \left(t^{\frac{N-M+1}{2}-j} v\right)^{\pm1} } , \\
\prod_{i < j}^{N-M} \Gpq{pq t^{-1} \left(t^{\frac{N-M+1}{2}-i} v\right)^{\pm1} \left(t^{\frac{N-M+1}{2}-j} v\right)^{\pm1}}
\end{array}\right.
\end{align}
where the first line correspond to $\mathsf{\tilde O_H}$, the second line corresponds to $R_i$, and the third line corresponds to $(\mathsf{O_H})^{\pm,\pm}_{M+i,M+j}$ in \eqref{eq:def fields}.}
\begin{align}
\label{eq:generalized}
&\prod_{i = 1}^{N-M} \Gpq{t^{-i+1} c^2} \prod_{i = 1}^{N-M} \Gpq{t^i} \oint \udl{\vec z_M} \Gd_M(\vec{z};t) \prod_{i = 1}^N \Gpq{t^\frac{N-M+1}{2} v^{\pm1} z_i^{\pm1}} \nonumber \\
&\quad \times \mathcal I_{FE[USp(2 N)]}(\vec z,t^\frac{N-M-1}{2} v,\dots,t^{-\frac{N-M-1}{2}} v;\vec x;t;c) \mathcal I_{FE[USp(2 M)]}(\vec z;\vec y;t;(pq/t)^\frac12 c^{-1}) \nonumber \\
&= \prod_{i = 1}^N \prod_{j = 1}^M \Gpq{(pq/t)^\frac12 x_i^{\pm1} y_j^{\pm1}} \prod_{i = 1}^N \Gpq{t^{-\frac{N-M-1}{2}} c v^{\pm1} x_i^{\pm1}} \prod_{j = 1}^M \Gpq{(pq)^\frac12 t^\frac{N-M}{2} c^{-1} v^{\pm1} y_j^{\pm1}}\nonumber\\
&\equiv \mathcal{I}^{(N,M)}_\bigtriangledown\left(\vec x;\vec y;v;t;c t^{-\frac{N-M}{2}}\right)\,,
\end{align}
where we have defined
\begin{align}
\label{eq:tri}
\mathcal{I}^{(N,M)}_\bigtriangledown \left(\vec x;\vec y;v;t;c\right) = \prod_{i = 1}^N \prod_{j = 1}^M \Gpq{(pq/t)^\frac12 x_i^{\pm1} y_j^{\pm1}} \prod_{i = 1}^N \Gpq{t^\frac12 c v^{\pm1} x_i^{\pm1}} \prod_{j = 1}^M \Gpq{(pq)^\frac12 c^{-1} v^{\pm1} y_j^{\pm1}} .
\end{align}

Also in this case, as we will see, the proof of the duality is based on the iterative application of the  IP duality. Indeed for $N=M=1$ the duality in  Figure \ref{fig:triangle} is the star-triangle duality corresponding to IP for 
$N_c=1$ and $N_f=3$ or Seiberg duality for $SU(2)$ with $6$ chirals.
 
The first step of the derivation consists in applying  iteratively the IP duality along the quiver starting from the leftmost $USp(2)$ node  until we reach the $(N-2)$-th gauge node, which leads to the second quiver in Figure \ref{fig:NS5 dual 1}.
Note that the $SU(2)$ flavor node with fugacity $x_1$ is now attached to the $(N-2)$-th gauge node and the $(N-1)$-th gauge node, both of which do not have antisymmetric fields.

We then apply IP duality on the $(N-1)$-th gauge node, doing so the $R_i$ fields in green becomes massive and disappear, while only two of the $\tilde Q^{(N-1,2)}_i$ fields remain massless,  which are denoted by a black line connecting the $(N-1)$-th gauge node and the $SU(2)_v$ flavor node in the third  line of Figure \ref{fig:NS5 dual 1}.
Indeed we can see that after the dualisation the fugacities of the $\tilde Q^{(N-1,2)}_i$ fields lead to a telescopic canellation
\begin{align}
&\prod_{i = 1}^{N-M} \prod_{j = 1}^{N-1} \Gpq{t^\frac12 \left(t^{\frac{N-M+1}{2}-i} v\right)^{\pm1} \left(z^{(N-1)}_j\right)^{\pm1}} \nonumber \\
&\rightarrow \prod_{i = 1}^{N-M} \prod_{j = 1}^{N-1} \Gpq{(pq/t)^\frac12 \left(t^{\frac{N-M+1}{2}-i} v\right)^{\pm1} \left(z^{(N-1)}_j\right)^{\pm1}} =
 \prod_{j = 1}^{N-1} \Gpq{(pq)^\frac12 t^{-\frac{N-M}{2}} v^{\pm1} \left(z^{(N-1)}_j\right)^{\pm1}} \,.
\end{align}

Now we apply the IP duality on the remaining gauge nodes sequentially arriving at the last diagram in Figure \ref{fig:NS5 dual 1}.

\begin{figure}
\centering
\includegraphics[width=0.98\textwidth]{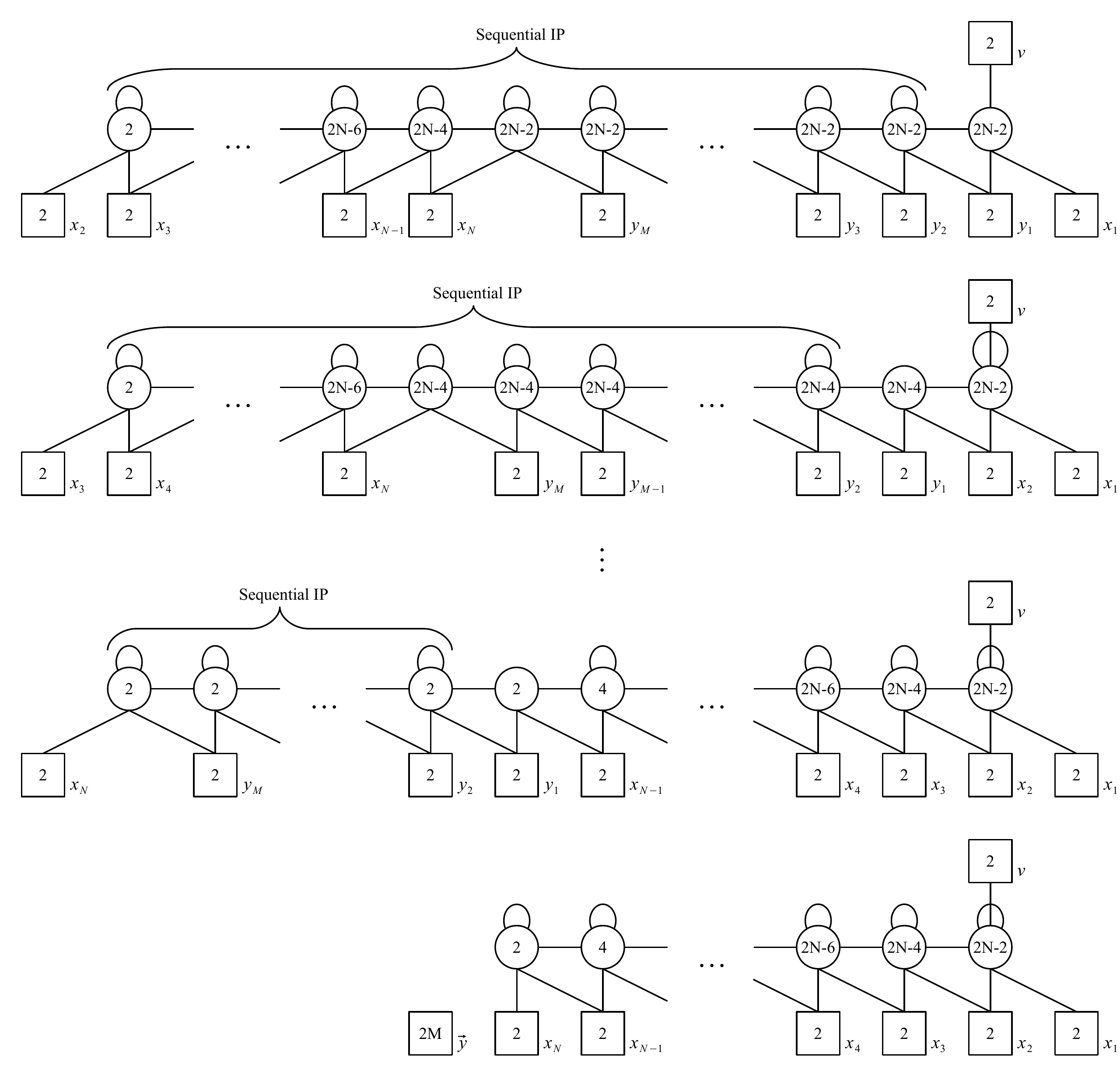}
\caption{The subsequent steps for the derivation of the duality associated to the gluing of two $S$-walls of different lengths with the insertion of 2 chirals. We avoid drawing gauge singlets not to clutter the picture. Some of these would connect the $USp(2M)_y$ flavor node to some of the other flavor nodes.}
\label{fig:NS5 dual 2} 
\end{figure}

The next steps are shown in Figure \ref{fig:NS5 dual 2}. We first apply the IP duality along the quiver starting from the left and up to the second to last gauge node, arriving at the second line of Figure \ref{fig:NS5 dual 2}. 
Then we start again from the leftmost node and apply IP sequentially, but this time stopping at the third to last node. We repeat this procedure for $i = 4,\dots,N-1$, applying the IP duality from the first gauge node to the $i$-th last node, obtaining the third diagram.
We now notice that the initial part of the tail consists of a series of $SU(2)$ gauge nodes which sequentially confine once we go through it with the IP duality, and we eventually reach the last diagram. 

Implementing these manipulations at the level of the index, we find that the l.h.s.~of \eqref{eq:generalized} is equal to
\begin{align}
\label{eq:NS5 dual}
&\Gpq{pq/t}^N \prod_{i < j}^N \Gpq{pq t^{-1} x_i^{\pm1} x_j^{\pm1}} \prod_{i = 1}^{N} \Gpq{t^i} \prod_{i = 1}^{N-1} \Gpq{t^{M-i} c^2} \nonumber \\
&\times \prod_{i = 1}^N \prod_{j = 1}^M \Gpq{(pq/t)^\frac12 x_i^{\pm1} y_j^{\pm1}} \prod_{j = 1}^M \Gpq{(pq)^\frac12 t^\frac{N-M}{2} c^{-1} v^{\pm1} y_j^{\pm1}} \nonumber 
\end{align}
\begin{align}
&\times \prod_{i = 1}^N \Gpq{t^{i-\frac{N-M+1}{2}} c v^{\pm1} x_1^{\pm1}} \oint \udl{\vec z_{N-1}} \Gd_{N-1}(z) \prod_{j = 1}^{N-1} \Gpq{(pq)^\frac{1}{2} t^{-\frac{N}{2}} v^{\pm1} z_j^{\pm1}} \nonumber \\
&\times\Gpq{(pq)^\frac12 t^{-\frac{M-1}{2}} c^{-1} x_1^{\pm1} z_j^{\pm1}}  \mathcal I_{FE[USp(2 N-2)]}(\vec z;x_N,\dots,x_2;pq/t;(pq)^{-\frac12} t^\frac{M+1}{2} c)
\end{align}
for $N > 1$. If $N = M = 1$, the identity \eqref{eq:equal ranks} is nothing but the IP duality itself \eqref{IPconf}.

We now observe that the last diagram of Figure \ref{fig:NS5 dual 2} is the mass-deformed $E[USp(2 N)]$ theory dual to a WZ model discussed in \cite{Hwang:2020wpd}. 
Indeed comparing with the recursive definition \eqref{indexFEN} of $\mathcal I_{FE[USp(2 N)]}$ and the relation \eqref{eq:FEvsE} between $\mathcal I_{FE[USp(2 N)]}$ and $\mathcal I_{E[USp(2 N)]}$, we find that the last two lines of \eqref{eq:NS5 dual} are nothing but the index $\mathcal I_{E[USp(2 N)]}$ with specialized $\vec y$ variables
\begin{align}
&\mathcal I_{E[USp(2 N)]} \left(t^\frac{N-1}{2} v,\dots, t^{-\frac{N-1}{2}} v;x_N,\dots,x_1;pq/t;t^\frac{M}{2}c\right) \nonumber \\
&= \prod_{i = 1}^N \Gpq{t^{i-\frac{N-M+1}{2}} c v^{\pm1} x_1^{\pm1}} \oint \udl{\vec z_{N-1}} \Gd_{N-1}(z) \prod_{j = 1}^{N-1} \Gpq{(pq)^\frac{1}{2} t^{-\frac{N}{2}} v^{\pm1} z_j^{\pm1}} \nonumber \\
&\times\Gpq{(pq)^\frac12 t^{-\frac{M-1}{2}} c^{-1} x_1^{\pm1} z_j^{\pm1}}  \mathcal I_{FE[USp(2 N-2)]}(\vec z;x_N,\dots,x_2;pq/t;(pq)^{-\frac12} t^\frac{M+1}{2} c) \,.
\end{align}
By the sequential application of the IP duality, this theory was shown in \cite{Hwang:2020wpd} to be dual to a WZ model. The corresponding index identity is given by
\begin{align}
&\mathcal I_{E[USp(2 N)]} \left(t^\frac{N-1}{2} v,\dots, t^{-\frac{N-1}{2}} v;x_N,\dots,x_1;pq/t;t^\frac{M}{2}c\right) \nonumber \\
&= \Gpq{t^M c^2} \Gpq{t}^N \prod_{n < m}^N \Gpq{t x_n^{\pm1} x_m^{\pm1}} \prod_{i = 1}^N \frac{\Gpq{t^{-\frac{N-M-1}{2}} c v^{\pm1} x_i^{\pm1}}}{\Gpq{t^{M-i+1} c^2} \Gpq{t^i}}
\end{align}
which
leads to the identity \eqref{eq:generalized}.

Also in this case although we proved the identity \eqref{eq:generalized} by gauging the manifest $USp(2 M)$ symmetry,  using the spectral duality of $FE[USp(2 M)]$ we can show that this identity holds for any combination of $USp(2 M)$, either manifest or emergent; i.e. one can either swap $(\vec z,t^\frac{N-M-1}{2} v,\dots,t^{-\frac{N-M-1}{2}} v)$ and $\vec x$ in the left block as follows:
\begin{align}
\label{eq:generalized with CH}
&\prod_{i = 1}^{N-M} \Gpq{t^{-i+1} c^2} \prod_{i = 1}^{N-M} \Gpq{t^i} \oint \udl{\vec z_M} \Gd_M(z,t) \prod_{i = 1}^N \Gpq{t^\frac{N-M+1}{2} v^{\pm1} z_i^{\pm1}} \nonumber \\
&\quad \times \mathcal I_{FE[USp(2 N)]}(\vec x;\vec z,t^\frac{N-M-1}{2} v,\dots,t^{-\frac{N-M-1}{2}} v;t;c) \mathcal I_{FE[USp(2 M)]}(\vec z;\vec y;t;(pq/t)^\frac12 c^{-1}) \nonumber \\
&= \mathcal{I}^{(N,M)}_\bigtriangledown \left(\vec x;\vec y;v;t;c t^{-\frac{N-M}{2}}\right)
\end{align}
or swap $\vec z$ and $\vec y$ in the right block, or both, without changing the identity.

\subsection{Gluing with $2L$ fundamental chirals for $L > 1$}

\begin{figure}[t]
\centering
\includegraphics[width=\textwidth]{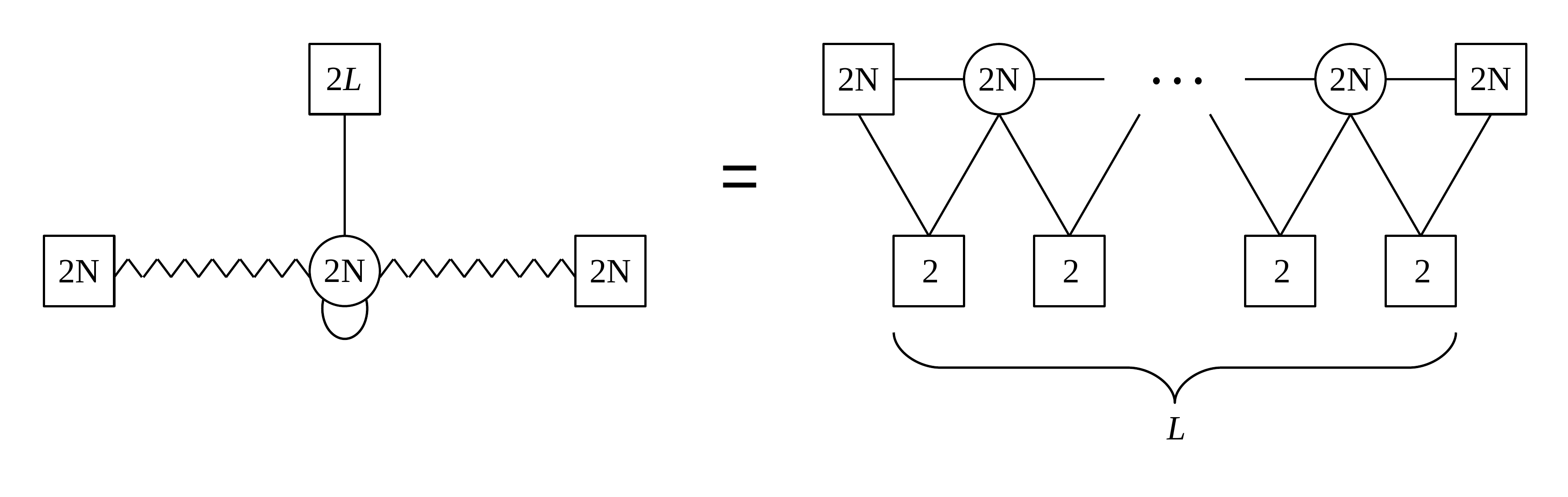}
\caption{The dual of the gluing of two $S$-walls with the insertion of $2L$ fundamental chirals fields. }
\label{fig:multi triangle}
\end{figure}

We now consider theory $\mathcal{T}_g^{2L}$ obtained by gluing two $S$-walls with the insertion of $2L$ fundamental fields $P^j$ for $j = 1,\dots,2 L$, which are rotated by an extra $USp(2 L)_v$ global symmetry. Similarly to the $L = 1$ case, we introduce the superpotential
\begin{align}
\mathrm{Tr}_z \left[\Phi \cdot \left(\mathsf{O_H}^L-\mathsf{O_H}^R\right)\right]+ \mathrm{Tr}_z \left[\mathrm{Tr}_v \left(P^2\right) \cdot \left(\mathsf{O_H}^L+\mathsf{O_H}^R\right)\right]\,,
\end{align}
where $\mathrm{Tr}_z$ and $\mathrm{Tr}_v$ are taken over the gauge $USp(2 N)_z$ and the global $USp(2 L)_v$ respectively. Once we integrate out the massive fields $\Phi$ and $\mathsf{O_H}^L-\mathsf{O_H}^R$, the remaining superpotential can be written as
\begin{align}
\label{eq:flavour braid}
\delta \mathcal W = \mathrm{Tr}_z \left[\mathrm{Tr}_v \left(P^2\right) \cdot A\right]\,,
\end{align}
where $A$ is defined by $A = \mathsf{O_H}^L+\mathsf{O_H}^R$.
The global symmetry of $\mathcal{T}_g^{2 L}$ is given
\begin{align}
USp(2 N)_x \times USp(2 N)_y \times USp(2 L)_v \times U(1)_c \times U(1)_t \,.
\end{align}
As we are about to see, $\mathcal{T}_g^{2L}$  admits a dual description given by the linear quiver  with $L-1$ gauge nodes shown in Figure \ref{fig:multi triangle}.

\begin{figure}[t]
\centering
\includegraphics[width=\textwidth]{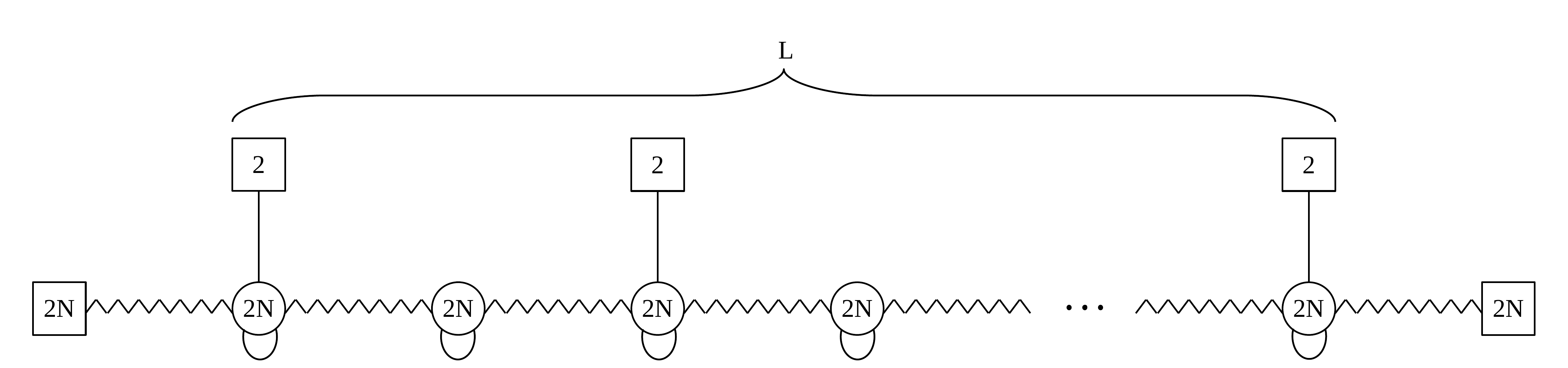}
\caption{The auxiliary dual frame for the gluing of two $S$-walls with $2L$ chiral fields obtained by the insertion of $L-1$ identity walls.}
\label{fig:decomposed flavor}
\end{figure}

This duality can be derived from the dualities we have obtained for $L = 0, \, 1$. 
The idea is to first split the  $2 L$ chirals into $L$ doublets by inserting $L-1$ Identity walls,
corresponding to pairs of $S$-walls glued together, as shown in Figure \ref{fig:decomposed flavor}.
This dual frame can also be regarded as a sequence of $L$ blocks formed by two $S$-walls glued with the insertion of a doublet of fundamental chirals.
We can then apply the duality of Figure \ref{fig:triangle} $L$ times to reach the dual frame on the right of Figure \ref{fig:multi triangle}.

In terms of the supersymmetric index, the duality can be expressed as follows:
\begin{align}
\label{2l2}
&  \oint \udl{\vec z_N} \Gd_N(\vec{z};t) \mathcal I_{FE[USp(2 N)]}(\vec z;\vec x;t;c) 
 \prod_{i = 1}^N \prod_{j= 1}^L \Gpq{t^\frac12 v_j^{\pm1} z_i^{\pm1}}
\mathcal I_{FE[USp(2 N)]}(\vec z;\vec y;t;(pq/t)^\frac{L}{2} c^{-1}) \nonumber \\
&=\oint   \udl{\vec w^{(1)}_N} \Gd_N(w^{(1)},t) \cdots  \oint  \udl{\vec w^{(L-1)}} \Gd_N(w^{(L-1)},t) \prod_{i = 1}^L \mathcal{I}^{(N,N)}_\bigtriangledown \left(\vec w^{(i-1)};\vec w^{(i)};v_i;t;(pq/t)^{-\frac{i-1}{2}} c\right)\,,
\end{align}
where $\vec w^{(0)}\equiv \vec x$ and $\vec w^{(L)}\equiv \vec y$ and $\mathcal{I}^{(N,M)}_\bigtriangledown \left(\vec x;\vec y;v;t;c\right)$ was defined in \eqref{eq:tri}.
Note that the $U(1)_c$ fugacity of the right block is fixed to $(pq/t)^\frac{L}{2} c^{-1}$  due to the anomaly cancelation and the superpotential \eqref{eq:flavour braid}. 

Again, while we have derived the duality for the theory with the Lagrangian gluing, this can be further extended to several non-Lagrangian dual frames with the gauging of emergent $USp(2 N)$ symmetries. Those can be achieved by using the spectral duality of each $FE[USp(2 N)]$ block; e.g. one can simply exchange $\vec z$ and $\vec x$ among the arguments of the left block without affecting the r.h.s.~using the spectral duality \eqref{flipselfduality}:
\begin{align}
& \oint \udl{\vec z_N} \Gd_N(z,t) \mathcal I_{FE[USp(2 N)]}(\vec x;\vec z;t;c) 
 \prod_{i = 1}^N \prod_{j= 1}^L \Gpq{t^\frac12 v_j^{\pm1} z_i^{\pm1}}
\mathcal I_{FE[USp(2 N)]}(\vec z;\vec y;t;(pq/t)^\frac{L}{2} c^{-1}) \nonumber \\
&= \oint   \udl{\vec w^{(1)}_N} \Gd_N(w^{(1)},t) \cdots  \oint  \udl{\vec w^{(L-1)}} \Gd_N(w^{(L-1)},t) \prod_{i = 1}^L \mathcal{I}^{(N,N)}_\bigtriangledown \left(\vec w^{(i-1)};\vec w^{(i)};v_i;t;(pq/t)^{-\frac{i-1}{2}} c\right)\,.
\end{align}
 In this case, we gauge the diagonal combination of the emergent $USp(2 N)$ of the left block and the manifest $USp(2 N)$ of the right block, with interaction
\begin{align}
\mathrm{Tr}_z \left[\Phi \cdot \left(\mathsf C^L-\mathsf{O_H}^R\right)\right]+ \mathrm{Tr}_z \left[\mathrm{Tr}_v \left(P^2\right) \cdot \left(\mathsf C^L+\mathsf{O_H}^R\right)\right] ,
\end{align}
where $\Phi$ and $\mathsf C^L-\mathsf{O_H}^R$ are massive and can be integrated out. In addition, one can also use the spectral duality of the right block, which simply exchanges $\vec z$ and $\vec y$ among its arguments.

The duality that we derived in this section, schematically represented in Figure \ref{fig:multi triangle}, is in fact one particular instance of the $4d$ mirror duality discussed in \cite{Hwang:2020wpd}. Specifically, up to gauge singlets, $\mathcal{T}_g^{2L}$, the theory on the left, coincides with $E_{\gr}^{\gs}[USp(2NL)]$ where $\gr=[(L-1)^N,1^N]$ and $\gs=[N^L]$, while the theory on the right can be identified with its mirror dual $E_{\gs}^{\gr}[USp(2N)]$. Hence, we found a derivation of this particular $4d$ mirror duality by iterative application of the IP duality. This leads us to wonder whether it is possible to derive all the mirror dualities, both in $4d$ and in $3d$, by iterative application of some more fundamental duality, like the IP duality in $4d$.


This actually turns out to be true, as it will be shown  in an upcoming paper \cite{prl},
where the dualities we derived in this section will play the role of basic duality moves that we apply locally 
in linear quivers to determine their mirror dual.
We will then need our dualities for generic values of the parameter $N,M,L$ which might include
cases in which our theories are not flowing to an interacting SCFT.
Actually determining the range of parameters for which our dual quiver theories flow to an interacting IR SCFT is quite complicated compared for example
with the case of Seiberg duality, which relates interacting theories if the number of colors and flavors is such that both theories are in the conformal window. Remember that in Seiberg duality, for a fixed value of, say, the number of colors, the range of the number of flavors that defines the conformal window is determined by requiring that the beta-function is negative for one of the two extreme of the range and that all the gauge invariant operators have their superconformal dimension above the unitarity bound for the other extreme. 
In our  theories determining the precise conformal window is more complicated for various reasons. First, the requirement on the negative beta-functions at each gauge node is not a necessary condition anymore. This is because several interactions are involved in our quiver theory, both gauge and superpotential interactions. 
The requirement on the dimension of gauge invariant operators is still a necessary condition, but verifying it is much more complicated than in Seiberg duality since the structure of the gauge invariant operators, as we saw, is more convoluted and moreover we have abelian symmetries that can mix with the R-symmetry in the IR.
 For these reasons, determining the precise conformal window for our dualities   is tricky and  we don't attempt this.
 In addition, for the reasons above, when our theories are coupled to other theories (as it will happen when we will use these dualities as basic duality moves) we would have to discuss again the IR dynamics as new interactions are generated.

\section{ $3d$ dualities and $SL(2,\mathbb{Z})$ relations}
\label{3dlimits}

In this section we will show how the basic properties of the $S$ and $T$ generators of $SL(2,\mathbb{Z})$
can be reinterpreted as dualities involving the $3d$ $S$-wall theory.
In particular we will show how to recover the $3d$ field theory realisations of the relations $S^2=-1$, $SS^{-1}=1$ and $T^{-1} S T=S^{-1} T S$. 
As we already mentioned we can equivalently identify the $3d$ $S$-wall with $T[SU(N)]$ or $FT[SU(N)]$ by just modifying the notion of gluings of walls, so we will always work with $FT[SU(N)]$.

%

\subsection{Gluing $3d$ $S$-walls and Identity walls}
\label{deltalimit}

In order to prove that gluing two $3d$ $S$-walls we get an Identity wall, we can either take the $3d$ limit followed by various real mass deformations of the $4d$ result or we can follow an iterative procedure as in Section \ref{psi}.
Here we will discuss in detail the first approach by showing how the various limits can be taken at the level indices and partition functions, giving just a quick description of the second approach.

\begin{figure}[t]
\centering
\includegraphics[width=1\textwidth]{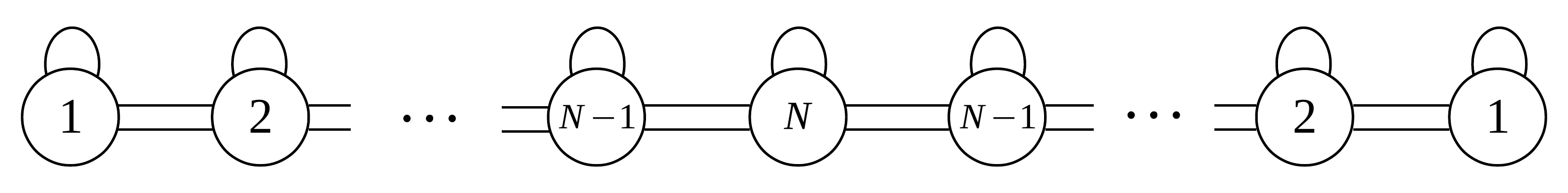}
\caption{The quiver representation of the Lagragian gluing of two $FT[U(N)]$ theories. 
Round nodes denote $U(n)$ gauge symmetries. Pairs of lines between adjacent nodes denote chiral fields in the bifundamental and anti-bifundamental representation of the two nodes symmetries, while arcs denote chiral fields in the adjoint representation of the corresponding node symmetry.}
\label{fig:3ddelta}
\end{figure}

For the iterative procedure we consider the Lagrangian gluing of two $3d$ $S$-walls, that is we gauge the manifest
$U(N)$ symmetries of two $FT[U(N)]$ tails as depicted in Figure \ref{fig:3ddelta}.

Notice that, since there is no flavor node in the quiver, an overall $U(1)$ in the gauge group is redundant and can be re-absorbed. The standard choice is to re-absorbe the $U(1)$ part of the middle $U(N)$ node, so to turn it into a $U(N)/U(1)=SU(N)/\mathbb{Z}_N$ node (we will see this at the level of partition functions later). In the following discussion we will neglect this redundancy in our description of the theory and keep the middle node to be $U(N)$.

The  fundamental duality in this case is  the Aharony duality\footnote{The Aharony duality relates $U(N_c)$ with $N_f$ flavors and no superpotential to $U(N_c-N_f)$ with $N_f$ flavors and $N_f^2+2$ singlets $X^a{}_b$, $S^{\pm1}$ flipping the dual mesons and monopoles $\mathcal{W}=X^a{}_bq_a\tilde{q}^b+S^-\mathfrak{M}^++S^+\mathfrak{M}^-$.} \cite{Aharony:1997gp}
which can be applied iteratively starting from the leftmost $U(1)$ node. The dualisation of each node among other things has the effect of removing the adjoint at the adjacent nodes so the dualisation can be iterated proceeding from the left to the right of the quiver.
As we pass the central $U(N)$ node the ranks start decreasing, and when we reach the right $U(2)$ node
this will only see two flavors and confine. So the original quiver splits into two $FT[U(N-1)]$ blocks gauged together 
and a decoupled $U(1)$ theory with a BF  coupling of the form:
\begin{align}
\int d^3 x d^2 \theta d^2\bar \theta \, \hat V \Sigma-\left(\int d^3 x d^2 \theta \, i \Phi \hat \Phi+ c.c. \right)
\end{align}
where $\Sigma$ and $\Phi$ are the dynamical linear and chiral multiplets inherited from the $\mathcal N = 4$ vector for the decoupled gauge $U(1)$, while $\hat V$ and $\hat \Phi$ are the background 
$\mathcal N = 2$ vector and chiral multiplets inherited from the background $\mathcal N = 4$ vector for a $U(1)$ global symmetry which is a combination of the topological symmetries
of the leftmost and of the rightmost $U(1)$ nodes of the original theory.
This is the analogue of what happens also in $4d$, where at the end of the first iteration of the IP duality we got a decoupled $SU(2)$ gauge theory with 4 chirals with the $SU(2)^2$ symmetry acting on them, originating from the leftmost and the rightmost $SU(2)$ symmetries of the saw of Figure \ref{fig:delta1}.
In $3d$ integrating over the dynamical fields will lead to a functional delta function setting the background fields to zero \cite{Kapustin:1999ha}
and so identifying the topological symmetries of the first and last nodes.
Iterating this procedure as in the $4d$ case we can split the theory obtained by gluing two $FT[U(N)]$ tails
into the product of $N$ $U(1)$ theories with BF couplings. The functional integration over these $N$ abelian nodes
yields the functional delta function identifying the topological symmetries of the left and of the right $FT[U(N)]$ tails.

We can alternatively obtain this result starting from the $4d$ gluing of two  $FE[USp(2N)]$ theories that gives an Identity wall and implementing the following steps which reduce the   $4d$ $FE[USp(2N)]$ theory to the  $3d$ $FT[U(N)]$ theory.

 We first compactify on a circle   so to get a $3d$ $\mathcal{N}=2$ quiver gauge theory that looks exactly as $FE[USp(2N)]$, but with the addition of a monopole superpotential that is dynamically generated in the reduction \cite{Aharony:2013dha,Aharony:2013kma}. Specifically, the fundamental monopole of each $USp$ gauge node is turned on in the superpotential, with the effect of breaking some abelian symmetries that were anomalous in $4d$. We will refer to the resulting theory as $FE[USp(2N)]^{3d}$. 
 
We then consider a combination of a Coulomb branch VEV that breaks all the gauge groups from $USp$ to $U$ combined with a real mass deformation so to follow the theory to a vacuum where half of the matter chiral fields remains massless. This gives the $3d$ $\mathcal{N}=2$ theory called $FM[U(N)]$ that was studied in \cite{Pasquetti:2019tix}\footnote{This theory was originally introduced in \cite{Pasquetti:2019tix} from a completely different perspective, that is exploiting a relation between $3d$ $\mathcal{N}=2$ dualities and $2d$ CFT correlators in the free field realisation \cite{Pasquetti:2019uop}. From this perspective, the $FM[SU(N)]$ theory corresponds to the $3d$ gauge theory avatar of the kernel function used to manipulate the free field correlators of Liouville theory in \cite{Fateev:2007qn}.}. In particular in this theory  the fundamental monopoles of magnetic charge $\pm1$ of every unitary gauge node is turned on in the superpotential.

Finally, we consider a further real mass deformation for what used to be the $U(1)_c$ symmetry in $4d$, which has the effect of giving mass to all of the fields of the saw and of lifting the monopole superpotential. 
The result is that now we have two $FT[U(N)]$ theories glued together yielding an Identity wall operator identifying
the topological symmetries of the two tails.

Let's see how these series of limits is implemented at the level of supersymmetric partition functions.
The starting point is the $4d$ delta-identity \eqref{dfhh}, which we report again here for reference
\begin{align}
\label{dfhhbis}
\mathcal{I}^N_{g} &=\oint\udl{\vec{z}_N}\Gd_N(\vec{z};t)\mathcal{I}_{FE[USp(2N)]}(\vec{z};\vec{x};c;t)\mathcal{I}_{FE[USp(2N)]}(\vec{z};\vec{y};c^{-1};t) \nonumber \\
&= \frac{\prod_{i=1}^N 2\pi i x_i}{\Gd_N(\vec x;t)} \sum_{\sigma \in S_N} \sum_\pm \prod_{i=1}^N \delta\left(x_i-y_{\sigma(i)}^{\pm1}\right) .
\end{align}
The first step consists of taking the limit from the $\mathbb{S}^3\times\mathbb{S}^1$ partition function of $FE[USp(2N)]$ to the $\mathbb{S}^3_b$ partition function of the theory obtained from circle compactification (see \cite{Kapustin:2009kz,Jafferis:2010un,Hama:2010av,Hama:2011ea} and Appendix \ref{S3bpf} for our conventions on the squashed three-sphere partition function). For this purpose, we redefine the parameters of \eqref{dfhhbis} as follows:
\be
\label{para}
&&z_{L/R,a}^{(n)}=\e^{2\pi irZ_{L/R,a}^{(n)}},\quad n=1,\cdots,N-1,\,\,a=1,\cdots,n,\nn\\
&&x_i=\e^{2\pi irX_i},\quad y_i=\e^{2\pi irY_i},\quad z_i=\e^{2\pi irZ_i},\quad i=1,\cdots,N,\nn\\
&&c=\e^{2\pi ir\Gd},\quad t=\e^{2\pi ir(iQ-2m_A)},\quad p=\e^{-2\pi rb},\quad q=\e^{-2\pi rb^{-1}}\,,
\ee
where $r$ is interpreted as the radius of $\mathbb{S}^1$ and all the new parameters in capital letters are taken to live in $\left[-\frac{1}{2r},\frac{1}{2r}\right]$, so that in the $r\to0$ limit we recover the real axis which is the standard domain for parameters of the $\mathbb{S}^3_b$ partition function. Moreover, in order to conform with the conventions already appering in the literature, we renamed the $U(1)_c$ symmetry in $3d$ with $U(1)_\Gd$. At the level of the integrand of the $\mathbb{S}^3\times\mathbb{S}^1$ partition function, the limit is taken using the following property that relates the elliptic gamma function to the double-sine function, in terms of which the contribution of $3d$ $\mathcal{N}=2$ multiplets to the $\mathbb{S}^3_b$ partition function can be written:
\be
\lim_{r\to0}\Gc_e\left(\e^{2\pi irx};p=\e^{-2\pi rb},q=\e^{-2\pi rb^{-1}}\right)=\e^{-\frac{i\pi}{6r}\left(i\frac{Q}{2}-x\right)}\sbfunc{i\frac{Q}{2}-x}\,,
\ee
where $Q=b+b^{-1}$. Using this, one can show that \cite{Pasquetti:2019hxf}
\be
\lim_{r\to 0}\mathcal{I}_{FE[USp(2N)]}(\vec{x};\vec{y};t;c)=C_N\,\mathcal{Z}_{FE[USp(2N)]^{3d}}(\vec{X};\vec{Y};m_A;\Gd)\,,
\ee
where we defined the $\mathbb{S}^3_b$ partition function of $FE[USp(2N)]^{3d}$ as
\be
&&\mathcal{Z}_{FE[USp(2N)]^{3d}}(\vec{X};\vec{Y};\Gd;m_A)=\sbfunc{-i\frac{Q}{2}+2\Gd}\sbfunc{i\frac{Q}{2}-2m_A}^N\times\nn\\
&&\times\prod_{i<j}^N\sbfunc{i\frac{Q}{2}\pm X_i\pm X_j-2m_A}\prod_{i=1}^N\sbfunc{i\frac{Q}{2}\pm Y_N\pm X_i-\Gd}\times\nn\\
&&\times\int\frac{\prod_{a=1}^{N-1}\udl{Z^{(N-1)}_a}}{2^{N-1}(N-1)!}\frac{\prod_{a=1}^{N-1}\sbfunc{Y_N\pm Z_a^{(N-1)}-m_A+\Gd}\prod_{i=1}^N\sbfunc{Z_a^{(N-1)}\pm X_i+m_A}}{\prod_{a=1}^{N-1}\sbfunc{i\frac{Q}{2}\pm2 Z^{(N-1)}_a}\prod_{a<b}^{N-1}\sbfunc{i\frac{Q}{2}\pm Z^{(N-1)}_a\pm Z^{(N-1)}_b}}\times\nn\\
&&\times\mathcal{Z}_{FE[USp(2N-2)]^{3d}}(Z_1^{(N-1)},\cdots,Z_{N-1}^{(N-1)};Y_1,\cdots,Y_{N-1};m_A;\Gd+m_A-i\frac{Q}{2})
\ee
and the divergent prefactor is
\be
C_N&=&\left[r\left(\e^{-2\pi rb};\e^{-2\pi rb}\right)_\infty\left(\e^{-2\pi rb^{-1}};\e^{-2\pi rb^{-1}}\right)_\infty\right]^{\frac{N(N-1)}{2}}\times\nn\\
&\times&\exp\left[-\frac{i\pi}{6r}\left(i\frac{Q}{4}N(9N-1)-2N(2N-1)m_A-2N\Gd\right)\right]\,.
\ee
Instead, the limit of the contribution of a $USp(2N)$ vector multiplet together with that of a $USp(2N)$ antisymmetric chiral is
\be
\lim_{r\to 0}\frac{\Gd_N(\vec{z};t)}{\left[(p;p)(q;q)\right]^N}&=&\exp\left[-\frac{i\pi}{6r}\left(2N(2N-1)m_A+i\frac{Q}{2}N(1-4N)\right)\right]\times\nn\\
&\times&\frac{\sbfunc{-i\frac{Q}{2}+2m_A}^N\prod_{i<j}^N\sbfunc{-i\frac{Q}{2}\pm Z_i\pm Z_j+2m_A}}{\prod_{i=1}^N\sbfunc{i\frac{Q}{2}\pm2 Z_i}\prod_{i<j}^N\sbfunc{i\frac{Q}{2}\pm Z_i\pm Z_j}}\,.\nn\\
\ee
Combining these results, we find that the $3d$ limit of \eqref{dfhhbis} yields
\be\label{3ddeltaeusp}
&&C\int\frac{\prod_{i=1}^N\udl{Z_i}}{2^N N!}\frac{\sbfunc{-i\frac{Q}{2}+2m_A}^N\prod_{i<j}^N\sbfunc{-i\frac{Q}{2}\pm Z_i\pm Z_j+2m_A}}{\prod_{i=1}^N\sbfunc{i\frac{Q}{2}\pm2 Z_i}\prod_{i<j}^N\sbfunc{i\frac{Q}{2}\pm Z_i\pm Z_j}}\times\nn\\
&&\times\mathcal{Z}^{3d}_{FE[USp(2N)]}(\vec{Z};\vec{X};\Gd;m_A)\mathcal{Z}^{3d}_{FE[USp(2N)]}(\vec{Z};\vec{Y};-\Gd;m_A)=\nn\\
&&=\frac{\prod_{i=1}^N\sbfunc{i\frac{Q}{2}\pm 2X_i}\prod_{i<j}^N\sbfunc{i\frac{Q}{2}\pm X_i\pm X_j}}{\sbfunc{-i\frac{Q}{2}+2m_A}^N\prod_{i<j}^N\sbfunc{-i\frac{Q}{2}\pm X_i\pm X_j+2m_A}}\sum_{\gs\in S_N}\sum_{\pm}\prod_{i=1}^N\gd(X_i\pm Y_{\gs(i)})\,.\nn\\
\ee
The overall divergent prefactor coming from both sides of the identity
\be
C=\left[r\left(\e^{-2\pi rb};\e^{-2\pi rb}\right)_\infty\left(\e^{-2\pi rb^{-1}};\e^{-2\pi rb^{-1}}\right)_\infty\right]^{N(N+1)}\exp\left[-\frac{i\pi}{6r}\left(i\frac{Q}{2}N(N+1)\right)\right]\underset{r\to0}{\longrightarrow}1\nn\\
\ee
turns out to be trivial as one can easily show using the following asymptotic behaviour of the $q$-Pochhammer symbol \cite{Fredenhagen:2004cj}:
\be
\lim_{\gb\to 0}(\e^{-2\gb};\e^{-2\gb})_\infty=\sqrt{\frac{\pi}{\gb}}\exp\left[-\frac{\pi^2}{12\gb}\right]\,.
\ee
The identity \eqref{3ddeltaeusp} we obtained is the analogue of \eqref{dfhhbis} for the theory associated to the circle compactification of $FE[USp(2N)]$.

We now consider the combination of Coulomb branch VEV and real mass deformation that makes us flow to the $FM[U(N)]$ theory of \cite{Pasquetti:2019tix}. This limit can be implemented at the level of the $\mathbb{S}^3_b$ partition function considering the following scaling of the parameters:
\be\label{rmMUN}
X_i\to X_i+s,\quad Y_i\to Y_i+s,\quad Z_i\to Z_i+s,\quad Z_{L/R,a}^{(n)} \to Z_{L/R,a}^{(n)}+s, \quad s\to+\infty
\ee
and using the following property of the double-sine function:
\be
\lim_{x\to\pm\infty}\sbfunc{x}=\e^{\pm i\frac{\pi}{2}x^2}\,.
\ee
Using this, one can show that \cite{Pasquetti:2019hxf}
\be \label{3dlimiteusp}
\mathcal{Z}_{FE[USp(2N)]^{3d}}(\vec{X};\vec{Y};m_A;\Gd)&\to& K_N \prod_{j=1}^N\sbfunc{-i\frac{Q}{2}+2\Gd+2(j-1)\left(m_A-i\frac{Q}{2}\right)}\times\nn\\
&\times&\mathcal{Z}_{FM[U(N)]}(\vec{X};\vec{Y};m_A;\Gd)\,,
\ee
where we defined the $\mathbb{S}^3_b$ partition function of the $FM[U(N)]$ theory as 
\be
&&\mathcal{Z}_{FM[U(N)]}(\vec{X};\vec{Y};\Gd;m_A)=\prod_{i=1}^Ns_b\left(i\frac{Q}{2}\pm(X_i-Y_N)-\Gd\right)\prod_{i,j=1}^N\sbfunc{i\frac{Q}{2}\pm(X_i-X_j)-2m_A}\times\nn\\
&&\times\int\frac{\prod_{a=1}^{N-1}\udl{Z_a^{(N-1)}}}{(N-1)!}\frac{\prod_{a=1}^{N-1}s_b\left(\pm(Z_a^{(N-1)}-Y_N)+\Gd-m_A\right)\prod_{i=1}^Ns_b\left(\pm(Z_a^{(N-1)}-X_i)+m_A\right)}{\prod_{a<b}^{N-1}s_b\left(i\frac{Q}{2}\pm(Z_a^{(N-1)}-Z_b^{(N-1)})\right)}\times\nn\\
&&\times \mathcal{Z}_{FM[U(N-1)]}\left(Z_1^{(N-1)},\cdots,Z_{N-1}^{(N_1)};Y_1,\cdots,Y_{N-1};m_A;\Gd+m_A-i\frac{Q}{2}\right)\nn\\
\label{pfmsun}
\ee
and the divergent prefactor is
\be
K_N=\exp\left\{2\pi i\left[i\frac{Q}{2}-\Gd+(N-1)\left(i\frac{Q}{2}-m_A\right)\right]\left[2Ns+\sum_{i=1}^N(X_i+Y_i)\right]\right\}\,.
\ee
The partition function of $FM[SU(N)]$ can be obtained from that of $FM[U(N)]$ by just imposing the tracelessness conditions $\sum_{i=1}^NX_i
=\sum_{i=1}^NY_i=0$. Instead, the limit of the contribution of a $USp(2N)$ vector multiplet together with that of a $USp(2N)$ antisymmetric chiral is
\be \label{3dlimitvector}
&&\frac{\sbfunc{-i\frac{Q}{2}+2m_A}^N\prod_{i<j}^N\sbfunc{-i\frac{Q}{2}\pm Z_i\pm Z_j+2m_A}}{\prod_{i=1}^N\sbfunc{i\frac{Q}{2}\pm2 Z_i}\prod_{i<j}^N\sbfunc{i\frac{Q}{2}\pm Z_i\pm Z_j}}\nn\\
&&\qquad\to\exp\left[2\pi i\left(2(N-1)m_A-iNQ\right)\left(Ns+\sum_{i=1}^NZ_i\right)\right]\frac{\prod_{i,j=1}^N\sbfunc{-i\frac{Q}{2}\pm(Z_i-Z_j)+2m_A}}{\prod_{i<j}^N\sbfunc{i\frac{Q}{2}\pm(Z_i-Z_j)}}\,.\nn\\
\ee
This is the contribution of a $U(N)$ vector multiplet and a $U(N)$ adjoint chiral multiplet. Indeed, as we mentioned before, the effect of this limit is of breaking all the symplectic groups, both of gauge and flavor symmetries, down to their unitary subgroups. Combining these results, we find that after such a limit \eqref{3ddeltaeusp} reduces to
\be\label{3ddeltaMUN}
&&K\prod_{j=1}^N\sbfunc{-i\frac{Q}{2}\pm 2\Gd+2(j-1)\left(m_A-i\frac{Q}{2}\right)}\times\nn\\
&&\times\int\frac{\prod_{i=1}^{N}\udl{Z_i}}{N!}\frac{\sbfunc{-i\frac{Q}{2}+2m_A}^{N}\prod_{i<j}^{N}\sbfunc{-i\frac{Q}{2}\pm (Z_i- Z_j)+2m_A}}{\prod_{i<j}^{N}s_b\left(i\frac{Q}{2}\pm(Z_i-Z_j)\right)}\times\nn\\
&&\times\mathcal{Z}_{FM[U(N)]}(\vec{Z};\vec{X};m_A;\Gd)\mathcal{Z}_{FM[U(N)]}(\vec{Z};\vec{Y};m_A;-\Gd)=\nn\\
&&=\frac{\prod_{i<j}^N\sbfunc{i\frac{Q}{2}\pm (X_i- X_j)}}{\sbfunc{-i\frac{Q}{2}+2m_A}^N\prod_{i<j}^N\sbfunc{-i\frac{Q}{2}\pm (X_i- X_j)+2m_A}}\sum_{\gs\in S_N}\prod_{i=1}^N\gd(X_i-Y_{\gs(i)})\,.\nn\\
\ee
The divergent prefactors on the two sides of the identity cancel out yielding a finite result, but we still have an overall prefactor
\be
K=\exp\left[2\pi i\left((N-1)m_A-\Gd-i\frac{Q}{2}N\right)\left(\sum_{i=1}^NX_i-Y_i\right)\right]=1\,,
\ee
which turns out to be trivial once we enforce the contraint $X_i=Y_{\gs(i)}$ due to the delta. The identity \eqref{3ddeltaMUN} we obtained is the analogue of \eqref{dfhhbis} for the $FM[U(N)]$ theory.

We remark at this point that one could have equivalently considered the limit in which the $Y_i$ parameters are sent to $-\infty$, namely $Y_i\to Y_i-s$. In this case, we would have obtained the same identity \eqref{3ddeltaMUN} but with the exchange $Y_i\to -Y_i$. The two choices correspond to two different embeddings of $U(N)$ inside $USp(2N)$ which are related by an element of the Weyl group of $USp(2N)$ which is not in the Weyl group of its $U(N)$ subgroup. This observation will be important in a moment.

We conclude the series of limits by considering the real mass deformation with respect to the parameter $\Gd$
\be\label{rmTUN}
\Gd\to\Gd+s,\qquad s\to+\infty\,,
\ee
after which, at it was shown in \cite{Pasquetti:2019tix}, the partition function of $FM[U(N)]$ reduces to that of $FT[U(N)]$\footnote{The same series of limits can be used to flow from $E[USp(2N)]$ to $T[SU(N)]$. In that case, the $U(1)_t$ symmetry is interpreted as the $U(1)$ axial symmetry that is the commutant of the $U(1)_R$ R-symmetry of $\mathcal{N}=2$ inside the $SU(2)_H\times SU(2)_C$ R-symmetry of $\mathcal{N}=4$ \cite{Tong:2000ky}. More precisely, the embedding is such that $U(1)_R=U(1)_H+U(1)_C$, where $U(1)_H\subset SU(2)_H$ and $U(1)_C\subset SU(2)_C$, and $U(1)_t$ is identified with the opposite combination.}
\be
\mathcal{Z}_{FM[U(N)]}(\vec{X};\vec{Y};\Gd;m_A)\to\Omega_N\e^{-i\pi\left(\sum_{i=1}^NX_i^2+Y_i^2\right)}\mathcal{Z}_{FT[U(N)]}(\vec{X};\vec{Y};m_A)
\ee
where the prefactor is
\be
\Omega_N&=&\exp\left\{i\pi\left[\frac{1}{12} N \left(-12 (\Delta+s)^2-8 m_A^2 (N-2) (N-1)+4 i m_A (N-1) ((2 N-1) Q+\right.\right.\right.\nn\\
&&\quad\left.\left.\left.+6 i (\Delta+s) )+\left(2 N^2+1\right) Q^2+12 i (\Delta+s)  N Q\right)\right]\right\}
\ee
and we defined the $\mathbb{S}^3_b$ partition function of $FT[U(N)]$ as
\be
&&\mathcal{Z}_{FT[U(N)]}(\vec{X};\vec{Y};m_A)=\e^{2\pi iY_N\sum_{i=1}^NX_i}\prod_{i,j=1}^N\sbfunc{i\frac{Q}{2}\pm(X_i-X_j)-2m_A}\times\nn\\
&&\quad\times\int\frac{\prod_{a=1}^{N-1}\udl{Z_a}}{(N-1)!}\e^{-2\pi i Y_N\sum_{a=1}^{N-1}Z_a^{(N-1)}}\frac{\prod_{a=1}^{N-1}\prod_{i=1}^Ns_b\left(\pm(Z_a^{(N-1)}-X_i)+m_A\right) }{\prod_{a<b}^{N-1}s_b\left(i\frac{Q}{2}\pm(Z_a^{(N-1)}-Z_b^{(N-1)})\right)} \times\nn\\
&&\quad\times\mathcal{Z}_{FT[U(N-1)]}\left(Z_1^{(N-1)},\cdots,Z_{N-1}^{(N-1)};Y_1,\cdots,Y_{N-1};m_A;\right)\, .\nn\\
\label{pftsun}
\ee
Again, the partition function of $FT[SU(N)]$ can be obtained from that of $FT[U(N)]$ by just imposing the tracelessness conditions $\sum_{i=1}^NX_i
=\sum_{i=1}^NY_i=0$, or equivalently, by gauging the diagonal $U(1)$ symmetries of both $U(N)_X$ and $U(N)_Y$. This is because the matter part is independent of such diagonal $U(1)_X \times U(1)_Y \subset U(N)_X \times U(N)_Y$, while $U(1)_X$ and $U(1)_Y$ have a BF coupling between them, which can be easily seen by shifting the variables as follows:
\begin{align}
\begin{aligned}
Z_a^{(n)} \quad &\longrightarrow \quad Z_a^{(n)} + w \,, \\
X_i \quad &\longrightarrow \quad \tilde X_i + w \,, \\
Y_i \quad &\longrightarrow \quad \tilde Y_i + u
\end{aligned}
\end{align}
where $\tilde X_i$ and $\tilde Y_i$ are defined such that $\sum_{i = 1}^N \tilde X_i = \sum_{i = 1}^N \tilde Y_i = 0$. Therefore, $w$ and $u$ parametrize $U(1)_X$ and $U(1)_Y$ respectively. Since the representation of each matter is either bifundamental or adjoint, $w$ in the matter part of the partition function completely cancels out. Indeed, $w$ and $u$ only appear in the exponential factors corresponding to the BF couplings; collecting those exponentials, we get
\begin{align}
&e^{2 \pi i \left[Y_N \sum_{i = 1}^N X_i+\left(Y_{N-1}-Y_N\right) \sum_{a = 1}^{N-1} Z^{(N-1)}_a+\dots+\left(Y_1-Y_2\right) Z^{(1)}_1\right]} \nonumber \\
&\longrightarrow \quad e^{2 \pi i \left[N w u+\left(\tilde Y_{N-1}-\tilde Y_N\right) \sum_{a = 1}^{N-1} Z^{(N-1)}_a+\dots+\left(\tilde Y_1-\tilde Y_2\right) Z^{(1)}_1\right]} \,.
\end{align}
The first term in the exponent corresponds to the BF coupling between $U(1)_X$ and $U(1)_Y$, whereas the remaining terms show the BF coupling between each gauge node and the associated topological symmetry, which is nothing but the FI term. If we focus on the BF coupling between $U(1)_X$ and $U(1)_Y$, the exponential factor gives rise to a delta function $\delta(N w)$ once it is integrated over $u$. 
This delta can be removed by further integrating over $Nw$. What we are left with is exactly the partition function of $FT[SU(N)]$.


We also use
\be
\prod_{i=1}^N\sbfunc{-i\frac{Q}{2}\pm 2\Gd+2(i-1)\left(m_A-i\frac{Q}{2}\right)}\to\exp\left[2\pi i\left(2(N-1)m_A-iNQ\right)\left(s+\Gd\right)\right]\,,\nn\\
\ee
to see that after such a limit \eqref{3ddeltaMUN} reduces to
\be\label{3ddeltaTUN}
&&\int\frac{\prod_{i=1}^{N}\udl{Z_i}}{N!}\frac{\sbfunc{-i\frac{Q}{2}+2m_A}^{N}\prod_{i<j}^{N}\sbfunc{-i\frac{Q}{2}\pm (Z_i- Z_j)+2m_A}}{\prod_{i<j}^{N}s_b\left(i\frac{Q}{2}\pm(Z_i-Z_j)\right)}\times\nn\\
&&\times\mathcal{Z}_{FT[U(N)]}(\vec{Z};\vec{X};m_A)\mathcal{Z}_{FT[U(N)]}(\vec{Z};-\vec{Y};m_A)=\nn\\
&&=\frac{\prod_{i<j}^N\sbfunc{i\frac{Q}{2}\pm (X_i- X_j)}}{\sbfunc{-i\frac{Q}{2}+2m_A}^N\prod_{i<j}^N\sbfunc{-i\frac{Q}{2}\pm (X_i- X_j)+2m_A}}\sum_{\gs\in S_N}\prod_{i=1}^N\gd(X_i-Y_{\gs(i)})\,,\nn\\
\ee
where again we simplified an overall prefactor which becomes trivial after imposing the contraint $X_i=Y_{\gs(i)}$ due to the delta. Notice that this identity corresponds to the gluing of two $FT[U(N)]$, while that of $FT[SU(N)]$ can be obtained by gauging the diagonal $U(1) \subset U(N)_X \times U(N)_Y$. Again we implement the shifts of variables
\begin{align}
\begin{aligned}
Z_{L,R,a}^{(n)} \quad &\longrightarrow \quad Z_{L,R,a}^{(n)} + v \,, \qquad n = 1, \dots, N-1 \,, \\
Z_{i} \quad &\longrightarrow \quad \tilde Z_{i}+ v \,, \\
X_i \quad &\longrightarrow \quad \tilde X_i + w \,, \\
Y_i \quad &\longrightarrow \quad \tilde Y_i + u
\end{aligned}
\end{align}
where $\sum_{i = 1}^N \tilde Z_{i} = \sum_{i = 1}^N \tilde X_i = \sum_{i = 1}^N \tilde Y_i = 0$ is satisfied. Note that $u, \, v$ and $w$ parametrize the diagonal $U(1)$'s of $U(N)_X$, $U(N)_Y$ and the gauged $U(N)_Z$ respectively. For the same reason we explained for a single $FT[U(N)]$, the matter part of the l.h.s. is independent of $u, \, v$ and $w$, which only appear in the exponential factors. Collecting the exponential factors depending on them, we get
\begin{align}
e^{2 \pi i N v (w-u)} \,.
\end{align}
Thus, once we gauge the anti-diagonal combination of $U(1)_X$ and $U(1)_Y$, i.e., integrate over $w-u$, we get a delta function $\delta(N v)$. Also note that the original $N$-dimensional integration can be decomposed into the one-dimensional integration over $v$ and the $(N-1)$-dimensional integration over $\tilde Z_i$ with a constraint $\sum_{i = 1}^{N} \tilde Z_i = 0$; more precisely,
\begin{align}
\label{eq:measure}
\prod_{i=1}^{N}\udl{Z_i} = N \, \udl v \prod_{i=1}^{N-1}\udl{\tilde Z_i} \,.
\end{align}
Since $v$ only appears in the delta function we got, we can explicitly perform the $v$-integration. Then $\delta(N v)$ gives $1/N$, which cancels the Jacobian determinant $N$ in \eqref{eq:measure}. The remaining integral is then exactly the gluing of two $FT[SU(N)]$.

One can perform the same operation on the right hand side shifting $X_i \rightarrow \tilde X_i + w, \, Y_i \rightarrow \tilde Y_i + u$ and integrating over $w-u$. The shift only affects the product of the delta functions, which can be written as
\begin{align}
\prod_{i=1}^N\gd(X_i-Y_{\gs(i)}) = \frac{1}{N} \, \gd(w-u) \prod_{i=1}^{N-1}\gd(\tilde X_i-\tilde Y_{\gs(i)}) \,,
\end{align}
where $\gd(w-u)$ becomes one if we perform the integration over $w-u$. Therefore, the final duality we get is
\begin{align}
\label{3ddeltaTSUN}
&\int\frac{\prod_{i=1}^{N-1}\udl{\tilde Z_i}}{N!}\frac{\sbfunc{-i\frac{Q}{2}+2m_A}^{N}\prod_{i<j}^{N}\sbfunc{-i\frac{Q}{2}\pm (\tilde Z_i- \tilde Z_j)+2m_A}}{\prod_{i<j}^{N}s_b\left(i\frac{Q}{2}\pm(\tilde Z_i-\tilde Z_j)\right)}\times\nn\\
&\times\mathcal{Z}_{FT[SU(N)]}\left(\vec{\tilde Z};\vec{\tilde X};m_A\right)\mathcal{Z}_{FT[SU(N)]}\left(\vec{\tilde Z};-\vec{\tilde Y};m_A\right)=\nn\\
&=\frac{\prod_{i<j}^N\sbfunc{i\frac{Q}{2}\pm (\tilde X_i- \tilde X_j)}}{\sbfunc{-i\frac{Q}{2}+2m_A}^N\prod_{i<j}^N\sbfunc{-i\frac{Q}{2}\pm (\tilde X_i- \tilde X_j)+2m_A}} \frac{1}{N} \sum_{\gs\in S_N}\prod_{i=1}^{N-1} \gd(\tilde X_i-\tilde Y_{\gs(i)})
\end{align}
with $\sum_{i = 1}^N \tilde Z_{i} = \sum_{i = 1}^N \tilde X_i = \sum_{i = 1}^N \tilde Y_i = 0$.
%
This duality can be interpreted as the relation $S^{-1}S=1$ for the $S$ generator of $SL(2,\mathbb{Z})$ once we identify either $FT[SU(N)]$ or $T[SU(N)]$ with the $S$-wall \cite{Gulotta:2011si,Assel:2014awa}. 

There is another relation satisfied by the $S$ element, namely $S^2=-1$. One can interpret it as \eqref{3ddeltaTSUN} where we redefined $Y_i\to- Y_i$, that is
\begin{align}
&\int\frac{\prod_{i=1}^{N-1}\udl{\tilde Z_i}}{N!}\frac{\sbfunc{-i\frac{Q}{2}+2m_A}^{N}\prod_{i<j}^{N}\sbfunc{-i\frac{Q}{2}\pm (\tilde Z_i- \tilde Z_j)+2m_A}}{\prod_{i<j}^{N}s_b\left(i\frac{Q}{2}\pm(\tilde Z_i-\tilde Z_j)\right)}\times\nn\\
&\times\mathcal{Z}_{FT[SU(N)]}\left(\vec{\tilde Z};\vec{\tilde X};m_A\right)\mathcal{Z}_{FT[SU(N)]}\left(\vec{\tilde Z};\vec{\tilde Y};m_A\right)=\nn\\
&=\frac{\prod_{i<j}^N\sbfunc{i\frac{Q}{2}\pm (\tilde X_i- \tilde X_j)}}{\sbfunc{-i\frac{Q}{2}+2m_A}^N\prod_{i<j}^N\sbfunc{-i\frac{Q}{2}\pm (\tilde X_i- \tilde X_j)+2m_A}} \frac{1}{N} \sum_{\gs\in S_N}\prod_{i=1}^{N-1} \gd(\tilde X_i+\tilde Y_{\gs(i)}) \,.
\end{align}
We already pointed out before that this other identity can be obtained by considering a different scaling in the limit that made us flow from $FE[USp(2N)]^{3d}$ to $FM[U(N)]$ and that the two choices are related by an element of the Weyl group of $USp(2N)$. Hence, while in $3d$ the identities associated to $SS^{-1}=1$ and $S^2=-1$ differ by acting with complex conjugation on one of the two $SU(N)$ global symmetries, in $4d$ they collapse on a single equation since such action is an element of the $USp(2N)$ global symmetry.

\subsection{Braid duality and the $S^{-1} T S = T^{-1} S T$ relation}
\label{3dlimitbraid}

\begin{figure}[t]
\centering
\includegraphics[width=1\textwidth]{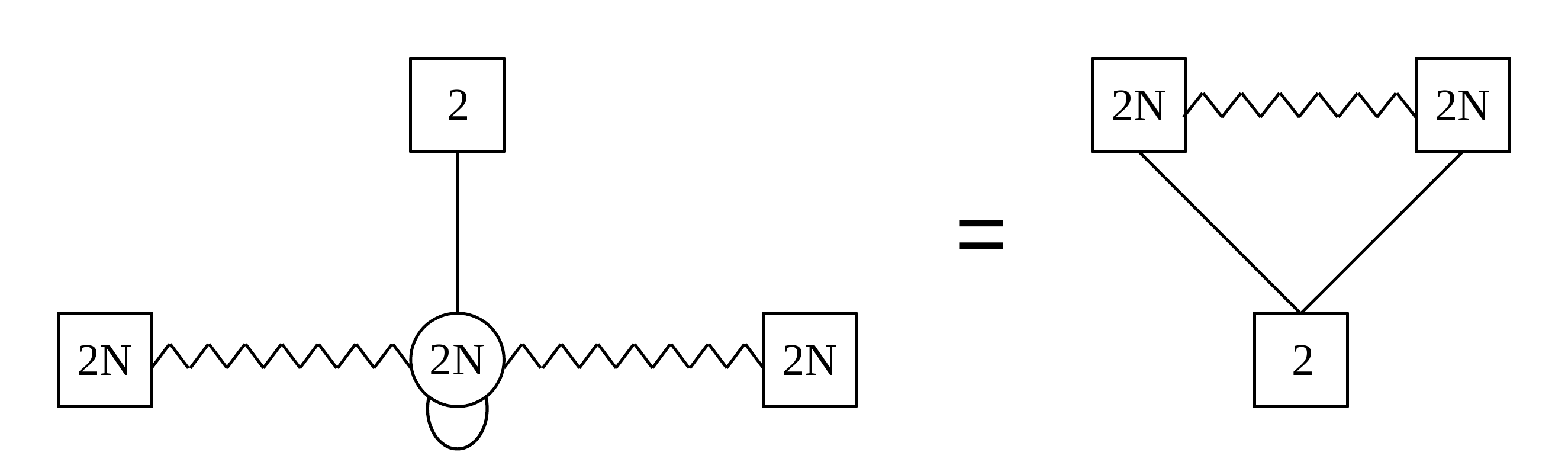}
\caption{A schematic representation of the braid duality. We remind the reader that the key difference with the duality of Figure \ref{fig:triangle} is that the doublet of chirals in the theory on the left is not interacting with the antisymmetric.}
\label{fig:braid}
\end{figure}

In order to derive the $S^{-1} T S = T^{-1} S T$ relation between the $S$ and $T$ generators of $SL(2,\mathbb{Z})$
we consider the braid duality depicted in Figure \ref{fig:braid} which was discussed in \cite{Pasquetti:2019hxf}.
Here we glue two $FE[USp(2N)]$  blocks by  gauging a diagonal combination of one $USp(2N)$ symmetry from each of the two blocks.

By now we have understood that it is equivalent to consider the manifest or the emergent symmetry, but for definiteness let us consider the Lagrangian gauging. In the process we also add, as usual, an antisymmetric chiral field which couples to the antisymmetric operators of the two blocks
\be\label{ptbraid}
\gd\mathcal{W}=\mathrm{Tr}_z \left[\Phi \cdot \left(\mathsf{O_H}^L-\mathsf{O_H}^R\right)\right]
\ee
and a doublet of fundamental chirals $(P^+,P^-)$ which are not involved in any superpotential interaction.
This is the key difference w.r.t.~the case considered in Figure \ref{fig:triangle}. As a consequence, while the ``$U(1)_t$" symmetries of the two blocks are still identified by the superpotential \eqref{ptbraid}, the ``$U(1)_c$" symmetries both survive as independent symmetries, which we call $U(1)_c$ and $U(1)_d$. The full global symmetry of the theory is thus
\be
USp(2N)_x\times USp(2N)_y\times U(1)_t\times U(1)_c\times U(1)_d\,,
\ee
where as usual the two $USp(2N)$ are enhanced from the symmetries of the saws of the two $FE[USp(2N)]$ blocks if we consider the Lagrangian gluing.

The  dual theory is one $FE[USp(2N)]$ block coupling to two sets of $2\times 2N$ chirals $L$ and $R$ via the superpotential
\be
\label{spn}
\gd\mathcal{W}=\epsilon_{\ga\gb}\Tr_z\left[L^\ga\Pi R^\gb\right]\,,
\ee
where we recall that $\Pi$ is the operator in the bifundamental representation of $USp(2N)_x\times USp(2N)_y$ symmetries of $FE[USp(2N)]$ and $\ga$, $\gb$ are $SU(2)_v$ flavor indices. Hence $L$ and $R$ are chirals in the bifundamental of $USp(2N)_x\times SU(2)_v$ and $USp(2N)_y\times SU(2)_v$ respectively.
We refer the reader to \cite{Pasquetti:2019hxf} for a detailed discussion of the mapping of the gauge invariant operators
in the braid duality.


The index identity corresponding to the braid  duality appeared in  \cite{2014arXiv1408.0305R} and  is given by
\begin{align} \label{flipbraidrelation}
	&\oint d \vec{z}_N \Delta_N (\vec{z};t) \mathcal{I}_{FE[USp(2N)]} (\vec{z};\vec{x};t;c) \mathcal{I}_{FE[USp(2N)]} (\vec{z};\vec{y};t;d) \prod_{n=1}^N \Gamma_e \big( (pq)^\frac{1}{2} c^{-1} d^{-1} v^{\pm 1} z_n^{\pm 1} \big) =\nonumber \\
 &= \prod_{n=1}^N \Gamma_e \big( (pq)^\frac{1}{2} d^{-1} v^{\pm 1} x_n^{\pm 1} \big) \Gamma_e \big( (pq)^\frac{1}{2} c^{-1} v^{\pm 1} y_n^{\pm 1} \big) \mathcal{I}_{FE[USp(2N)]} (\vec{x};\vec{y};t;cd).
\end{align}

From the braid duality we can flow to the duality of Figure \ref{fig:triangle} by turning on the following interaction on the l.h.s.:
\begin{align}
\mathrm{Tr}_z \left[P^+ P^- \cdot \left(\mathsf{O_H}^L+\mathsf{O_H}^R\right)\right]\,,
\end{align}
which, as we discussed in Subsection \ref{g2chir}, has the effect of breaking one combination of $U(1)_c$ and $U(1)_d$, that at the level of fugacities in the index is reflected in the constraint
\begin{align}\label{eq:sepc}
c d = (p q/t)^\frac12\,.
\end{align}
On the r.h.s.~this deformation is mapped to a VEV for the singlet $\gb_{N-1}$ of $FE[USp(2N)]$, which makes the theory flow to a simple WZ model. This was discussed in \cite{Pasquetti:2019hxf} from the field theory perspective, while the associated index identity was proven in Proposition 2.10 of \cite{2014arXiv1408.0305R}
\begin{align}
&\mathcal{I}_{FE[USp(2N)]}(\vec{x};\vec{y};t;(pq/t)^\frac12) = \prod_{i = 1}^N \prod_{j = 1}^N \Gpq{(pq/t)^\frac12 x_i^\pm y_j^\pm}\,.
\end{align}
Using this identity inside \eqref{flipbraidrelation} with the specialisation \eqref{eq:sepc} we precisely recover \eqref{eq:equal ranks}.

Now starting from the braid duality in Figure \ref{fig:braid} we take the $3d$ limit discussed in previous section.
In particular we first compactify on a circle and obtain on the l.h.s.~two 
$FE[USp(2N)]^{3d}$ blocks glued with the insertions of a doublet of chirals (not involved in the superpotential)
and on the r.h.s.~an $FE[USp(2N)]^{3d}$ block  coupled to the two sets of  chirals $L$ and $R$ via the superpotential
\eqref{spn}.

We then take the  Coulomb branch VEV that breaks all the gauge groups from $USp$ to $U$. On the l.h.s.~we get  two $FM[U(N)]$ blocks glued by the gauging of a common $U(N)$ symmetry with the insertion of one flavor (again not involved in the superpotential), while on the r.h.s.~an $FM[U(N)]$ block coupled to the two sets of  flavors  $L$, $\tilde{L}$ and $R$, $\tilde{R}$ via the superpotential terms
$L \Pi R$ and $\tilde L \tilde \Pi \tilde L$,
where $\Pi$ and $\tilde \Pi$ are in the bifundamental and anti/bifundamental representations of the $U(N)\times U(N)$
symmetries of $FM[U(N)]$ discussed in   \cite{Pasquetti:2019tix}.

Finally, we consider two further real mass deformations for what used to be the $U(1)_c$ and $U(1)_d$ symmetries in $4d$. This is done in two steps. We first give mass to the combination $U(1)_c- U(1)_d$. This has the effect on the l.h.s.~of giving mass to all of the fields of the saw and of lifting the monopole superpotential so to get the gluing of two $FT[U(N)]$, while still keeping massless the flavor at the middle $U(N)$ node. On the r.h.s., instead, we are giving mass to the flavors $L$, $\tilde{L}$ and $R$, $\tilde{R}$ while keeping the $FM[U(N)]$ block unchanged.

The last step is the real mass deformation for the remaining $U(1)_c+ U(1)_d$ symmetry. On the l.h.s.~side this is a mass deformation associated to the axial symmetry of the flavor at the middle $U(N)$ node, which results in a CS level $+1$ or $-1$, depending on the sign of the mass, for such gauge node. On the r.h.s.~the deformation is the one that we already mentioned in the previous subsection that makes $FM[U(N)]$ flow to $FT[U(N)]$, with background CS couplings for its global $U(N)$ symmetries. Hence, we recovered the duality relating the gluing of two $FT[U(N)]$ with a CS level $\pm1$ and a single $FT[U(N)]$ theory with background CS levels for the two $U(N)$ global symmetries at level $\mp1$.

Recalling the interpretation of the $T$ generator as an integer shift of the Chern-Simons level, one can interpret such duality as a $3d$ field theoretic version of the $SL(2,\mathbb{Z})$ relation $S^{-1} T S = T^{-1} S T$. 
This naturally leads us to interpret the braid relation \eqref{flipbraidrelation} as a $4d$ realisation of this $SL(2,\mathbb{Z})$ relation, which confirms our identification of $E[USp(2N)]$ or $FE[USp(2N)]$ with the $S$ element. Moreover, this result allows us to identify also the representative in $4d$ of the $T$ generator, which seems to correspond to the addition of a doublet of chiral fields. We plan to investigate this aspect further in a future work \cite{wip}.

%
%
%
%

Let us now consider the $3d$ limit of the braid duality we just described at the level of supersymmetric partition functions. We start from the index identity \eqref{flipbraidrelation} and, as in the previous subsection, we first take the limit from the $\mathbb{S}^3 \times \mathbb{S}^1$ partition function of $FE[USp(2N)]$ to the $\mathbb{S}_b^3$ partition function of the theory obtained after circle compactification. In this case, we will use the following redefinition of the parameters of \eqref{flipbraidrelation}:
\begin{align}
&z_{L/R,a}^{(n)}= \e^{2\pi irZ_{L/R,a}^{(n)}},\quad n=1,\cdots,N-1,\,\,a=1,\cdots,n, \nonumber \\
&x_i=\e^{2\pi irX_i},\quad y_i=\e^{2\pi irY_i},\quad z_i=\e^{2\pi irZ_i},  \quad i=1, \cdots,N, \nonumber \\
&c=\e^{2\pi ir \Delta_1},\quad d=\e^{2\pi ir \Delta_2} , \quad t=\e^{2\pi ir(iQ-2m_A)}, \nonumber \\
&v = \e^{2 \pi ir V}, \quad p=\e^{-2\pi rb},\quad q=\e^{-2\pi rb^{-1}}.
\end{align}

Using some of the results of the previous subsection, we obtain the following result for the $3d$ limit of \eqref{flipbraidrelation}:
\begin{align} \label{3dbraideusp}
	&C \int \frac{\prod_{i=1}^N d Z_i}{2^N N!} \frac{s_b (-i \frac{Q}{2} + 2 m_a)^N \prod_{i<j}^N s_b (-i \frac{Q}{2} \pm Z_i \pm Z_j + 2 m_A)}{\prod_{i=1}^N s_b(i\frac{Q}{2} \pm 2 Z_i) \prod_{i<j}^N s_b (i \frac{Q}{2} \pm Z_i \pm Z_j)}  \times \nonumber \\
	&\times  \mathcal{Z}_{FE[USp(2N)]^{3d}} (\vec{Z};\vec{X}; m_A; \Delta_1)  \mathcal{Z}_{FE[USp(2N)]^{3d}} (\vec{Z};\vec{Y}; m_A; \Delta_2)  \times \nonumber \\
	&\times \prod_{i=1}^N s_b (\Delta_1 + \Delta_2 \pm V \pm Z_i) = \prod_{i=1}^N s_b (\Delta_2 \pm V \pm X_i) \prod_{i=1}^N s_b (\Delta_1\pm V \pm Y_i) \nonumber \\
	&\times \mathcal{Z}_{FE[USp(2N)]^{3d}} (\vec{X};\vec{Y}; m_A; \Delta_1+\Delta_2).
\end{align}
The divergent prefactor $C$ coming from both sides of the identity 
\begin{equation}
	C = \big[ r 	\big( \e^{-2\pi rb}; \e^{-2\pi r b} \big)_\infty \big( \e^{-2\pi rb^{-1}}; \e^{-2\pi r b^{-1}} \big)_\infty\big]^\frac{N(N+1)}{2} \exp \big[-\frac{i \pi}{6 r} \big ( i \frac{Q}{2} \frac{N(N+1)}{2}\big) \big]\underset{r\to0}{\longrightarrow} 1
\end{equation}
is again trivial, as expected. 

Now we can consider the combination of Coulomb branch VEV and real mass deformation that leads us to the braid relation for the $FM[U(N)]$ theory. This is achieved by the following shift of the parameters:
\begin{equation}
Z_i \rightarrow Z_i +s,\quad Z_{L/R,a}^{(n)} \rightarrow Z_{L/R,a}^{(n}+s,\quad X_i \rightarrow X_i +s, \quad Y_i \rightarrow Y_i +s, \quad V\rightarrow V +s, \quad s \rightarrow \infty.
\end{equation}
Again using some of the results of the previous subsection, we find that \eqref{3dbraideusp} reduces to 
\begin{align} \label{3dbraidfmu}
	&K \prod_{j=1}^N s_b \big(-i \frac{Q}{2} + 2 \Delta_1 + 2(j-1) (m_a -i \frac{Q}{2}) \big) \prod_{j=1}^N s_b \big(-i \frac{Q}{2} + 2 \Delta_2 + 2(j-1) (m_a -i \frac{Q}{2}) \big) \times \nonumber \\
	&\int \frac{\prod_{i=1}^N d Z_i}{N!} \frac{s_b (-i \frac{Q}{2} + 2 m_a)^N \prod_{i<j}^N s_b (-i \frac{Q}{2} \pm (Z_i - Z_j) + 2 m_A)}{\prod_{i<j}^N s_b (i \frac{Q}{2} \pm (Z_i - Z_j))} \mathcal{Z}_{FM[U(N)]} (\vec{Z};\vec{X};m_A;\Delta_1) \times \nonumber \\
	&\times \mathcal{Z}_{FM[U(N)]} (\vec{Z};\vec{Y};m_A;\Delta_2) \prod_{i=1}^N s_b \big(\Delta_1 + \Delta_2 \pm (V- Z_i) \big) = \tilde{K} \prod_{i=1}^N s_b \big(\Delta_2 \pm (V- X_i) \big) \times \nonumber \\
	&\times \prod_{i=1}^N s_b \big(\Delta_1 \pm (V- Y_i) \big) \prod_{j=1}^N s_b \big(-i \frac{Q}{2} + 2 (\Delta_1+\Delta_2) + 2(j-1) (m_a -i \frac{Q}{2}) \big) \times \nonumber \\
	&\times  \mathcal{Z}_{FM[U(N)]} (\vec{X};\vec{Y};m_A;\Delta_1+\Delta_2).
\end{align}
The prefactors $K$ and $\tilde{K}$ obtained collecting all exponentials on each side respectively
\begin{align}
K &= \exp \big\{ 2\pi i \big[ 2N V (\Delta_1 + \Delta_2) - 2\Delta_1 \sum_{i=1}^N X_i -   2\Delta_2 \sum_{i=1}^N Y_i +\nn\\
&- (2 (N-1) m_A - iN Q) (2Ns+ \sum_{i=1}^N X_i +\sum_{i=1}^N Y_i ) \big] \big\}=\tilde{K}.
\end{align}
turn out to be equal, so that they cancel out between the two sides of the equation giving a finite result. We can also notice that in equation \eqref{3dbraidfmu} the parameter $V$ is now redundant, and can be reabsorbed into a shift of the parameters of the non-abelian global symmetries and of the gauge symmetries $X_i, \, Y_i, \, Z_i$ and $Z_{L/R,a}^{(n)}$. Hence we will set $V=0$ in the following. 

At this point, we can perform two further mass deformations that lead us to the $FT[U(N)]$ theory. First, we consider the limit
\begin{equation}
	\Delta_1 \rightarrow \Delta_1 + s, \quad \Delta_2 \rightarrow \Delta_2 - s, \quad s \rightarrow \infty,
\end{equation}
such that the combination $\Delta_1 + \Delta_2$ remains finite. In this way, the $FM[U(N)] (\vec{X};\vec{Y};m_A;\Delta_1+\Delta_2)$ on the r.h.s.~of \eqref{3dbraidfmu} remains unaffected by the limit, as well as the flavor glued to the the two $FM[U(N)] (\vec{Z};\vec{X};m_A;\Delta_1)$ and $FM[U(N)] (\vec{Z};\vec{Y};m_A;\Delta_2)$ on the l.h.s., which instead turn into $FT[U(N)]$'s. Again recycling some of the result of the previous subsection, we obtain the following identity:
\begin{align} \label{3dbraidmix}
 &\Omega \int  \frac{\prod_{i=1}^N d Z_i}{N!} \frac{s_b (-i \frac{Q}{2} + 2 m_a)^N \prod_{i<j}^N s_b (-i \frac{Q}{2} \pm (Z_i - Z_j) + 2 m_A)}{\prod_{i<j}^N s_b (i \frac{Q}{2} \pm (Z_i - Z_j))} \mathcal{Z}_{FT[U(N)]} (\vec{Z};\vec{X};m_A) \times \nonumber \\
	&\times \mathcal{Z}_{FT[U(N)]} (\vec{Z};-\vec{Y};m_A) \prod_{i=1}^N s_b \big(\Delta_1 + \Delta_2 \pm Z_i \big) = \tilde{\Omega} \prod_{j=1}^N s_b \big(-i \frac{Q}{2} + 2 (\Delta_1+\Delta_2) + 2(j-1) (m_a -i \frac{Q}{2}) \big) \times \nonumber \\
	&\times  \mathcal{Z}_{FM[U(N)]} (\vec{X};\vec{Y};m_A;\Delta_1+\Delta_2)
\end{align}
Again, it turns out that the two exponential prefactors $\Omega$ and $\tilde{\Omega}$
\begin{equation}
	\Omega = \exp \big\{ \pi i \big[  N \Delta_1^2 - N \Delta_2^2 + 2 N (\Delta_1+\Delta_2) s - \sum_{i=1}^N (X_i^2-Y_i^2) \big] \big\}=\tilde{\Omega}.
\end{equation}
coming from the l.h.s.~and the r.h.s.~respectively are the same, so that they cancel out. 

Lastly, we can take a final mass deformation to reduce also the the partition function of the $FM[U(N)]$ on the r.h.s.~to that of $FT[U(N)]$ and which has also the effect on the l.h.s.~of intergating out the flavor. This is achieved by shifting the parameters as
\begin{equation}
	\Delta_1 \rightarrow \Delta_1 +s, \quad \Delta_2 \rightarrow \Delta_2 + s, \quad s \rightarrow \infty. 
\end{equation}
Performing this last limit, equation \eqref{3dbraidmix} becomes 
\begin{align}\label{3dbraidftu}
&\e^{-i \pi \{ [N(N-1) m_A^2 - i \frac{Q}{2} m_A (N-1) + \frac{Q^2}{8} ]\} } \int \frac{\prod_{i=1}^N d Z_i}{N!} \frac{s_b (-i \frac{Q}{2} + 2 m_a)^N \prod_{i<j}^N s_b (-i \frac{Q}{2} \pm (Z_i - Z_j) + 2 m_A)}{\prod_{i<j}^N s_b (i \frac{Q}{2} \pm (Z_i - Z_j))} \times \nonumber \\
&\times e^{i \pi \sum_{i=1}^N Z_i^2}  \mathcal{Z}_{FT[U(N)]} (\vec{Z};\vec{X};m_A)  \mathcal{Z}_{FT[U(N)]} (\vec{Z};-\vec{Y};m_A) = e^{ -i \pi \sum_{i=1}^N (X_i^2+Y_i^2)} \mathcal{Z}_{FT[U(N)]} (\vec{X};\vec{Y};m_A)\,.
\end{align}
Notice that the quadratic exponentials correspond to CS terms, hence what we have found is the $\mathbb{S}^3_b$ integral identity relating the gluing of two $FT[U(N)]$ with a CS level $-1$ and a single $FT[U(N)]$ theory with background CS levels for the two $U(N)$ global symmetries at level $+1$.

Again we can translate this identity to the one written in terms of $FT[SU(N)]$ by gauging one global $U(1)$ symmetry. We first make the shift of variables
\begin{align}
\begin{aligned}
Z_{L/R,a}^{(n)} \quad &\longrightarrow \quad Z_{L/R,a}^{(n)} + v \,, \qquad n = 1, \dots, N-1 \,, \\
Z_{i} \quad &\longrightarrow \quad \tilde Z_{i}+ v \,, \\
Z_{a}^{(n)} \quad &\longrightarrow \quad Z_{a}^{(n)} + w \,, \qquad \quad n = 1, \dots, N-1 \,, \\
X_i \quad &\longrightarrow \quad \tilde X_i + w \,, \\
Y_i \quad &\longrightarrow \quad \tilde Y_i + u
\end{aligned}
\end{align}
where $Z_{a}^{(n)}$ without the subscript $L$ or $R$ in the third line is an integration variable on the right hand side. Then the factors depending on $u, \, v$ and $w$ are
\begin{align}
\label{eq:lhs}
e^{i \pi N \left[v^2+2 v (w-u)\right]}
\end{align}
on the left hand side and
\begin{align}
\label{eq:rhs}
e^{-i \pi N (w-u)^2}
\end{align}
on the right hand side. Then we gauge the anti-diagonal $U(1) \subset U(N)_X \times U(N)_Y$ by integrating over $w-u$. Namely, on the left hand side, we have
\begin{align}
&\int d(w-u) \int N dv \, e^{i \pi N \left[v^2+2 v (w-u)\right]} \int \frac{\prod_{i=1}^N d \tilde Z_i}{N!} \, \dots \nonumber \\
&= \int N dv \, e^{i \pi N v^2} \delta(N v) \int \frac{\prod_{i=1}^N d \tilde Z_i}{N!} \, \dots = \int \frac{\prod_{i=1}^N d \tilde Z_i}{N!} \, \dots \,,
\end{align}
where the remaining integrand denoted by $\dots$ is exactly in the same form as the l.h.s. of \eqref{3dbraidftu} but with $\tilde X_i, \, \tilde Y_i, \, \tilde Z_i$ instead of $X_i, \, Y_i, \, Z_i$. On the right hand side, we have
\begin{align}
\int d(w-u) \, e^{-i \pi N (w-u)^2} \dots = \sqrt{\frac{1}{i N}} \, \dots\,,
\end{align}
where $\dots$ again denotes the remaining integrand, which is in the same form as the r.h.s. of \eqref{3dbraidftu} but with $\tilde X_i, \, \tilde Y_i, \, \tilde Z_i$ instead of $X_i, \, Y_i, \, Z_i$. Thus, the duality we get is
\begin{align}
&\e^{-i \pi \{ [N(N-1) m_A^2 - i \frac{Q}{2} m_A (N-1) + \frac{Q^2}{8} ]\} } \int \frac{\prod_{i=1}^N d \tilde Z_i}{N!} \frac{s_b (-i \frac{Q}{2} + 2 m_a)^N \prod_{i<j}^N s_b (-i \frac{Q}{2} \pm (\tilde Z_i - \tilde Z_j) + 2 m_A)}{\prod_{i<j}^N s_b (i \frac{Q}{2} \pm (\tilde Z_i - \tilde Z_j))} \times \nonumber \\
&\times e^{i \pi \sum_{i=1}^N \tilde Z_i^2}  \mathcal{Z}_{FT[SU(N)]} \left(\vec{\tilde Z};\vec{\tilde X};m_A\right)  \mathcal{Z}_{FT[SU(N)]} \left(\vec{\tilde Z};-\vec{\tilde Y};m_A\right) \nonumber \\
&= \sqrt{\frac{1}{i N}} e^{ -i \pi \sum_{i=1}^N (\tilde X_i^2+\tilde Y_i^2)} \mathcal{Z}_{FT[SU(N)]} \left(\vec{\tilde X};\vec{\tilde Y};m_A\right)\,.
\end{align}

\section{Conclusions}

In this paper we investigated some new properties of the $E[USp(2N)]$ theory, or equivalently of its $FE[USp(2N)]$ variant. Specifically, we considered the gluing of two $E[USp(2N)]$ blocks by gauging a common $USp(2N)$ symmetry with the addition of one antisymmetric and $2L$ fundamental matter chiral fields. We found some dualities for the resulting theories, which we derived from iterative applications of the  Intriligator--Pouliot duality. This plays for us the role of fundamental duality, from which we derive all others.

 The case $L=0$ is somewhat special, in the sense that we showed that the resulting theory has a quantum deformed moduli space with chiral symmetry breaking and that its index takes the form of a delta-function. We interpreted it as the Identity wall which identifies the two surviving $USp(2N)$ of each $E[USp(2N)]$ block.
 
 For higher $L$, instead, we found a dual frame consisting of a linear quiver with bifundamental matter and the characteristic saw structure. For $L=0,1$ we also generalized the results to the case in which we glue two blocks of different lengths, that is a $E[USp(2M)]$ theory with a $E[USp(2N)]$ theory subjected to a deformation that breaks $USp(2N)\to USp(2M)\times SU(2)$, where $M < N$.

We then focused on the $3d$ version of some of the $4d$ dualities discussed above which now involve the $T[SU(N)]$  quiver theory, the $3d$ $S$-wall.
We  showed how these $3d$ dualities correspond to the relations $S^2=-1$, $S^{-1}S=1$ and $T^{-1} S T=S^{-1} T S$, 
where $S$ and $T$ are the $SL(2,\mathbb{Z})$ generators.
These observations lead us to conjecture that  $E[USp(2N)]$ can also be interpreted as an  $S$-wall in $4d$.

In $3d$ we can use the  Type IIB brane set-up to understand that  $T[SU(N)]$ is a domain wall interpolating between two copies of the $4d$ $\mathcal{N}=4$ $SU(N)$ Super-Yang-Mills (SYM) at two different values of the gauge coupling which are related by the $S$ action $\tau\to-1/\tau$. 
Indeed,  $T[SU(N)]$ can be engineered as a system of $N$ D3-branes suspended between D5 and NS5-branes, of which we can think as imposing boundary conditions on the SYM theory living on the $N$ D3's \cite{Gaiotto:2008ak}.

 It would be interesting to have a similar explanation for why $E[USp(2N)]$ possesses all the properties of an $S$-duality domain wall in $4d$. In this case we miss an interpretation similar to the one that we have for $T[SU(N)]$, primarily because we lack a brane realisation of the $E[USp(2N)]$ theory. It may be possible, though, to achieve a domain wall interpretation of $E[USp(2N)]$ by exploiting its E-string origin. Indeed, $E[USp(2N)]$ was originally introduced in \cite{Pasquetti:2019hxf} because it plays a crucial role in the study of the compactifications of the $6d$ $\mathcal{N}=(1,0)$ rank-$N$ E-string SCFT \cite{Ganor:1996mu,Seiberg:1996vs} on Riemann surfaces with fluxes. Specifically, the theory obtained by adding two octets of chiral fields to $E[USp(2N)]$ with some superpotential interaction corresponds to the compactification of E-string on a tube with a specific value of flux that breaks the $E_8$ global symmetry in $6d$ down to $U(1)\times E_7$. Since the compactification of E-string to $5d$ on a circle with a suitable holonomy flows to the $5d$ $\mathcal{N}=1$ $USp(2N)$ gauge theory with one antisymmetric and 8 fundamental hypers \cite{Ganor:1996pc}, we may expect that $E[USp(2N)]$ with the octet fields is a domain wall interpolating between two copies of this $5d$ gauge theory. Indeed, this was first noticed in \cite{Kim:2017toz} for the rank-one case: the $E[USp(2)]$ theory, which is just an $SU(2) \times SU(2)$ bifundamental, with two octets is a domain wall interpolating two $5d$ theories corresponding to the $6d$ rank-one E-string theory compactified on a circle with holonomies of opposite signs. This point of view may be useful for obtaining an interpretation of $E[USp(2N)]$ as an $S$-duality wall.

Another important observation that leads to interesting developments is that the duality for the gluing of two $E[USp(2N)]$ with $2L$ chirals that we derived in this paper coincides with one of the $4d$ mirror dualities of \cite{Hwang:2020wpd}, specifically the one between $E_{\gr}^{\gs}[USp(2N)]$ and $E^{\gr}_{\gs}[USp(2N)]$ with $\gr=[(L-1)^N,1^N]$ and $\gs=[N^L]$. This means that we provided a field theory derivation for this instance of mirror symmetry by iterative applications of the Intriligator--Pouliot duality. Reducing to $3d$, we can get a similar derivation for the mirror duality between $T_{\gr}^{\gs}[SU(N)]$ and $T^{\gr}_{\gs}[SU(N)]$ with $\gr=[(L-1)^N,1^N]$ and $\gs=[N^L]$ by sequentially applying the Aharony duality.

 A natural question is then whether we can derive other mirror dualities in a similar fashion, that is by iterating some more fundamental dualities. In an upcoming paper \cite{prl} we will show that this is possible in the case of linear quivers, that is for the $E_{\gr}^{\gs}[USp(2N)]$ theories in $4d$ and the $T_{\gr}^{\gs}[SU(N)]$ in $3d$. More precisely, we will present an algorithm to obtain the mirror dual of a given $E_{\gr}^{\gs}[USp(2N)]$ quiver by locally dualising the fields using two basic duality moves. These duality moves are strictly related to the dualities that we derived in this paper and
 can be interpreted as local actions of the  $S$ element of  $SL(2,\mathbb{Z})$. This will strengthen  our interpretation of $E[USp(2N)]$ as an $S$-duality domain wall.
  
%
%

\section*{Acknowledgements}

We would like to thank S.~Bajeot, S.~Benvenuti, P.~B.~Genolini, S.~S.~Razamat and D.~Tong for useful discussions. C.H. is partially supported by the STFC consolidated grants ST/P000681/1, ST/T000694/1. S.P. and M.S. are partially supported by the  INFN.   M.S. is also partially supported by the University of Milano-Bicocca grant 2016-ATESP0586 and by the MIUR-PRIN contract 2017CC72MK003.

\appendix

\section{Supersymmetric partition functions conventions}

\subsection{$\mathbb{S}^3_b$ partition function}
\label{S3bpf}

The partition function of a $3d$ $\mathcal{N}=2$ theory on the three-sphere was first computed using localization techniques in \cite{Kapustin:2009kz}. The set-up considered was that of the theory on a round sphere, namely with trivial squashing parameter $b=1$, and with canonical assignment of R-charges to the chiral fields, namely R-charge $\frac{1}{2}$. This result was later generalized in \cite{Jafferis:2010un} to the case of generic R-charges and in \cite{Hama:2010av,Hama:2011ea} to the case of a squashed sphere $\mathbb{S}^3_b$, which can be parametrized as
\be
b^2|z_1|+\frac{1}{b^2}|z_2|=1,\qquad z_1,z_2\in\mathbb{C}\,.
\ee
The result is a matrix integral with the following form:
\be
Z(\vec{m},\eta,k)=\frac{1}{|\mathcal{W}|}\int_{-\infty}^{+\infty}\prod_{a=1}^{\mathrm{rk}G}\udl{s_a}Z_{cl}\,Z_{vec}\,Z_{chir}\,.
\label{S3bpartitionfunction}
\ee
where $|\mathcal{W}|$ is the dimension of the Weyl group associated to the gauge group $G$. The different contributions to the integrand of \eqref{S3bpartitionfunction} are:
\begin{itemize}
\item the contribution from the classical action of CS and BF interactions
\be
Z_{cl}=\e^{2\pi i\eta\sum_{a=1}^{\mathrm{rk}G}s_a}\e^{-i\pi k\sum_{a=1}^{\mathrm{rk}G}s_a^2}\,,
\ee
where $\text{rk}G$ is the rank of the gauge group $G$ and we denoted with $k$ the CS level and with $\eta$ the FI parameter;
\item the contribution of the $\mathcal{N}=2$ vector multiplet
\be
Z_{vec}=\frac{1}{\prod_{\ga>0}\sbfunc{i\frac{Q}{2}+\ga(s)}},\qquad Q=b+b^{-1}\,,
\ee
where $\alpha$ are the positive roots of the gauge algebra $\frak g$ and we are using the short-hand notations
\be
\ga(s)=\prod_{a=1}^{\text{rk}G}\ga_a s_a\,;
\ee
\item the contribution of an $\mathcal{N}=2$ chiral field transforming in some representation $\mathcal{R}_G$ and $\mathcal{R}_F$ of the gauge and the flavour symmetry respectively and with $R$-charge $r$
\be
Z_{chir}=\prod_{\gr\in\mathcal{R}_G}\prod_{\tilde{\gr}\in\mathcal{R}_F}\sbfunc{i\frac{Q}{2}-\gr(s)-\tilde{\gr}(m)-i\frac{Q}{2}r},\qquad Q=b+b^{-1}\,,
\ee
where $\rho$ and $\tilde \rho$ are the weights of $\mathcal{R}_G$ and $\mathcal{R}_F$ respectively.
\end{itemize}

In all the previous definitions appeared the double-sine function that can be defined as
\be
\sbfunc{x}=\prod_{n,m=0}^\infty\frac{nb+mb^{-1}+\frac{Q}{2}-ix}{nb+mb^{-1}+\frac{Q}{2}+ix},\qquad Q=b+b^{-1}\,.
\ee

\subsection{$\mathbb{S}^3\times\mathbb{S}^1$ partition function}
\label{S3S1pf}

In this appendix we briefly summarize the basic notion of the $\mathbb{S}^3\times\mathbb{S}^1$ partition function of an $\mathcal{N}=1$ theory, which is also known as the $4d$ supersymmetric index. This coincides with the superconformal index \cite{Kinney:2005ej,Romelsberger:2005eg,Dolan:2008qi} when computed with the superconformal R-symmetry; see also \cite{Rastelli:2016tbz} for a more comprehensive review.  We follow closely the exposition of the latter reference.

The index of a $4d$ $\mathcal{N}=1$ SCFT is a refined Witten index of the theory quantized on $\mathbb{S}^3\times {\mathbb R}$,
\be
\mathcal{I}=\Tr(-1)^F e^{-\beta \delta} e^{-\mu_i \mathcal{F}_i}, \qquad  \delta=\half \left\{\mathcal{Q},\mathcal{Q}^{\dagger}\right\}\,,
\ee
where $\mathcal{Q}$ is one of the Poincar\'e supercharges, $\mathcal{Q}^{\dagger}=\mathcal{S}$ is the conjugate conformal supercharge, $\mathcal{F}_i$ are $\mathcal{Q}$-closed conserved charges and $\mu_i$ are their chemical potentials. All the states contributing to the index with non-vanishing weight have $\delta=0$, which makes the index independent of $\beta$.

For $\mathcal{N}=1$ SCFTs, the supercharges are 
\be
\left\{\mathcal{Q}_{\alpha},\,\mathcal{S}^{\alpha} = \mathcal{Q}^{\dagger\alpha}, \,\widetilde{\mathcal{Q}}_{\dot{\alpha}},\,\widetilde{\mathcal{S}}^{\dot{\alpha}} = \widetilde{\mathcal{Q}}^{\dagger\dot{\alpha}}\right\}\,,
\ee
where $\alpha=\pm$ and $\dot{\alpha}=\dot{\pm}$ are respectively the $SU(2)_1$ and $SU(2)_2$ indices of the isometry group $SO(4)=SU(2)_1 \times SU(2)_2$ of $\mathbb{S}^3$.
For definiteness, let us choose $\mathcal{Q}=\widetilde{\mathcal{Q}}_{\dot{-}}$. With this particular choice, it is common to define the index as
\be
\mathcal{I}\left(p,q\right)=\Tr(-1)^F p^{j_1 + j_2 +\half r} q^{j_2 - j_1 +\half r}\,.
\ee
where $p$ and $q$ are fugacities associated with the supersymmetry preserving squashing of the $\mathbb{S}^3$ \cite{Dolan:2008qi}. Indeed, even if the dimension of the bosonic part of the $4d$ $\mathcal{N}=1$ superconformal algebra is four, the number of independent fugacities that we can turn on in the index is two because of the constraints $\gd=0$ and $[\mathcal{F}_i,\mathcal{Q}]=0$. We should then make a choice for which combinations of the bosonic generators that satisfy these requirements to take, and we shall use $\pm j_1+j_2+\frac{R}{2}$, where $j_1$ and $j_2$ are the Cartan generators of $SU(2)_1$ and $SU(2)_2$, and $R$ is the generator of the $U(1)_R$ $R$-symmetry.  

The index counts gauge invariant operators that can be constructed from modes of the fields. The latter are usually referred to as "letters" in the literature. The single-letter index for a vector multiplet and a chiral multiplet $\chi(\mathcal{R})$ transforming in the $\mathcal{R}$ representation of the gauge and flavour group and with R-charge $R$ is
\be
i_V \left(p,q,U\right) & = & \frac{2pq-p-q}{(1-p)(1-q)} \chi_{adj}\left(U\right), \nonumber\\
i_{\chi(r)}\left(p,q,U,V\right) & = & 
\frac{(pq)^{\half R} \chi_{\mathcal{R}} \left(U,V\right) - (pq)^{\frac{2-R}{2}} \chi_{\bar{\mathcal{R}}} \left(U,V\right)}{(1-p)(1-q)}\,,
\ee
where $\chi_{\mathcal{R}} \left(U,V\right)$ and $\chi_{\bar{\mathcal{R}}} \left(U,V\right)$ are the characters of $\mathcal{R}$ and the conjugate representation $\bar{\mathcal{R}}$, with $U$ and $V$ gauge and flavour group matrices, respectively.

The index can then be obtained by symmetrizing of all of such letters into words and then projecting them to the gauge invariant ones by integrating over the Haar measure of the gauge group. This takes the general form
\be
\mathcal{I} \left(p,q,V\right)=\int \left[\udl{U}\right] \prod_{a} \PE\left[i_a\left(p,q,U,V\right)\right]\,,
\ee
where $a$ labels the different multiplets in the theory, and $\PE[i_a]$ is the plethystic exponential of the single-letter index of the $a$-th multiplet, responsible for generating the symmetrization of the letters. It is defined as
\be
\PE\left[i_a\left(p,q,U,V\right)\right] = \exp \left[  \sum_{k=1}^{\infty} \frac{1}{k} i_a\left(p^k,q^k,U^k,V^k\right) \right]\,.
\ee

For definiteness, let us discuss a specific example of the $SU(N_c)$ gauge group.  The contribution of a chiral superfield in the fundamental representation $\mathbf{N_c}$ or anti-fundamental representation $\bar{\mathbf{N_c}}$ of $SU(N_c)$ with $R$-charge $R$ can be written as follows
\be
\PE\left[i_{\chi(\mathbf{N_c})} \left(p, q,U\right)\right] &= \prod_{a=1}^{N_c} \Gamma_e \left( (pq)^{\frac{R}{2}}  z_a \right) , \quad 
\PE\left[i_{\chi(\bar{\mathbf{N_c}})}  \left(t, y,U\right)\right] = \prod_{a=1}^{N_c} \Gamma_e \left( (pq)^{\frac{R}{2}}  z^{-1}_a \right)  \,,\nn\\
\ee
where $\{z_a\}$, with $a=1,...,N_c$ and $\prod_{a=1}^{N_c} z_a=1$, are the fugacities parametrizing the Cartan subalgebra of $SU(N_c)$ and the elliptic gamma function is defined as
\be
\Gamma_e(z)\equiv \Gamma_e\left(z;p,q\right)  = \prod_{n,m=1}^\infty \frac{1-p^{n} q^{m} z^{-1}}{1- p^{n+1}  q^{m+1} z}\,.
\ee
We will also use the shorthand notation
\be
\Gamma_e \left(u z^{\pm n} \right)=\Gamma_e \left(u z^{n} \right)\Gamma_e \left(u z^{-n} \right)\,.
\ee
On the other hand, the contribution of the vector multiplet in the adjoint representation of $SU(N_c)$, together with the $SU(N_c)$ Haar measure, can be written as
\be
\frac{\kappa^{N_c-1}}{N_c!} \oint_{\mathbb{T}^{N_c}} \prod_{a=1}^{N_c-1} \frac{\udl{z}_a}{2\pi i z_a} \prod^{N_c}_{a\neq b} \frac1{\Gamma_e(z_a z_b^{-1})} \cdots\,,
\ee
where the dots denote that it will be used in addition to the full matter multiplets transforming in representations of the gauge group. The integration contour is taken over a unitary circle in the complex plane for each element of the maximal torus of the gauge group and $\kappa$ is the index of a $U(1)$ free vector multiplet defined as
\be
\kappa = (p; p)_{\infty}( q ; q)_{\infty}\,,
\ee
where we recall the definition of the $q$-Pochhammer symbol $(a;b) = \prod_{n=0}^\infty \left( 1-ab^n \right)$.  

In case of a $USp(2N_c)$ gauge group, instead, the contribution of a chiral multiplet in the fundamental representation and with R-charge $R$ is
\be
\PE\left[i_{\chi(\mathbf{N_c})} \left(p, q,U\right)\right] &= \prod_{a=1}^{N_c} \Gamma_e \left( (pq)^{\frac{R}{2}}  z_a^{\pm1} \right) \,,
\ee
while the full contribution of the vector multiplet in the adjoint representation together with the matching Haar measure and the projection to gauge singlets can be written as
\be
\frac{\kappa^{N_c}}{2^{N_c}N_c !} \oint_{\mathbb{T}^{N_c}} \prod_{a=1}^{N_c} \frac{\udl{z_i}}{2\pi i z_a} \prod^{N_c}_{a<b} \frac1{\Gamma_e(z_a^{\pm1}z_b^{\pm1})}\prod_{a=1}^{N_c} \frac1{\Gamma_e(z_a^{\pm2})}\cdots\,.
\ee

We conclude by mentioning one important property enjoyed by the elliptic gamma function that we used extensively in the main text
\be
\Gpq{x}\Gpq{pq\,x^{-1}}=1\,.
\ee
This is the manifestation in the index of the fact that two chirals $X$ and $Y$ interacting quadratically with the superpotential $\mathcal{W}=XY$ are massive and can be integrated out at low energies.


\section{The Intriligator--Pouliot duality}
\label{IP}

The Intriligator--Pouliot duality \cite{Intriligator:1995ne} relates a $USp(2N_c)$ gauge theory with $2N_f$ fundamental chirals and no superpotential $\mathcal{W}=0$ to a $USp(2N_f-2N_c-4)$ gauge theory with $2N_f$ fundamental chirals, $N_f(2N_f-1)$ singlets antisymmetric matrix $X_{ab}$ and superpotential $\hat{\mathcal{W}}=X^{ab}q_aq_b$.

This implies the following identity between the supersymmetric indices of the dual theories, which was proven in Theorem 3.1 of \cite{2003math......9252R}
\begin{align}
&\oint\udl{\vec{u}_{N_c}}\Gd_{N_c}(\vec{u}_{N_c}) \prod_{a=1}^{N_c}\prod_{i=1}^{2N_f}\Gpq{x_i u_a^{\pm1}}=\nn\\
&\qquad=\prod_{i<j}^{2N_f}\Gpq{x_ix_j}\oint\udl{\vec{u}_{N_f-N_c-2}} \Gd_{N_f-N_c-2}(\vec{u}_{N_f-N_c-2})\prod_{a=1}^{N_f-N_c-2}\prod_{i=1}^{2N_f}\Gpq{(pq)^{1/2} x_i^{-1} u_a^{\pm1}}\, ,\nn\\
\label{IP}
\end{align}
which holds provided that
\be
\prod_{i=1}^{2N_f}x_i=(pq)^{N_f-N_c-1}\,.
\label{balancingIP}
\ee

Notice that for $N_c=N$ and $N_f=N+2$ the dual theory is a WZ model of $(N+2)(2N+3)$ chiral fields and the identity \eqref{IP} reduces to \cite{10.1155/S1073792801000526}
\be
\oint\udl{\vec{u}_{N}}\Gd_{N}(\vec{u}_{N}) \prod_{a=1}^{N}\prod_{i=1}^{2N+4}\Gpq{x_iu_a^{\pm1}}=\prod_{i<j}^{2N+4}\Gpq{x_ix_j}\, ,\nn\\
\label{IPconf}
\ee
with the condition
\be
\prod_{i=1}^{2N+4}x_i=pq\, .
\label{balancingIPconf}
\ee

\section{The iterative proof of \eqref{4ddelta} for arbitrary $N$}
\label{sec:induction}

In this appendix, we are going to provide a more rigorous proof of the delta-function property \eqref{4ddelta} of the $E[USp(2 N)]$ block of arbitrary length:
\begin{align}
\label{eq:delta_app}
\mathcal{I}^{N}_{g} &= \Gpq{pq c^{\pm 2}} \oint\udl{\vec{z}_N}\Gd_N(\vec{z};pq/t)\mathcal{I}_{E[USp(2N)]}(\vec{z};\vec{x};t;c)\mathcal{I}_{E[USp(2N)]}(\vec{z};\vec{y};t;c^{-1}) \nonumber \\
&= \frac{\prod_{j=1}^N 2\pi i x_j}{\Gd_N(\vec x;t)} \sum_{\sigma \in S_N} \sum_\pm \prod_{i=1}^N \delta\left(x_i-y_{\sigma(i)}^{\pm1}\right) \,.
\end{align}
capturing all the expected poles.

\begin{figure}[t]
\centering
\includegraphics[width=\textwidth]{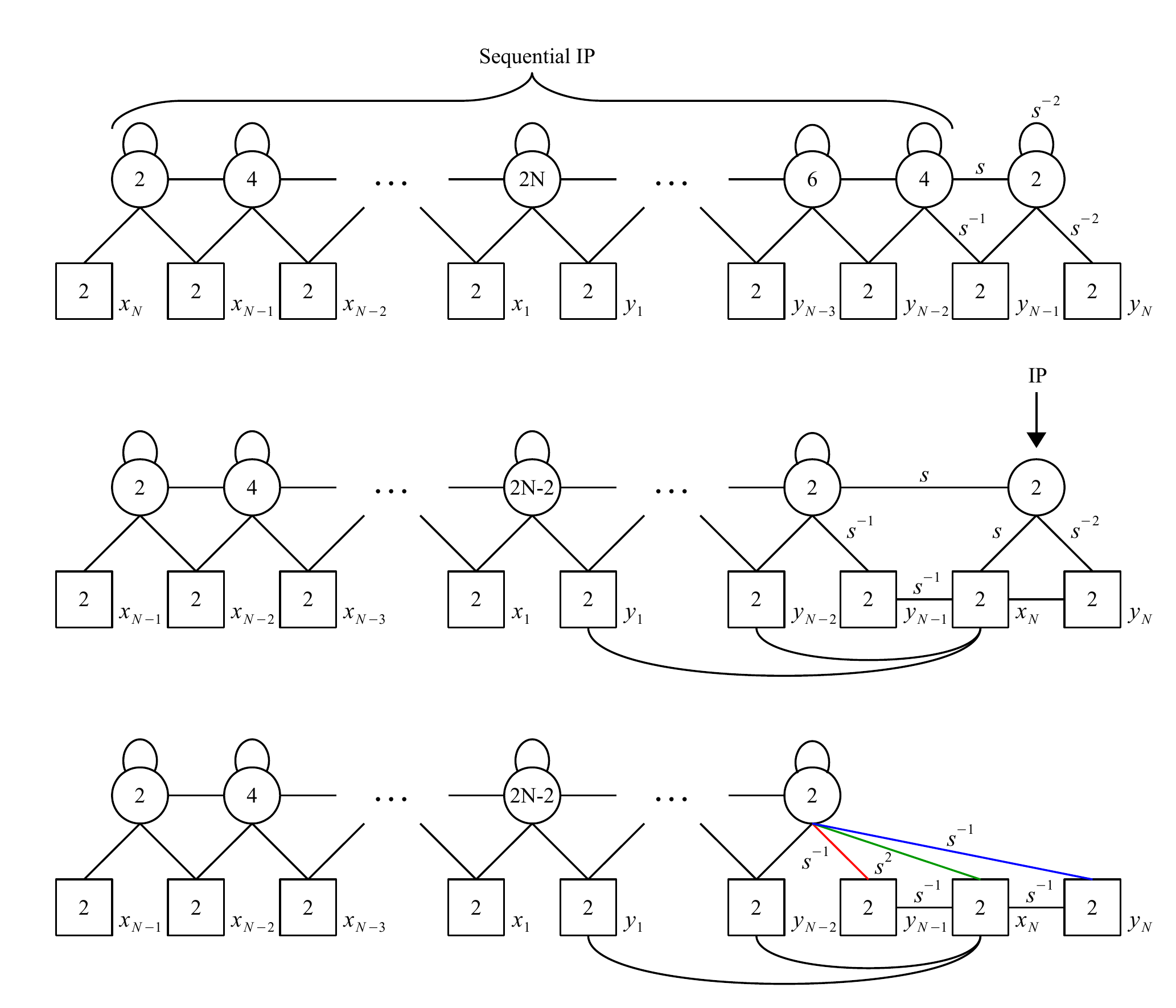}
\caption{Derivation of the delta-function property in the presence of the regulator $s$. For simplicity, we omit the gauge singlets excepts those between $SU(2)_{x_N}$ and $SU(2)_{y_i}$, which contain poles we are interested in.}
\label{fig:delta2} 
\end{figure}

The l.h.s.~of the identity \eqref{eq:delta_app} can be represented by the quiver diagram shown in Figure \ref{fig:delta1}, or equivalently, the first diagram in Figure \ref{fig:delta2} where we have used the permutation symmetry $S_N \subset USp(2 N)_x$ for later convenience so that the $\prod_{i = 1}^N SU(2)_{x_i}$ UV symmetry is embedded in the $USp(2 N)_x$ IR symmetry as shown in the figure and similarly for $\prod_{i = 1}^N SU(2)_{y_i} \subset USp(2 N)_y$.

Note that the r.h.s.~of \eqref{eq:delta_app} vanishes for generic $\vec x$ and $\vec y$, while it is singular at $x_i = y_{\sigma(i)}^{\pm1}$. Thus, we are going to examine the behavior of the l.h.s.~at $x_i = y_{\sigma(i)}^{\pm1}$ as well as at generic $\vec x$ and $\vec y$. We first focus on $x_N$ and examine the behavior of the index around $x_N = y_N^{\pm1}$. For this purpose, we apply the IP duality sequentially along the quiver from the left. Naively, this results in the last gauge node disconnected from the others because once the IP duality is applied on the second last node, its rank becomes zero and there is no remaining field between the last gauge node and the other gauge nodes of the quiver. There would have been a bifundamental field between the last node and the third last node, which is however massive and should be integrated out. However, this is not completely correct because, as we have seen in Subsubsection \ref{sec:N=2} for the $N = 2$ case, we lose some information of the poles if we blindly ignore such a massive bifundamental field. Thus, we have to be careful when we apply the IP duality on the gauge node whose dual rank is zero. For this reason, here we slightly deform the quiver by introducing a regulator $s$ as shown in Figure \ref{fig:delta2}, which has to be sent to $1$ at the end. This $s$ can be regarded as a fugacity for a fictitious $U(1)_s$ symmetry that is broken by the superpotential term
\begin{align}
\mathrm{Tr}_{1}^R \mathrm{Tr}_{2}^R A^{(1)}_R Q^{(1,2)}_RQ^{(1,2)}_R \,.
\end{align}
where $R$ in super/subscripts denotes the right $E[USp(2 N)]$ block.
In other words, the presence of $s$ means that we are considering the theory without this superpotential term. Our procedure will give us a duality for such a theory, which we will then deform restoring the aforementioned superpotential term, which is done in the index by sending $s$ to $1$.

Once we apply the IP duality sequentially until the second last node, the result is almost the same as that without $s$ except the fact that now the bifundamental field between the last gauge node and the third last is not massive anymore, as shown in the second quiver diagram in Figure \ref{fig:delta2}, because its mass term was mapped to the superpotential term we removed in the original theory, which breaks $U(1)_s$. Then we apply the IP duality on the last node and obtain the third quiver diagram with gauge singlets:
\begin{align}
\label{eq:singlet}
&\Gpq{pq s^2} \Gpq{s^{-1} x_N^{\pm1} y_N^{\pm1}} \Gpq{s^{-1} x_N^{\pm1} y_{N-1}^{\pm1}} \prod_{j = 1}^{N-2} \Gpq{x_N^{\pm1} y_j^{\pm1}} \nonumber \\
&\Gpq{p q t^{-1}} \prod_{j = 1}^{N-1} \Gpq{pq t^{-1} x_j^{\pm1} x_N^{\pm1}} \Gpq{t^{N-1} c^2 s^2} \Gpq{pq t^{-N+1} c^{-2}} 
\end{align}
where the last two factors cancel each other if we take $s \rightarrow 1$. We immediately see that this vanishes for generic $x_N$ due to the first factor, while it becomes singular when $x_N$ approaches $y_N^{\pm1}$ due to the second factor. More precisely, the first two factors give, using another form of the result \eqref{funddelta} of \cite{Spiridonov:2014cxa},
\begin{align}
\lim_{s \rightarrow 1} \Gpq{pq s^2} \Gpq{s^{-1} x_N^{\pm1} y_N^{\pm1}} = \frac{\Gpq{x_N^{\pm2}}}{(p;p)_\infty (q;q)_\infty} \left[\delta\left(X_N+Y_N\right)+\delta\left(X_N-Y_N\right)\right]\,,
\end{align}
where $x_n = e^{2 \pi i X_n}, \, y_n = e^{2 \pi i Y_n}$.
In addition, the blue and green lines in the third diagram of Figure \ref{fig:delta2}, whose index contributions are
\begin{gather}
\Gpq{(pq)^\frac12 t^{-\frac{N-1}{2}} c^{-1} s^{-1} z^{(2 N-3)}{}^{\pm1} y_N^{\pm1}}, \\
\Gpq{(pq)^\frac12 t^{\frac{N-1}{2}} c s^2 z^{(2 N-3)}{}^{\pm1} x_N^{\pm1}}
\end{gather}
respectively, become massive at $x_N = y_N^{\pm1}$ for $s \to 1$. The resulting index of the gauge part is then nothing but $\mathcal I^{N-1}_{g}(\vec x^{(N-1)};\vec y^{(N-1)})$. Thus, the entire index around $x_N = y_N^{\pm1}$ can be written as
\begin{align}
\left.\mathcal I^{N}_g (\vec x^{(N)};\vec y^{(N)})\right|_{x_N \approx y_N^{\pm1}} &= \frac{\Gpq{x_N^{\pm2}}}{(p;p)_\infty (q;q)_\infty} \Gpq{p q t^{-1}} \prod_{j = 1}^{N-1} \Gpq{pq t^{-1} x_j^{\pm1} x_N^{\pm1}} \nonumber \\
&\quad \times \prod_{j = 1}^{N-1} \Gpq{x_N^{\pm1} y_j^{\pm1}} \delta\left(X_N \mp Y_N\right) \mathcal I^{N}_{g} (\vec x^{(N-1)};\vec y^{(N-1)})\,,
\end{align}
where now for clarity we are explicitly writing the dependence on the fugacities for the non-abelian symmetries of the theory $\mathcal{T}_g$, and the factors other than $\mathcal I^{N-1}_{g}(\vec x^{(N-1)};\vec y^{(N-1)})$ come from the gauge singlet contributions \eqref{eq:singlet}.

Similarly, one can also examine the index around $x_N = y_{N-1}^{\pm1}$. In this case, the third factor in \eqref{eq:singlet} is singular and can be written as
\begin{align}
\lim_{s \rightarrow 1} \Gpq{pq s^2} \Gpq{s^{-1} x_N^{\pm1} y_{N-1}^{\pm1}} = \frac{\Gpq{x_N^{\pm2}}}{(p;p)_\infty (q;q)_\infty} \left(\delta\left(X_N+Y_{N-1}\right)+\delta\left(X_N-Y_{N-1}\right)\right)
\end{align}
where $y_{N-1} = e^{2 \pi i Y_{N-1}}$. Moreover, the green and red lines in Figure \ref{fig:delta2}, whose index contributions are
\begin{gather}
\Gpq{(pq)^\frac12 t^{\frac{N-1}{2}} c s^2 z^{(2 N-3)}{}^{\pm1} x_N^{\pm1}}, \\
\Gpq{(pq)^\frac12 t^{-\frac{N-1}{2}} c^{-1} s^{-1} z^{(2 N-3)}{}^{\pm1} y_{N-1}^{\pm1}}
\end{gather}
respectively, become massive at $x_N = y_{N-1}^{\pm1}$ for $s \to 1$. The gauge part is then given by $\mathcal I^{N-1}_{g}(\vec x_{N-1};\vec y_{N-2};y_N)$, which leads to the following index around $x_N = y_{N-1}^{\pm1}$:
\begin{align}
\left.\mathcal I^{N}_g (\vec x^{(N)};\vec y^{(N)})\right|_{x_N \approx y_{N-1}^{\pm1}} &= \frac{\Gpq{x_N^{\pm2}}}{(p;p)_\infty (q;q)_\infty} \Gpq{p q t^{-1}} \prod_{j = 1}^{N-1} \Gpq{pq t^{-1} x_j^{\pm1} x_N^{\pm1}} \nonumber \\
&\quad \times \prod_{\substack{j = 1 \\ j \neq N-1}}^{N} \Gpq{x_N^{\pm1} y_j^{\pm1}} \delta\left(X_N \mp Y_{N-1}\right) \mathcal I^{N-1}_{g} (\vec x^{(N-1)};\vec y^{(N-2)},y_N) \,.
\end{align}
On the other hand, the behavior around $x_N = y_i^{\pm1}$ for other $i$ is not manifest in this frame. Instead, we can equivalently start from the initial quiver with $y_{N-1}$ swapped with $y_i$ by the permutation symmetry $S_N \subset USp(2 N)_y$. Then we again apply the IP duality along the quiver and find the behavior of $\mathcal I^{N}_g(\vec x;\vec y)$ around $x_N = y_i^{\pm1}$ as follows:
\begin{align}
\left.\mathcal I^{N}_g (\vec x^{(N)};\vec y^{(N)})\right|_{x_N \approx y_{i}^{\pm1}} &= \frac{\Gpq{x_N^{\pm2}}}{(p;p)_\infty (q;q)_\infty} \Gpq{p q t^{-1}} \prod_{j = 1}^{N-1} \Gpq{pq t^{-1} x_j^{\pm1} x_N^{\pm1}} \nonumber \\
&\quad \times \prod_{\substack{j = 1 \\ j \neq i}}^{N} \Gpq{x_N^{\pm1} y_j^{\pm1}} \delta\left(X_N \mp Y_i\right) \mathcal I^{N-1}_{g} (\vec x^{(N-1)};\vec y^{(N)}\setminus \{y_i\}) .
\end{align}
where $y_i = e^{2 \pi i Y_i}$.

Once we combine the contributions from different singularities, we obtain the following recursive relation between $\mathcal I^{N}_g$ and $\mathcal I_g^{N-1}$:
\begin{align}
\mathcal I^{N}_g (\vec x^{(N)};\vec y^{(N)}) &= \frac{\Gpq{x_N^{\pm2}}}{(p;p)_\infty (q;q)_\infty} \Gpq{p q t^{-1}} \prod_{j = 1}^{N-1} \Gpq{pq t^{-1} x_j^{\pm1} x_N^{\pm1}} \nonumber \\
&\quad \times \sum_{i = 1}^N \prod_{\substack{j = 1 \\ j \neq i}}^{N} \Gpq{x_N^{\pm1} y_j^{\pm1}} \delta\left(X_N \pm Y_i\right)  \mathcal I^{N-1}_{g} (\vec x^{(N-1)};\vec y^{(N)}\setminus \{y_i\}) \,.
\end{align}
Recall that we have already shown the identity \eqref{eq:delta_app} is true for $N = 1$. Now if we assume this is true for $N = n-1$, i.e.
\begin{align}
\mathcal I^{n-1}_{g}(\vec x^{(n-1)};\vec y^{(n-1)}) &= \prod_{i = 1}^{n-1} \frac{\Gpq{x_i^{\pm2}}}{(p;p)_\infty (q;q)_\infty} \Gpq{p q t^{-1}}^{n-1} \prod_{i < j}^{n-1} \Gpq{pq t^{-1} x_i^{\pm1} x_j^{\pm1}} \nonumber \\
&\quad \times \sum_{\sigma \in S^{n-1}} \left(\prod_{i > j}^{n-1} \Gpq{x_i^{\pm1} y_{\sigma(j)}^{\pm1}}\right) \left(\prod_{i = 1}^{n-1} \left[\delta(X_i\pm Y_{\sigma(i)})\right]\right) \,,
\end{align}
then $\mathcal I^{n}_g(\vec x^{(n)};\vec y^{(n)})$ is given by
\begin{align}
&\mathcal I^{n}_g(\vec x^{(n)};\vec y^{(n)}) =\nonumber \\
&= \prod_{i = 1}^{n} \frac{\Gpq{x_i^{\pm2}}}{(p;p)_\infty (q;q)_\infty} \Gpq{p q t^{-1}}^{n} \prod_{i < j}^{n} \Gpq{pq t^{-1} x_i^{\pm1} x_j^{\pm1}} \nonumber \\
&\quad \times \left(\prod_{j = 1}^{n-1} \Gpq{x_n^{\pm1} y_j^{\pm1}}\right) \delta(X_n\pm Y_n) \sum_{\sigma \in S^{n-1}} \left(\prod_{i > j}^{n-1} \Gpq{x_i^{\pm1} y_{\sigma(j)}^{\pm1}}\right) \left(\prod_{i = 1}^{N-1} \left[\delta\left(X_i\pm Y_{\sigma(i)}\right)\right]\right) \nonumber \\
&\quad +(y_n \leftrightarrow y_i \text{ for } i = 1,\dots,n-1) =\nonumber \\
&= \prod_{i = 1}^{n} \frac{\Gpq{x_i^{\pm2}}}{(p;p)_\infty (q;q)_\infty} \Gpq{p q t^{-1}}^{n} \prod_{i < j}^{n} \Gpq{pq t^{-1} x_i^{\pm1} x_j^{\pm1}} \sum_{\sigma \in S^{n}} \left(\prod_{i > j}^{n} \Gpq{x_i^{\pm1} y_{\sigma(j)}^{\pm1}}\right) \left(\prod_{i = 1}^{n} \left[\delta(X_i\pm Y_{\sigma(i)})\right]\right) \nonumber \\
&=\frac{\left[\prod_{i=1}^n 2\pi i x_i\right] \left[\prod_{i=1}^n \Gpq{x_i^{\pm2}}\right] \left[\prod_{i<j}^n\Gpq{x_i^{\pm1}x_j^{\pm1}}\right]}{\left[(p;p)_\infty (q;q)_\infty\right]^n\Gpq{t}^n \left[\prod_{i<j}^n\Gpq{t\,x_i^{\pm1}x_j^{\pm1}}\right]}\sum_{\sigma \in S_n} \left(\prod_{i=1}^n \left[\delta\left(x_i-y_{\sigma(i)}\right)+\delta\left(x_i-y_{\sigma(i)}^{-1}\right)\right]\right) \,,
\label{eq:delta3}
\end{align}
where for the last equality we have used \eqref{deltachange}.
Since \eqref{eq:delta3} is exactly the r.h.s.~of identity \eqref{eq:delta_app}, by mathematical induction this proves the identity for arbitrary $N$.

\bibliographystyle{ytphys}
\bibliography{refs}

\providecommand{\href}[2]{#2}\begingroup\raggedright\begin{thebibliography}{10}

\bibitem{Hwang:2020wpd}
C.~Hwang, S.~Pasquetti, and M.~Sacchi, ``{4d mirror-like dualities},''
  \href{http://dx.doi.org/10.1007/JHEP09(2020)047}{{\em JHEP} {\bfseries 09}
  (2020) 047}, \href{http://arxiv.org/abs/2002.12897}{{\ttfamily
  arXiv:2002.12897 [hep-th]}}.

\bibitem{Gaiotto:2008ak}
D.~Gaiotto and E.~Witten, ``{S-Duality of Boundary Conditions In N=4 Super
  Yang-Mills Theory},''
  \href{http://dx.doi.org/10.4310/ATMP.2009.v13.n3.a5}{{\em Adv. Theor. Math.
  Phys.} {\bfseries 13} no.~3, (2009) 721--896},
  \href{http://arxiv.org/abs/0807.3720}{{\ttfamily arXiv:0807.3720 [hep-th]}}.

\bibitem{Hanany:1996ie}
A.~Hanany and E.~Witten, ``{Type IIB superstrings, BPS monopoles, and
  three-dimensional gauge dynamics},''
  \href{http://dx.doi.org/10.1016/S0550-3213(97)00157-0}{{\em Nucl. Phys. B}
  {\bfseries 492} (1997) 152--190},
  \href{http://arxiv.org/abs/hep-th/9611230}{{\ttfamily arXiv:hep-th/9611230}}.

\bibitem{Pasquetti:2019hxf}
S.~Pasquetti, S.~S. Razamat, M.~Sacchi, and G.~Zafrir, ``{Rank $Q$ E-string on
  a torus with flux},''
  \href{http://dx.doi.org/10.21468/SciPostPhys.8.1.014}{{\em SciPost Phys.}
  {\bfseries 8} no.~1, (2020) 014},
  \href{http://arxiv.org/abs/1908.03278}{{\ttfamily arXiv:1908.03278
  [hep-th]}}.

\bibitem{Intriligator:1996ex}
K.~A. Intriligator and N.~Seiberg, ``{Mirror symmetry in three-dimensional
  gauge theories},'' \href{http://dx.doi.org/10.1016/0370-2693(96)01088-X}{{\em
  Phys. Lett. B} {\bfseries 387} (1996) 513--519},
  \href{http://arxiv.org/abs/hep-th/9607207}{{\ttfamily arXiv:hep-th/9607207}}.

\bibitem{Aharony:2013dha}
O.~Aharony, S.~S. Razamat, N.~Seiberg, and B.~Willett, ``{3d dualities from 4d
  dualities},'' \href{http://dx.doi.org/10.1007/JHEP07(2013)149}{{\em JHEP}
  {\bfseries 07} (2013) 149}, \href{http://arxiv.org/abs/1305.3924}{{\ttfamily
  arXiv:1305.3924 [hep-th]}}.

\bibitem{Aharony:2013kma}
O.~Aharony, S.~S. Razamat, N.~Seiberg, and B.~Willett, ``{3$d$ dualities from
  4$d$ dualities for orthogonal groups},''
  \href{http://dx.doi.org/10.1007/JHEP08(2013)099}{{\em JHEP} {\bfseries 08}
  (2013) 099}, \href{http://arxiv.org/abs/1307.0511}{{\ttfamily arXiv:1307.0511
  [hep-th]}}.

\bibitem{Benvenuti:2011ga}
S.~Benvenuti and S.~Pasquetti, ``{3D-partition functions on the sphere: exact
  evaluation and mirror symmetry},''
  \href{http://dx.doi.org/10.1007/JHEP05(2012)099}{{\em JHEP} {\bfseries 05}
  (2012) 099}, \href{http://arxiv.org/abs/1105.2551}{{\ttfamily arXiv:1105.2551
  [hep-th]}}.

\bibitem{Nishioka:2011dq}
T.~Nishioka, Y.~Tachikawa, and M.~Yamazaki, ``{3d Partition Function as Overlap
  of Wavefunctions},'' \href{http://dx.doi.org/10.1007/JHEP08(2011)003}{{\em
  JHEP} {\bfseries 08} (2011) 003},
  \href{http://arxiv.org/abs/1105.4390}{{\ttfamily arXiv:1105.4390 [hep-th]}}.

\bibitem{Gulotta:2011si}
D.~R. Gulotta, C.~P. Herzog, and S.~S. Pufu, ``{From Necklace Quivers to the
  F-theorem, Operator Counting, and T(U(N))},''
  \href{http://dx.doi.org/10.1007/JHEP12(2011)077}{{\em JHEP} {\bfseries 12}
  (2011) 077}, \href{http://arxiv.org/abs/1105.2817}{{\ttfamily arXiv:1105.2817
  [hep-th]}}.

\bibitem{Assel:2014awa}
B.~Assel, ``{Hanany-Witten effect and SL(2, $\mathbb{Z}$) dualities in matrix
  models},'' \href{http://dx.doi.org/10.1007/JHEP10(2014)117}{{\em JHEP}
  {\bfseries 10} (2014) 117}, \href{http://arxiv.org/abs/1406.5194}{{\ttfamily
  arXiv:1406.5194 [hep-th]}}.

\bibitem{Seiberg:1994bz}
N.~Seiberg, ``{Exact results on the space of vacua of four-dimensional SUSY
  gauge theories},'' \href{http://dx.doi.org/10.1103/PhysRevD.49.6857}{{\em
  Phys. Rev. D} {\bfseries 49} (1994) 6857--6863},
  \href{http://arxiv.org/abs/hep-th/9402044}{{\ttfamily arXiv:hep-th/9402044}}.

\bibitem{Spiridonov:2014cxa}
V.~P. Spiridonov and G.~S. Vartanov, ``{Vanishing superconformal indices and
  the chiral symmetry breaking},''
  \href{http://dx.doi.org/10.1007/JHEP06(2014)062}{{\em JHEP} {\bfseries 06}
  (2014) 062}, \href{http://arxiv.org/abs/1402.2312}{{\ttfamily arXiv:1402.2312
  [hep-th]}}.

\bibitem{prl}
C.~Hwang, S.~Pasquetti, and M.~Sacchi, ``{Rethinking mirror symmetry as a local
  duality on fields},'' \href{http://arxiv.org/abs/2110.11362}{{\ttfamily
  arXiv:2110.11362 [hep-th]}}.

\bibitem{Intriligator:1995ne}
K.~A. Intriligator and P.~Pouliot, ``{Exact superpotentials, quantum vacua and
  duality in supersymmetric SP(N(c)) gauge theories},''
  \href{http://dx.doi.org/10.1016/0370-2693(95)00618-U}{{\em Phys. Lett. B}
  {\bfseries 353} (1995) 471--476},
  \href{http://arxiv.org/abs/hep-th/9505006}{{\ttfamily arXiv:hep-th/9505006}}.

\bibitem{Garozzo:2020pmz}
I.~Garozzo, N.~Mekareeya, M.~Sacchi, and G.~Zafrir, ``{Symmetry enhancement and
  duality walls in 5d gauge theories},''
  \href{http://dx.doi.org/10.1007/JHEP06(2020)159}{{\em JHEP} {\bfseries 06}
  (2020) 159}, \href{http://arxiv.org/abs/2003.07373}{{\ttfamily
  arXiv:2003.07373 [hep-th]}}.

\bibitem{Hwang:2021xyw}
C.~Hwang, S.~S. Razamat, E.~Sabag, and M.~Sacchi, ``{Rank $Q$ E-String on
  Spheres with Flux},'' \href{http://arxiv.org/abs/2103.09149}{{\ttfamily
  arXiv:2103.09149 [hep-th]}}.

\bibitem{Pasquetti:2019tix}
S.~Pasquetti and M.~Sacchi, ``{3d dualities from 2d free field correlators:
  recombination and rank stabilization},''
  \href{http://dx.doi.org/10.1007/JHEP01(2020)061}{{\em JHEP} {\bfseries 01}
  (2020) 061}, \href{http://arxiv.org/abs/1905.05807}{{\ttfamily
  arXiv:1905.05807 [hep-th]}}.

\bibitem{Romelsberger:2005eg}
C.~Romelsberger, ``{Counting chiral primaries in N = 1, d=4 superconformal
  field theories},''
  \href{http://dx.doi.org/10.1016/j.nuclphysb.2006.03.037}{{\em Nucl. Phys. B}
  {\bfseries 747} (2006) 329--353},
  \href{http://arxiv.org/abs/hep-th/0510060}{{\ttfamily arXiv:hep-th/0510060}}.

\bibitem{Kinney:2005ej}
J.~Kinney, J.~M. Maldacena, S.~Minwalla, and S.~Raju, ``{An Index for 4
  dimensional super conformal theories},''
  \href{http://dx.doi.org/10.1007/s00220-007-0258-7}{{\em Commun. Math. Phys.}
  {\bfseries 275} (2007) 209--254},
  \href{http://arxiv.org/abs/hep-th/0510251}{{\ttfamily arXiv:hep-th/0510251}}.

\bibitem{Dolan:2008qi}
F.~A. Dolan and H.~Osborn, ``{Applications of the Superconformal Index for
  Protected Operators and q-Hypergeometric Identities to N=1 Dual Theories},''
  \href{http://dx.doi.org/10.1016/j.nuclphysb.2009.01.028}{{\em Nucl. Phys. B}
  {\bfseries 818} (2009) 137--178},
  \href{http://arxiv.org/abs/0801.4947}{{\ttfamily arXiv:0801.4947 [hep-th]}}.

\bibitem{Rastelli:2016tbz}
L.~Rastelli and S.~S. Razamat, ``{The supersymmetric index in four
  dimensions},'' \href{http://dx.doi.org/10.1088/1751-8121/aa76a6}{{\em J.
  Phys. A} {\bfseries 50} no.~44, (2017) 443013},
  \href{http://arxiv.org/abs/1608.02965}{{\ttfamily arXiv:1608.02965
  [hep-th]}}.

\bibitem{2014arXiv1408.0305R}
E.~M. {Rains}, ``{Multivariate Quadratic Transformations and the Interpolation
  Kernel},'' {\em arXiv e-prints} (Aug, 2014) arXiv:1408.0305,
  \href{http://arxiv.org/abs/1408.0305}{{\ttfamily arXiv:1408.0305 [math.CA]}}.

\bibitem{Aprile:2018oau}
F.~Aprile, S.~Pasquetti, and Y.~Zenkevich, ``{Flipping the head of $T[SU(N)]$:
  mirror symmetry, spectral duality and monopoles},''
  \href{http://dx.doi.org/10.1007/JHEP04(2019)138}{{\em JHEP} {\bfseries 04}
  (2019) 138}, \href{http://arxiv.org/abs/1812.08142}{{\ttfamily
  arXiv:1812.08142 [hep-th]}}.

\bibitem{Aharony:1997gp}
O.~Aharony, ``{IR duality in d = 3 N=2 supersymmetric USp(2N(c)) and U(N(c))
  gauge theories},''
  \href{http://dx.doi.org/10.1016/S0370-2693(97)00530-3}{{\em Phys. Lett. B}
  {\bfseries 404} (1997) 71--76},
  \href{http://arxiv.org/abs/hep-th/9703215}{{\ttfamily arXiv:hep-th/9703215}}.

\bibitem{Giacomelli:2020ryy}
S.~Giacomelli, N.~Mekareeya, and M.~Sacchi, ``{New aspects of Argyres--Douglas
  theories and their dimensional reduction},''
  \href{http://dx.doi.org/10.1007/JHEP03(2021)242}{{\em JHEP} {\bfseries 03}
  (2021) 242}, \href{http://arxiv.org/abs/2012.12852}{{\ttfamily
  arXiv:2012.12852 [hep-th]}}.

\bibitem{Gaiotto:2012xa}
D.~Gaiotto, L.~Rastelli, and S.~S. Razamat, ``{Bootstrapping the superconformal
  index with surface defects},''
  \href{http://dx.doi.org/10.1007/JHEP01(2013)022}{{\em JHEP} {\bfseries 01}
  (2013) 022}, \href{http://arxiv.org/abs/1207.3577}{{\ttfamily arXiv:1207.3577
  [hep-th]}}.

\bibitem{Kapustin:1999ha}
A.~Kapustin and M.~J. Strassler, ``{On mirror symmetry in three-dimensional
  Abelian gauge theories},''
  \href{http://dx.doi.org/10.1088/1126-6708/1999/04/021}{{\em JHEP} {\bfseries
  04} (1999) 021}, \href{http://arxiv.org/abs/hep-th/9902033}{{\ttfamily
  arXiv:hep-th/9902033}}.

\bibitem{Pasquetti:2019uop}
S.~Pasquetti and M.~Sacchi, ``{From 3$d$ dualities to 2$d$ free field
  correlators and back},''
  \href{http://dx.doi.org/10.1007/JHEP11(2019)081}{{\em JHEP} {\bfseries 11}
  (2019) 081}, \href{http://arxiv.org/abs/1903.10817}{{\ttfamily
  arXiv:1903.10817 [hep-th]}}.

\bibitem{Fateev:2007qn}
V.~A. Fateev and A.~V. Litvinov, ``{Multipoint correlation functions in
  Liouville field theory and minimal Liouville gravity},''
  \href{http://dx.doi.org/10.1007/s11232-008-0038-3}{{\em Theor. Math. Phys.}
  {\bfseries 154} (2008) 454--472},
  \href{http://arxiv.org/abs/0707.1664}{{\ttfamily arXiv:0707.1664 [hep-th]}}.

\bibitem{Kapustin:2009kz}
A.~Kapustin, B.~Willett, and I.~Yaakov, ``{Exact Results for Wilson Loops in
  Superconformal Chern-Simons Theories with Matter},''
  \href{http://dx.doi.org/10.1007/JHEP03(2010)089}{{\em JHEP} {\bfseries 03}
  (2010) 089}, \href{http://arxiv.org/abs/0909.4559}{{\ttfamily arXiv:0909.4559
  [hep-th]}}.

\bibitem{Jafferis:2010un}
D.~L. Jafferis, ``{The Exact Superconformal R-Symmetry Extremizes Z},''
  \href{http://dx.doi.org/10.1007/JHEP05(2012)159}{{\em JHEP} {\bfseries 05}
  (2012) 159}, \href{http://arxiv.org/abs/1012.3210}{{\ttfamily arXiv:1012.3210
  [hep-th]}}.

\bibitem{Hama:2010av}
N.~Hama, K.~Hosomichi, and S.~Lee, ``{Notes on SUSY Gauge Theories on
  Three-Sphere},'' \href{http://dx.doi.org/10.1007/JHEP03(2011)127}{{\em JHEP}
  {\bfseries 03} (2011) 127}, \href{http://arxiv.org/abs/1012.3512}{{\ttfamily
  arXiv:1012.3512 [hep-th]}}.

\bibitem{Hama:2011ea}
N.~Hama, K.~Hosomichi, and S.~Lee, ``{SUSY Gauge Theories on Squashed
  Three-Spheres},'' \href{http://dx.doi.org/10.1007/JHEP05(2011)014}{{\em JHEP}
  {\bfseries 05} (2011) 014}, \href{http://arxiv.org/abs/1102.4716}{{\ttfamily
  arXiv:1102.4716 [hep-th]}}.

\bibitem{Fredenhagen:2004cj}
S.~Fredenhagen and V.~Schomerus, ``{Boundary Liouville theory at c = 1},''
  \href{http://dx.doi.org/10.1088/1126-6708/2005/05/025}{{\em JHEP} {\bfseries
  05} (2005) 025}, \href{http://arxiv.org/abs/hep-th/0409256}{{\ttfamily
  arXiv:hep-th/0409256}}.

\bibitem{Tong:2000ky}
D.~Tong, ``{Dynamics of N=2 supersymmetric Chern-Simons theories},''
  \href{http://dx.doi.org/10.1088/1126-6708/2000/07/019}{{\em JHEP} {\bfseries
  07} (2000) 019}, \href{http://arxiv.org/abs/hep-th/0005186}{{\ttfamily
  arXiv:hep-th/0005186}}.

\bibitem{wip}
C.~Hwang, S.~Pasquetti, and M.~Sacchi {\em work in progress} .

\bibitem{Ganor:1996mu}
O.~J. Ganor and A.~Hanany, ``{Small E(8) instantons and tensionless noncritical
  strings},'' \href{http://dx.doi.org/10.1016/0550-3213(96)00243-X}{{\em Nucl.
  Phys. B} {\bfseries 474} (1996) 122--140},
  \href{http://arxiv.org/abs/hep-th/9602120}{{\ttfamily arXiv:hep-th/9602120}}.

\bibitem{Seiberg:1996vs}
N.~Seiberg and E.~Witten, ``{Comments on string dynamics in six-dimensions},''
  \href{http://dx.doi.org/10.1016/0550-3213(96)00189-7}{{\em Nucl. Phys. B}
  {\bfseries 471} (1996) 121--134},
  \href{http://arxiv.org/abs/hep-th/9603003}{{\ttfamily arXiv:hep-th/9603003}}.

\bibitem{Ganor:1996pc}
O.~J. Ganor, D.~R. Morrison, and N.~Seiberg, ``{Branes, Calabi-Yau spaces, and
  toroidal compactification of the N=1 six-dimensional E(8) theory},''
  \href{http://dx.doi.org/10.1016/S0550-3213(96)00690-6}{{\em Nucl. Phys. B}
  {\bfseries 487} (1997) 93--127},
  \href{http://arxiv.org/abs/hep-th/9610251}{{\ttfamily arXiv:hep-th/9610251}}.

\bibitem{Kim:2017toz}
H.-C. Kim, S.~S. Razamat, C.~Vafa, and G.~Zafrir, ``{E-String Theory on Riemann
  Surfaces},'' \href{http://dx.doi.org/10.1002/prop.201700074}{{\em Fortsch.
  Phys.} {\bfseries 66} no.~1, (2018) 1700074},
  \href{http://arxiv.org/abs/1709.02496}{{\ttfamily arXiv:1709.02496
  [hep-th]}}.

\bibitem{2003math......9252R}
E.~M. {Rains}, ``{Transformations of elliptic hypergometric integrals},'' {\em
  arXiv Mathematics e-prints} (Sep, 2003) math/0309252,
  \href{http://arxiv.org/abs/math/0309252}{{\ttfamily arXiv:math/0309252
  [math.QA]}}.

\bibitem{10.1155/S1073792801000526}
J.~F. van Diejen and V.~P. Spiridonov, ``{Elliptic Selberg integrals},'' {\em
  International Mathematics Research Notices} {\bfseries 2001} no.~20, (01,
  2001) 1083--1110.

\end{thebibliography}\endgroup

\end{document}